\documentclass[twocolumn, twocolappendix]{aastex701}

\usepackage{amsmath}
\usepackage{csquotes}
\usepackage{multirow}

\newcommand{\vect}[1]{\boldsymbol{#1}}
\newcommand{\kms}{km s$^{-1}$}

\usepackage{xcolor}
\definecolor{kcolor}{RGB}{150,0,250}

\shorttitle{LMC Halo-Disk Interaction}
\shortauthors{Rathore et al.}

\begin{document}

\title{The Role of Dark Matter in Driving Disk Perturbations: Application to LMC-SMC Like Galaxy Interactions} 

\author[orcid=0009-0009-0158-585X]{Himansh Rathore}
\affiliation{Department of Astronomy and Steward Observatory, University of Arizona, 933 North Cherry Avenue, Tucson, AZ 85721, USA}
\email[show]{himansh@arizona.edu}

\author[orcid=0000-0003-1183-701X]{Hayden R. Foote}
\affiliation{Department of Astronomy and Steward Observatory, University of Arizona, 933 North Cherry Avenue, Tucson, AZ 85721, USA}
\email[]{haydenfoote@arizona.edu}

\author[0000-0003-0715-2173]{Gurtina Besla}
\affiliation{Department of Astronomy and Steward Observatory, University of Arizona, 933 North Cherry Avenue, Tucson, AZ 85721, USA}
\email[]{gbesla@arizona.edu}

\author[0000-0003-3213-8736]{Silvio Varela-Lavin}
\affiliation{Departamento de F\'isica, Universidad Tecnica Federico Santa Maria, Avenida Espa\~na 1680, Valpara\'iso, Chile}
\email[]{silvio.varela@userena.cl}

\author[orcid=0000-0001-7107-1744, sname=Garavito-Camargo, gname=Nicol\'as]{Nicol\'as Garavito-Camargo}
\thanks{NASA Einstein Fellow}
\affil{Department of Astronomy, University of Maryland, 4296 Stadium Drive,
College Park, MD 20742-2421}
\affil{Department of Astronomy and Steward Observatory, University of Arizona, 933 North Cherry Avenue, Tucson, AZ 85721, USA}
\affil{Instituto de Astrof\'isica, Pontificia Universidad Cat\'olica de Chile, Av. Vicu\~na Mackenna 4860, 782-0436 Macul, Santiago, Chile}

\email{garavito@umd.edu}

\author[orcid=0000-0003-2660-2889, sname=Weinberg, gname=Martin]{Martin D. Weinberg}
\affil{Department of Astronomy, University of Massachusetts, Amherst, MA 01003-9305, USA}
\email{weinberg@astro.umass.edu}

\author[orcid=0000-0003-4232-8584, sname=G\'omez, gname=Facundo]{Facundo A. G\'omez}
\affil{Departamento de Astronom\'ia, Universidad de La Serena, Av. Raúl Bitrán 1305, La Serena, Chile}
\email{fagomez@userena.cl}

\author[orcid=0000-0003-2594-8052, sname=Daniel, gname=Kathryne]{Kathryne J. Daniel}
\affil{Department of Astronomy and Steward Observatory, University of Arizona, 933 North Cherry Avenue, Tucson, AZ 85721, USA}
\email{kjdaniel@arizona.edu}

\author[0000-0003-3922-7336]{Chervin F. P. Laporte}
\affiliation{LIRA, Observatoire de Paris, Universit\'e PSL, Sorbonne Universit\'e, Universit\'e Paris Cit\'e, CY Cergy Paris Universit\'e, CNRS, 92190 Meudon, France}
\affiliation{Institut de Ciencies del Cosmos (ICCUB), Universitat de Barcelona (IEEC-UB), Martí i Franquès 1, E-08028 Barcelona, Spain}
\affiliation{Kavli IPMU (WPI), UTIAS, The University of Tokyo, Kashiwa, Chiba 277-8583, Japan}
\email{Chervin.Laporte@obspm.fr}
\collaboration{all}{The EXP collaboration}

\begin{abstract}
Dark-matter (DM) models predict that satellite galaxies distort their host's DM halo. As such, both the satellite and the distorted halo can perturb the host's stellar disk. The LMC-SMC system is a nearby $\sim1:10$~mass-ratio binary with a well-constrained orbit and well-characterized LMC disk perturbations. Hence, the system offers unique opportunities to test DM theory, provided the impact of halo torques on the LMC's disk can be characterized along with the SMC's torques. We use halo Basis Function Expansions (BFEs) of an N-body simulation of an isolated LMC-SMC-like interaction to quantify the time-dependent torques exerted on the LMC's disk by the SMC and the LMC's DM halo. We find that the halo torques arise from the quadrupole halo distortion, and the inner (R$<6$~kpc) LMC disk warps are the most promising probe of this quadrupole. For the first time, we construct a BFE to quantify the LMC's disk perturbations resulting from the SMC and halo torques. We find that the LMC's disk is significantly perturbed prior to the Clouds' Milky Way infall, possessing warps with a mean vertical extent of $\sim1$~kpc. Finally, we apply multi-channel singular spectral analysis (mSSA) to the time-series of the disk BFE coefficients, halo quadrupole, and the LMC-SMC orbit. mSSA successfully separates the temporal evolution of the LMC warps into components correlated with the halo versus the SMC. Our framework identifies correlated evolution between the LMC's halo distortions and the LMC's disk perturbations, enabling the Clouds to be a precision laboratory for DM physics.
\end{abstract}
\keywords{\href{https://astrothesaurus.org/uat/903}{LMC (903)}; \href{https://astrothesaurus.org/uat/1468}{SMC (1468)}; \href{https://astrothesaurus.org/uat/1880}{Dark matter halos (1880)}; \href{https://uat.astrothesaurus.org/uat/600}{Galaxy interactions (600)}; \href{https://uat.astrothesaurus.org/uat/1965}{Computational methods (1965)}}

\section{Introduction} \label{sec:intro}
Dark matter (DM) theory predicts that satellite galaxies distort their host's DM halo from spherical symmetry \citep[e.g.][]{Chandrasekhar1943, White1983, TW1984, Choi2009, Sridhar2018, GC2019, Banik2021, Banik2022}. Hence, the stellar disk of the host can be affected by both the satellite's direct gravitational influence and the gravitational influence of the satellite-induced halo distortions \citep[e.g.][]{Weinberg1998, Vesperini2000, Gomez2016, Laporte2018, Poggio2021, Varela-Lavin2023, Wille2024, Hunt2026}. The LMC and SMC (\enquote{Clouds}) are the closest example of a $\sim1:10$~mass-ratio interacting galaxy pair \citep[e.g.][]{DB2012, Besla2012, Pardy2016, Lucchini2024, Arranz2024, Garver2026} with a reasonably well-understood LMC-SMC orbit \citep{Patel2020} and well-characterized LMC stellar disk perturbations \citep[e.g.][]{Olsen2002, Subramaniam2009, Haschke2012, Besla2016, Choi2018a, Choi2018b, Belokurov2019, Cusano2021, Saroon2022, Rathore2025a, Arranz2025, Oden2025}. In a companion paper \citep[][hereafter F26]{Foote2026}, we demonstrated that significant DM distortions (up to $60\%$ of the unperturbed DM density field) are expected in the LMC's halo due to the SMC. As such, the Clouds are a novel testbed for DM theory, provided the influence of the distorted LMC halo on the LMC stellar disk can be characterized along with the SMC's direct influence. Our goal is to build a theoretical framework based on N-body simulations that enables a rigorous understanding of how a primary galaxy's stellar disk is affected by torques from both satellite-induced halo distortions and the satellite directly. Ultimately, we aim to use this new framework to quantify the impact of DM distortions and identify signatures in the LMC disk evolution that can directly probe Cold Dark Matter (CDM) theory.

Early works by e.g. \cite{Weinberg1998, Weinberg1998a,
Vesperini2000} characterized the satellite-induced distortions in a primary's halo through a combination of linear perturbation theory \citep[e.g.][]{Kalnajs1977, Weinberg1991} and basis function expansions (BFEs) \citep[e.g.][]{CluttonBrock73, Ostriker1992, Earn1996, Weinberg1999, Lowing2011, GC2021}. They showed that the resulting non-spherical halo potential can drive warps with an observable amplitude in the primary's disk. However, perturbative techniques have limited applicability in regimes where perturbations (like the LMC halo distortions) can affect the perturbing agents (like the SMC, F26). In particular, the SMC's tidal truncation and mass loss need to be accounted for as it orbits the LMC.
 
Works by e.g. \cite{Gomez2016, Gomez2017, Gomez2021, Varela-Lavin2023, GarciaConde2024} used cosmological simulations to characterize the influence of the satellite-induced halo distortions on the primary's disk. Although cosmological simulations self-consistently solve Poisson's equation for the density and potential field, interpreting their results is challenging given the plethora of other effects that can simultaneously disturb a primary galaxy's disk (e.g., hierarchical assembly and baryonic processes). 

Surveys like Gaia \citep{Luri2021}, DECam-SMASH \citep{Nidever2017}, and SDSS \citep{Nidever2026} have enabled an increasingly precise quantification of the disturbed morphology and internal kinematics of the Clouds. So, to use the LMC-SMC system as a testbed for DM theory, we need frameworks that can be applied to {\em observationally tailored} simulations of the Clouds. In this work, we use BFEs of a high-resolution N-body simulation of an LMC-SMC-like interacting system to quantify the time-dependent torques applied to the LMC's disk by both the LMC's distorted DM halo and the SMC's direct gravitational force. 

Using halo BFEs, F26 quantified the SMC-induced LMC halo distortions through different spatial modes \citep[e.g.][]{GC2021, Lilleengen2023, Brooks2025, Darragh-Ford2025, Arora2026}. We found that the amplitude of halo distortions decays as a power law with increasing spatial multipole order. In particular, the dipole and quadrupole distortions are the most dominant. The dipole distortion arises from the LMC's response to the changing LMC-SMC barycenter, and the quadrupole response includes the SMC's dynamical friction wake in addition to the density perturbations in the inner ($R < 20$ kpc) halo. In this work, we will identify which of these modes in the DM halo contribute towards disturbing the LMC's stellar disk.

We will investigate the LMC's simulated, time-dependent stellar morphology using disk BFEs. Traditionally, the disturbed morphology of a disk is quantified using spatial Fourier decomposition in radial bins \citep[e.g.][]{AM2002, Khachaturyants2022, Lucey2023, Silva2023, Grand2023, Rathore2025b, Dolfi2026}. In Fourier-based methods, the in-plane and vertical perturbations of a disk need to be analyzed separately. Consequently, Fourier methods cannot encode the disk density and potential fluctuations in a self-consistent way. BFEs allow for a simultaneous analysis of the vertical and in-plane perturbations, where the resulting density and potential perturbations are self-consistent by design. However, it is challenging to identify a disk basis that is flexible enough to capture the diverse range of time-dependent in-plane and vertical morphological perturbations expected during galaxy interactions and collisions, with a reasonable number of expansion terms. In this work, we will construct a novel basis that can adequately encode the strongly perturbed LMC disk morphology in a high-resolution LMC-SMC N-body simulation that tracks their interaction history over $\approx 7$ Gyr. As such, this basis would provide a new means to understand the disk evolution in interacting galaxies.

To identify correlations between the LMC's halo distortions, SMC's orbit, and the LMC stellar disk's temporal response, we will use an unsupervised machine learning algorithm called multi-channel singular spectral analysis (mSSA, \citealt{Ghil1991, Golyandina2013, Weinberg2021}). mSSA decomposes a collection of time series into correlated principal components. mSSA has been used successfully to understand dynamical processes that drive the secular evolution of an isolated MW-like barred disk \citep[e.g.][]{Johnson2023, Hunt2026} and response of the MW halo to satellite accretion \citep[e.g.][]{Arora2025}. To the best of our knowledge, mSSA has not been applied to analyze disk perturbations in interacting galaxy systems before.

Following F26, we exclude the MW in this analysis in order to isolate the impact of LMC-SMC interactions on the LMC's halo and disk. Excluding the MW further allows us to place our results in context with other $\sim1:10$ mass ratio interacting galaxy pairs in the local universe. Such interacting systems are expected to be relatively common in isolation \citep[e.g.][]{Zaritsky1993, Zaritsky1997, Besla2018, Chamberlain2024}, whereas an LMC-SMC-MW-like system is rare \citep{Boylan-Kolchin2011}. Our goal is not to provide an exact match to the present-day LMC stellar disk, which requires both the influence of the MW and a recent direct collision between the LMC and SMC \citep{Choi2022}. Instead, we aim to: (i) establish a generic framework to characterize halo torques and satellite torques that can also be applied to the LMC-SMC-MW system; and (ii) understand the morphology of the LMC stellar disk prior to its infall into the MW, which is a critical initial condition to simulating the actual system.

The manuscript is organized as follows. Section \ref{sec:methods} describes: the N-body simulation setup; the procedure for constructing the LMC halo and disk BFEs; and the mSSA algorithm. Section \ref{sec:results} presents our framework, where we compute the torques applied by the LMC's distorted halo and the SMC on the LMC's disk, quantify the perturbations in the LMC disk with BFE coefficients, and infer dynamical correlations between the LMC-SMC orbit, LMC's halo distortions and the LMC's disk perturbations. In section \ref{sec:discussion}, we will discuss: implications for the pre-MW-infall and present day state of the observed Clouds; and avenues for using the Clouds as a testbed for DM theory. Throughout this manuscript, bold-italic mathematical font denotes vector quantities. Unit vectors are denoted by a hat symbol ($\mathbf{\hat{}}$) over a vector.

\section{Methods} \label{sec:methods}
This section presents the strategy employed to build a framework that can disentangle the influence of the SMC-induced LMC halo distortions from the SMC's direct influence on the LMC's stellar disk. First, we build a high resolution N-body simulation of the LMC-SMC interaction history in isolation over a $\approx 7$ Gyr timeline (section \ref{sec:sims}). The SMC-induced LMC halo distortions in the simulation will be characterized using halo BFEs (section \ref{sec:halo_BFEs}), which will enable us to compute the torques applied by the halo distortions on the LMC disk, and distinguish these halo torques from the SMC's torques. Then, a disk BFE will be developed to understand the perturbations in the LMC disk resulting from the combination of halo and SMC torques (section \ref{sec:disk_bfe}), which is also a novelty of this work as a suitable basis for a highly distorted disk has not been well-explored. Finally, the disk perturbations will be linked to the distorted halo and the SMC using mSSA (section \ref{sec:mssa_method}), enabling us to disentangle the influence of the halo from the SMC in driving the LMC disk's temporal evolution.

\subsection{N-body Simulation of the LMC-SMC Interaction}
\label{sec:sims}

In this section, we present a fiducial N-body simulation of the isolated LMC-SMC interaction. This simulation is part of the \texttt{MEGHA}\footnote{Megha is the Sanskrit word for \enquote{Clouds}.} simulations of the isolated LMC-SMC and LMC-SMC-MW interaction history (H. Rathore et al. 2026b {\em in prep}). The same isolated LMC-SMC simulation used here was also used in the companion study by F26. Further, the \texttt{MEGHA} simulations are motivated by the models developed by \cite{Besla2012} (hereafter B12).

\subsubsection{Initial LMC and SMC N-body Models}

The LMC and SMC are modeled with live DM halos and live stellar disks. The initial conditions for the LMC and SMC galaxies were generated using the code \texttt{MakeGalaxy}, which is a proprietary version of the publicly available code \texttt{MakeNewDisk}. \texttt{MakeGalaxy} approximates the phase space distribution of the DM and star particles through the moments of the collisionless Boltzmann equation \citep{Hernquist1993}.

The initial ($t = 0$) LMC and SMC DM halos are modeled with Hernquist profiles \citep{Hernquist1990}, with a halo mass of $1.76\times10^{11}$ M$_\odot$ and $1.995\times10^{10}$ M$_\odot$ respectively, and a halo scale radius of $21.4$~kpc and $7.3$~kpc respectively. The halos are initialized with velocity anisotropy $\beta = 0$.  The halo mass resolution is $4\times10^4$ M$_\odot$ per particle, and the halo gravitational softening length is $0.08$ kpc. The halo mass resolution is sufficient to resolve the SMC-induced density distortions in the LMC halo \citep{Weinberg1998}. Note that our halo resolution is more than an order of magnitude larger than that of B12 ($\sim10^6$ M${_\odot}$). A high DM resolution such as used here is necessary to avoid spurious warping of the disk due to Poisson fluctuations in the halo density field \citep{Weinberg1998, Sellwood2024}. The halo softening length is chosen to be the same as the value adopted in \cite{GC2019}.

The initial halo parameter choices yield halo properties that are reasonably consistent with the present day defined by B12 as well as observations \citep[e.g.][]{Penarrubia2016, Erkal2019, Watkins2024, DeLeo2024, Rathore2025b}. We refer the reader to F26 for a more in-depth comparison of the Clouds' halo properties in the simulation with the halo properties inferred from observations.

Following B12, the initial LMC and SMC stellar disks are modeled with an exponential density profile:
\begin{equation} \label{eq:disk_dens_profile}
    \rho_{\ast} (R, z) = \frac{M_d}{4\pi R_d^2 H_d}e^{-\frac{R}{R_d}} {\rm sech}^2\left(\frac{|z|}{H_d}\right)
\end{equation}
\noindent where $(R, z)$ are the cylindrical disk coordinates. The total stellar masses ($M_d$) for the LMC and SMC are chosen to be $3.6 \times 10^9$ M$_\odot$ and $5.22 \times 10^8$ M$_\odot$, respectively. The scale radii ($R_d$) for the LMC and SMC are 1.7 kpc and 1.1 kpc, respectively. The scale heights ($H_d$) of the LMC and SMC are 0.34 kpc and 0.22 kpc, respectively. 

The simulated LMC's initial disk parameters are consistent with the properties of the observed LMC disk \citep[e.g.][]{Marel2001, Marel2002, Choi2018a, Rathore2025a}. Since we do not include hydrodynamics, our simulated LMC disk mass represents the sum of the stellar ($2.5 \times 10^9$ M$_\odot$) and gas mass ($1.1 \times 10^9$ M$_\odot$) of the observed LMC. We adopt a disk mass resolution of $500$ M$_\odot$ per star particle, and a disk softening length of $0.05$ kpc. The chosen disk mass resolution ensures that the simulation resolves the phase space of the LMC disk at a level where comparisons with observations like Gaia DR3 \citep{Luri2021} are feasible (J. Hunt {\em priv. comm.}). Note that our resolution is a factor of 5 improvement over B12, who adopted a value of $2500$ M$_\odot$ per particle. Our chosen disk softening length implies that the LMC disk scale height is at least $5$ times the softening value, ensuring that the simulation adequately resolves the vertical structures in the disk.

Traditionally, hydrodynamic models of the SMC incorporate a gas disk with at least $\approx$3 times the mass and $\approx$3 times the scale radius of the SMC's stellar disk \citep[e.g.][]{Besla2012, Pardy2018, Lucchini2024}. In such models, most of the SMC's gas beyond the gas scale radius ends up in the SMC's $150^\circ$ long HI stream (MS, \citealt{Nidever2010}) due to the LMC-SMC interactions. As such, only the SMC's baryons present within the gas scale radius are bound to the SMC at present day (see B12 and \citealt{Rathore2026}). Hence, the modeled SMC disk mass matches the total baryonic mass present within the gas scale radius of the B12 SMC model. Note that our results are not expected to be sensitive to the exact setup of the simulated SMC disk, as the SMC's mass is dominated by DM even in its central regions \citep{Rathore2025b}. The initial SMC's scale radius and scale height are consistent with the values inferred for field dIrr galaxies in the local universe that have a similar stellar mass and gas fraction as the SMC \citep[e.g.][]{Swaters2002, Kreckel2011, Hunter2012}.

We have ensured that the properties of the LMC and SMC disks are convergent with respect to both the disk and halo resolution. In particular, the LMC bar's strength, semi-major axis length, and pattern speed do not change significantly if the halo and disk resolution were reduced by a factor of $\sim5$. The above bar properties were measured using a spatial Fourier decomposition of the simulated disk \citep{Rathore2025b}. For resolution convergence tests pertaining to the simulated halo properties, we refer the reader to F26.

\begin{figure*}
    \centering
    \includegraphics[width=\textwidth]{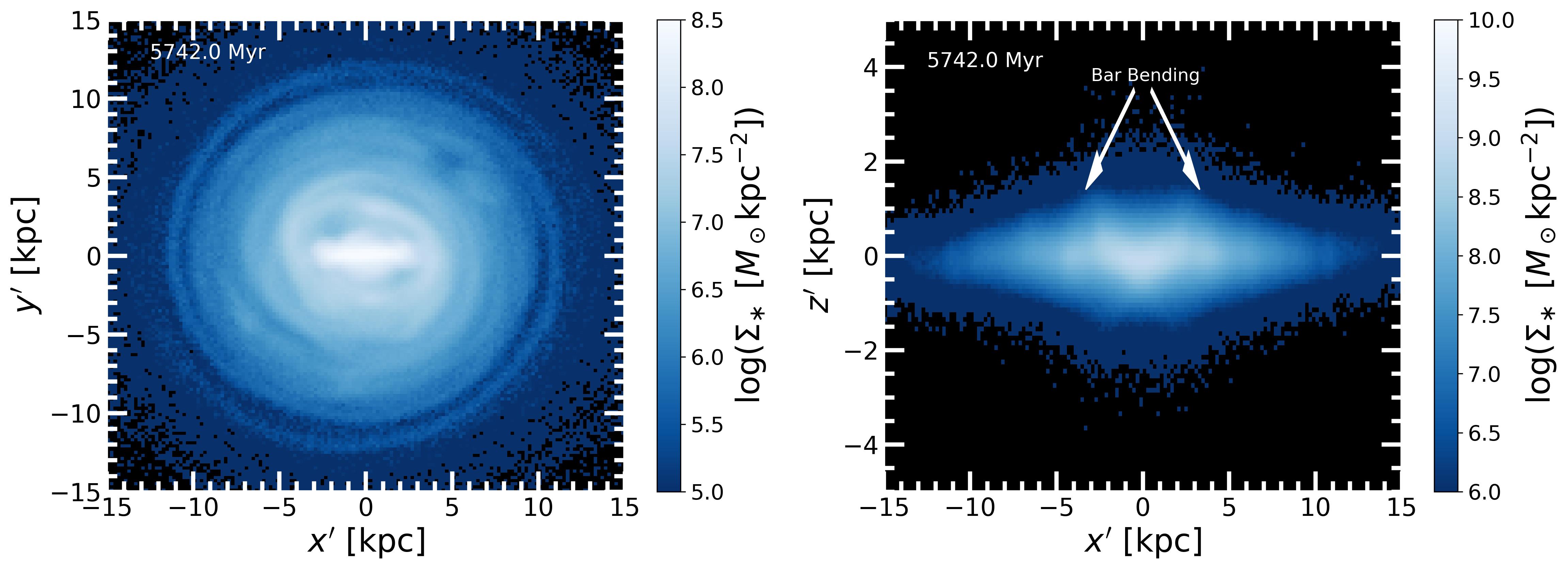}
    \includegraphics[width = \textwidth]{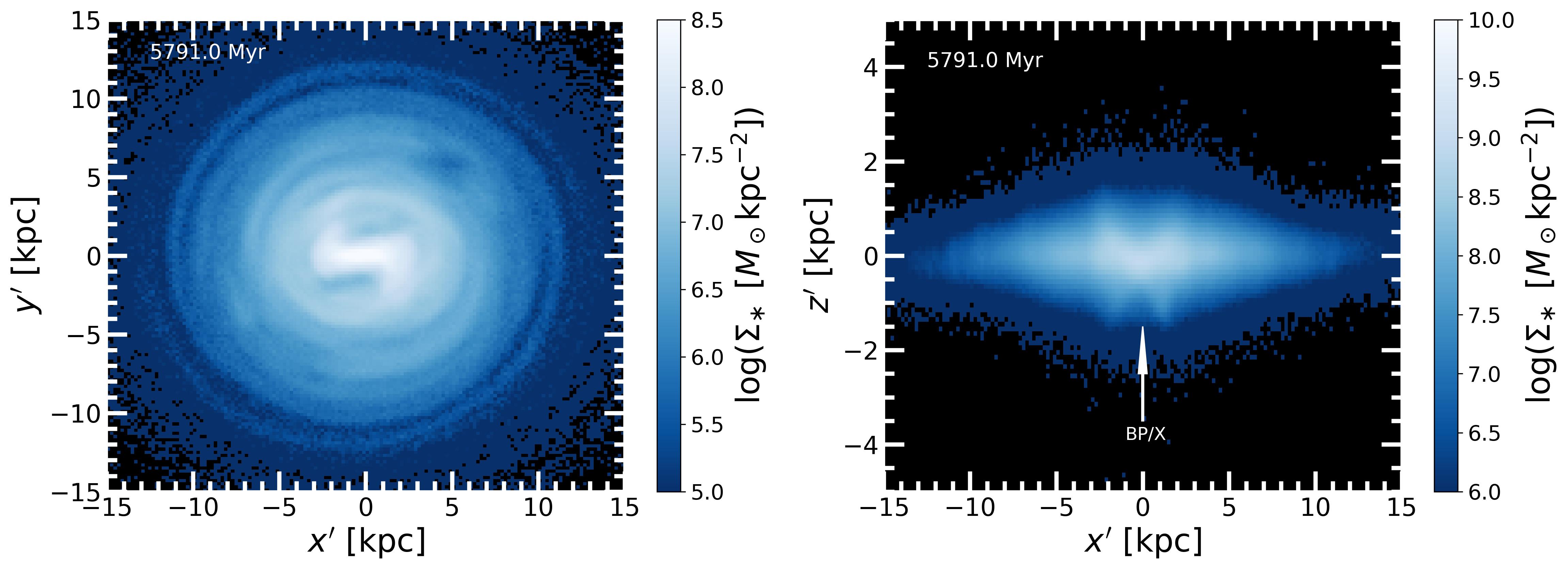}
    \caption{The stellar distribution of the simulated LMC in the LMC-only control simulation (without the SMC). Two epochs are shown: MW infall epoch as defined by B12 ({\em top panel}); and an epoch that is 50 Myr after ({\em bottom panel}). Both face-on and edge-on projections of the LMC bar cartesian frame are shown. The colorbar depicts the stellar surface density distribution. A prominent bar forms out of secular evolution, as seen in the face-on projection. Through buckling ({\em top panel}) and/or resonant interactions, a BP/X structure forms in the edge-on projection ({\em bottom panel}). Apart from the BP/X, no other vertical perturbations are visible indicating that the LMC disk evolves in a stable manner in isolation. This isolated LMC disk will serve as a benchmark for understanding the LMC disk perturbations in our LMC-SMC simulation.}
    \label{fig:isolmcsim}
\end{figure*}

The LMC disk is designed to form a bar due to secular growth followed by saturation of linear instabilities (section \ref{sec:disk_morphology} provides more details). The Radial Dispersion Factor parameter in \texttt{MakeGalaxy} sets the ratio of the azimuthal to radial velocity dispersions in the disk, thereby controlling the size of the bar. For the LMC disk, the Radial Dispersion Factor is set to 2 in order to get a bar length that is consistent with the LMC's observations ($4 - 5$ kpc, \citealt{Rathore2025a}). The simulated bar visibly thickens to form a boxy-peanut-X-shaped bulge (BP/X) at $t \sim 5$~Gyr, where multiple mechanisms \citep{SellwoodGerhard2020} have been proposed for BP/X formation including a buckling instability \citep[e.g.][]{Raha1991, Sellwood1994} and vertical resonances \citep[e.g.][]{AM2002, Quillen2014, BeraldoeSilva2023, McClure2025, Lokas2025}. Figure~\ref{fig:isolmcsim} shows the stellar distribution of the LMC disk evolved in isolation (without the SMC's influence) at an epoch that corresponds to the MW infall in B12 ($t \approx 5.7$ Gyr). The onset of BP/X growth \citep[e.g.][]{Athanassoula2005, Valluri2016, Tahmasebzadeh2024, Dattathri2024} can be clearly seen.

We have checked that our LMC and SMC disks evolve in a stable way by themselves (i.e., without their mutual influence on each other) for at least 7 Gyr (see also Figure \ref{fig:isolmcsim}). In particular, at a radius corresponding to $3R_d$, the disk surface density profiles, enclosed mass profiles, and rotation curves change by $<10\%$ of their initial values. For stability tests of the halo properties, we refer the reader to F26.

In the subsequent analyses, we will use the above simulation (Figure \ref{fig:isolmcsim}) of the LMC disk evolved in isolation as an LMC-only control simulation. Using this control simulation, we will evaluate the LMC halo's gravitational influence on the LMC's disk that would be expected due to Poisson fluctuations in the halo density field that result from a finite number of halo particles. As such, this control simulation will provide us with the noise floor that needs to be considered while interpreting the influence of the SMC-induced halo distortions on the LMC disk in the LMC-SMC simulation.

\begin{figure}
    \centering
    \includegraphics[width=\columnwidth]{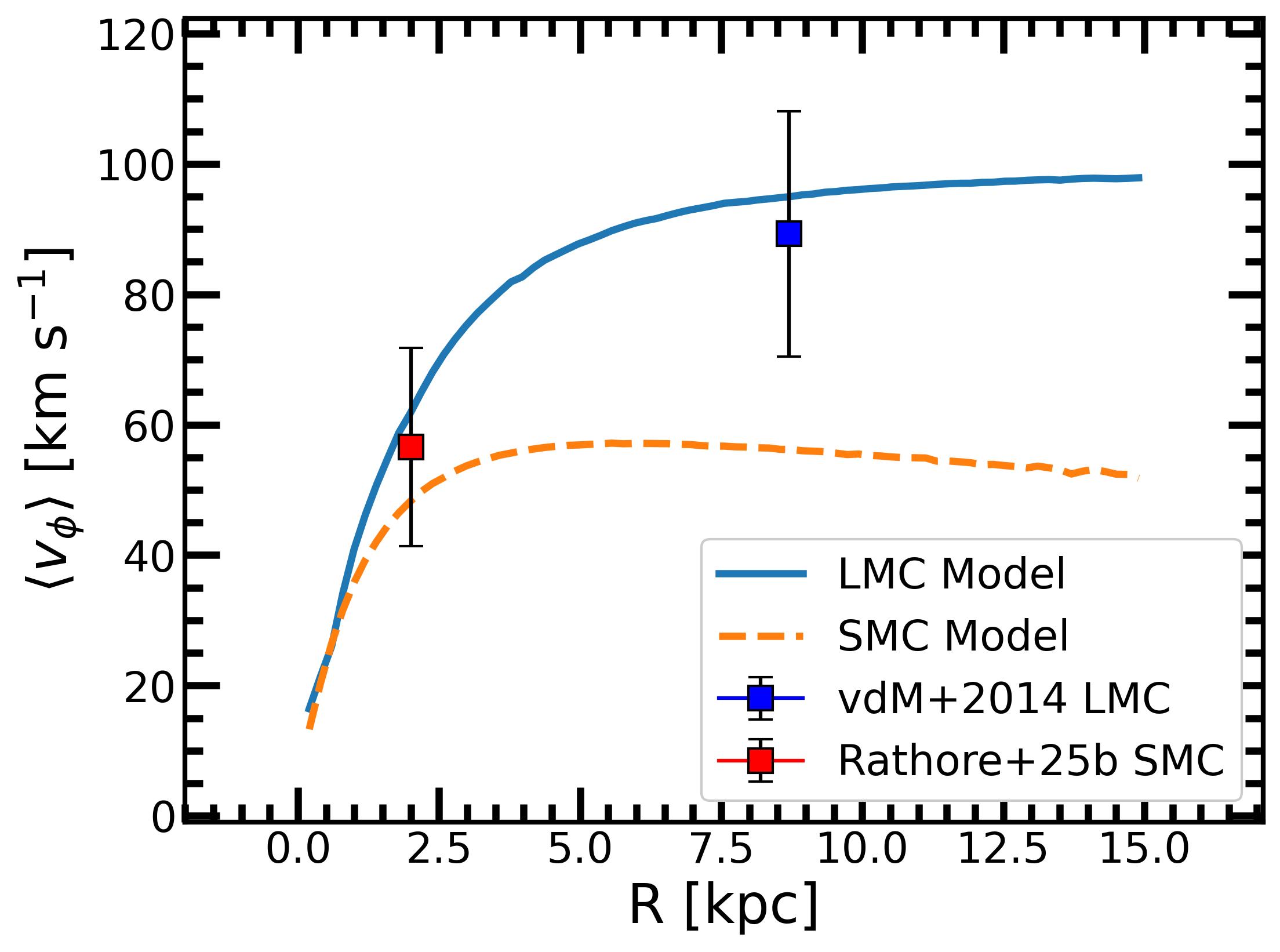}
    \caption{The initial rotation curve of the LMC and SMC N-body models. The average azimuthal velocity is plotted as a function of the radial coordinate. The LMC (SMC) model has a peak circular velocity of $\approx 98$ \kms ($\approx 57$ \kms). The blue square depicts the observed constraint on the LMC's rotation velocity ($89.3 \pm 18.8$ km s$^{-1}$) at $R = 8.7$ kpc as obtained by \cite{vdM2014}. The red square depicts the SMC's expected rotation velocity ($56.6 \pm 15.2$ km s$^{-1}$) at $R = 2$ kpc calculated using the mass profile constraints obtained by \cite{Rathore2025b} at the Clouds' MW infall. Our modeled initial rotation curves are consistent with the observational constraints.}
    \label{fig:rotcurve}
\end{figure}

The resulting rotation curves of the LMC and SMC galaxy models are shown in Figure \ref{fig:rotcurve}. The LMC's rotation curve is consistent with the observational constraint of \cite{vdM2014}, who obtained a rotation velocity of $89.3 \pm 18.8$ km s$^{-1}$ at $R = 8.7$ kpc. 

The observed SMC has a small ($\lesssim 10$ km s$^{-1}$) stellar rotation velocity \citep{Zivick2018, Zivick2021, Niederhofer2021, Dhanush2025, Vijaysree2026}. \cite{Rathore2026} showed that the SMC's lack of stellar rotation is explainable by the LMC-SMC interactions, which convert an initially disk-like SMC (rotation-to-dispersion ratio $> 4$) to a dispersion dominated galaxy (rotation-to-dispersion ratio $< 0.6$). As such, the current stellar kinematics of the SMC are in a high state of disequilibirum, and are not suitable to constrain the initial SMC's enclosed mass or equivalently the rotation curve. 

HI surveys of the SMC have revealed a velocity gradient of $60 - 100$ km s$^{-1}$ across the SMC's spatial extent, which has been previously interpreted as a sign of a rotating gas disk \citep{Stanimirovic1999, DiTeodoro2019}. However, recently, \cite{Rathore2026} showed that the SMC's HI velocity gradient is not due to a rotating gas disk, but is instead caused by radially outwards motion of gas driven by the LMC's tidal and hydrodynamic forces (see also \citealt{Murray2019, Nakano2025b}). Hence, like the SMC's stellar kinematics, its HI kinematics are also in a state of disequilibrium and cannot be used to constrain the initial SMC model. The inability of the SMC's internal kinematics to constrain the SMC's enclosed mass or the rotation curve poses a serious challenge to construct accurate N-body models of the SMC. Moreover, even Virial mass estimators (e.g. \citealt{Harris2006}) may not be applicable given the disequilibrium in this galaxy \cite{Rathore2026}. 

Recently, \cite{Rathore2025b} provided a way forward by constraining the SMC's enclosed mass by modeling the SMC's observable impact on the LMC bar. In particular, they showed that the SMC's torques on the LMC bar during the recent ($< 200$ Myr ago) LMC-SMC collision explains the observed LMC's $5^\circ - 15^\circ$ bar tilt \citep{Choi2018a} with respect to the disk plane. By modeling these torques, they constrained the SMC's mass enclosed mass within 2 kpc to be $M_{\rm SMC} (R < 2 \: {\rm kpc}) = 1.6\pm0.8 \times 10^9$~M$_\odot$ at MW infall.  F26 further showed that this \cite{Rathore2025b} constraint at MW infall is representative of the initial SMC's enclosed mass (within 2 kpc), and that this enclosed mass distribution remains spherically symmetric at least until the Clouds' MW infall. Using this mass constraint, we calculate the initial SMC's circular velocity at $R = 2$ kpc to be:
\begin{equation}
    v^{\rm SMC}_{\rm c} (2 \: {\rm kpc}) = \sqrt{\frac{G M_{\rm SMC} (R < 2 \: {\rm kpc})}{(2 \: {\rm kpc})}}
\end{equation}
\noindent which is $56.6 \pm 15.2$ km s$^{-1}$. Our simulated SMC rotation curve is consistent with this circular velocity.

\begin{table}[]
    \centering
    \caption{Parameters of the LMC and SMC N-body galaxy models.}
    \begin{tabular}{c c c}
        \hline
        \hline
        Parameter & LMC Model & SMC Model\\
        \hline
        DM Halo Profile & Hernquist & Hernquist \\
        Halo Mass [M$_\odot$] & $1.76 \times 10^{11}$ & $1.995 \times 10^{10}$ \\
        Halo Scale Radius [kpc] & 21.4 & 7.3 \\
        Halo Number of Particles & $4.4 \times 10^6$ & $5.0 \times 10^5$ \\ 
        Halo Mass Resolution [M$_\odot$] & $4 \times 10^{4}$ & $4 \times 10^{4}$ \\
        Halo Softening Length [kpc] & 0.08 & 0.08 \\
        Halo $\beta$ (velocity anisotropy) & 0 & 0\\
        Stellar Disk Profile & Exponential & Exponential \\
        Disk Mass [M$_\odot$] & $3.6 \times 10^9$ & $5.22 \times 10^8$ \\
        Disk Scale Length [kpc] & 1.7 & 1.1 \\
        Disk Scale Height [kpc] & 0.34 & 0.22 \\
        Disk Number of Particles & $7.2 \times 10^6$ & $1.044 \times 10^6$\\
        Disk Mass Resolution [M$_\odot$] & 500 & 500 \\
        Disk Radial Dispersion Factor & 2 & 2\\
        Disk Softening Length [kpc] & 0.05 & 0.05\\
        \hline
    \end{tabular}
    \tablenotetext{}{\textbf{Note:} {\em Left column:} the property of the galaxy model. {\em Middle column:} value for the LMC. {\em Right column:} value for the SMC. For a more detailed description of the parameters, consult section~\ref{sec:sims} of the main text. Several parameters in this table are also given in F26 and have been repeated here for the reader's convenience.}
    \label{tab:ic_params}
\end{table}

\subsubsection{Combined LMC $+$ SMC N-body Model}

To construct a simulation where the Clouds interact over time, the initial LMC and SMC galaxy models are combined using the publicly available code \texttt{combine\_ics} \citep{combine_ics}. The combined LMC+SMC model is then evolved using the \texttt{Gadget-4} N-body code\footnote{\url{https://wwwmpa.mpa-garching.mpg.de/gadget4/}} \citep{Springel2021}. Table \ref{tab:ic_params} summarizes the properties of the LMC and SMC galaxy models. Note that the MW is not included in this simulation.

Our LMC-SMC orbital parameters and the relative inclinations of the SMC and LMC disks are motivated by the B12 hydrodynamic simulations of the LMC-SMC-MW interaction history. \cite{Besla2010} and B12 showed that the observations of the MS \citep{Mathewson1974, Braun2004, Nidever2010} are best reproduced by a $1:10$ mass-ratio LMC-SMC encounter, where the LMC's total mass is $\sim 10^{11}$ M$_\odot$. B12 chose an eccentric LMC-SMC orbit, with an initial $e \approx 0.7$. Given the rate of orbital decay due to dynamical friction, a less eccentric orbit would result in a merger of the Clouds prior to their MW infall (see also \citealt{Kallivayalil2006, Kallivayalil2006b}). On the other hand, a more eccentric orbit will {\em not} give the required repeated LMC-SMC interactions needed to explain: the LMC-SMC bridge of HI gas and stars \citep{Kerr1957, Putman2003, Bruns2005, Zivick2019}; stellar counterpart to the MS \citep{Zaritsky2020, Chandra2023, Zaritsky2025}; and the observed mutually-elevated and coincident star-formation events in the Clouds over the past $\sim 6$ Gyr \citep{Noel2007, Noel2009, Weisz2013, Massana2022, Burhene2026}. The SMC's orbital plane and the inclination of its disk with respect to the orbital plane are chosen to ensure a prograde encounter that results in an MS of the required extent. Other authors \citep[e.g.][]{Pardy2018, TepperGarcia2019, Craig2021, Lucchini2024} have also used a similar LMC-SMC orbital setup as ours to understand the properties of the MS and the LMC-SMC bridge.

\begin{table}[]
    \centering
    \caption{Initial phase space and geometric setup of the SMC in the LMC-SMC simulation.}
    \begin{tabular}{c c}
        \hline
        \hline
        Parameter & Value\\
        \hline
        $\vect{R_{\rm SMC}^{i}}$ [kpc] & $(39.64, 44.86, 0)$\\
        $\vect{v_{\rm SMC}^{i}}$ [\kms] & $(-121.56, -0.97, 0)$\\
        $\vect{J_{\rm SMC}^{i}}$ & $(0.22, 0.32, 0.92)$\\
        \hline
    \end{tabular}
    \tablenotetext{}{\textbf{Note:} {\em Left column:} the geometric property of the SMC. {\em Right column:} the corresponding value in the LMC disk frame coordinates. See section \ref{sec:sims} of the main text for more details.}
    \label{tab:sim_params}
\end{table}

Table \ref{tab:sim_params} gives the initial geometric parameters like the SMC-LMC position vector ($\vect{R_{SMC}^i}$), the SMC-LMC relative velocity vector ($\vect{v_{SMC}^i}$), and the SMC's orbital angular momentum vector ($\vect{L_{SMC}^i}$). The above geometric parameters are specified in the LMC disk cartesian frame, wherein the origin is chosen to be the LMC's stellar density center. The directions of the $x, y, z$ axis of this frame are fixed as follows. First, the Galactocentric axes are determined using the observed angular momentum vector of the LMC disk \citep{Salem2015}. Then, a coordinate rotation is applied such that the resulting z-axis is aligned with the LMC disk angular momentum vector. Following \cite{Besla2010}, the SMC's disk angular momentum vector ($\vect{J_{SMC}^i}$) is chosen to be orthogonal to $\vect{L_{SMC}^i}$. Following \cite{Rathore2025b}, we also define a LMC bar cartesian frame ($x^\prime, y^\prime, z^\prime$). This frame is defined by applying a coordinate rotation to the LMC disk cartesian frame such that the $x$-axis is aligned with the LMC bar's major axis. More details on computing the position angle of the bar's major axis can be found in \cite{Rathore2025a}.

The stellar density centers of the simulated LMC and SMC are computed using an iterative shrinking sphere algorithm \citep[e.g.][]{Power2003, GC2019, Rathore2025b, Rathore2026}. First, a mass weighted average of the LMC and SMC stars' position vectors is calculated to obtain guess density centers for the LMC and SMC, respectively. Then, we construct a sphere of radius 10 kpc (5 kpc) centered at the LMC (SMC) guess density centers. The mass-weighted average position of the stars that reside inside the sphere gives a new estimate for the density center. The sphere is re-centered at this new density center, and its radius is shrunk by $30\%$. Then, the mass-weighted average position of the stars that reside inside the shrunken sphere gives the density center for the next iteration. The process is continued until the density centers obtained in two successive iterations differ by less than the stellar softening length. The systemic velocities of the simulated SMC and LMC are computed using a similar iterative shrinking sphere algorithm, but applied in the velocity space instead of the position space (see \citealt{Rathore2026} for more details). Following \cite{Gomez2016, Gomez2017, Rathore2025b}, we define the LMC's disk plane in the simulation using the disk angular momentum vector ($\vect{J_{\rm LMC}}$). The angular momentum vector of the LMC is computed by averaging the angular momenta of star particles that reside within 10 kpc of the LMC's stellar density center.

\begin{figure}
    \centering
    \includegraphics[width=0.49\textwidth]{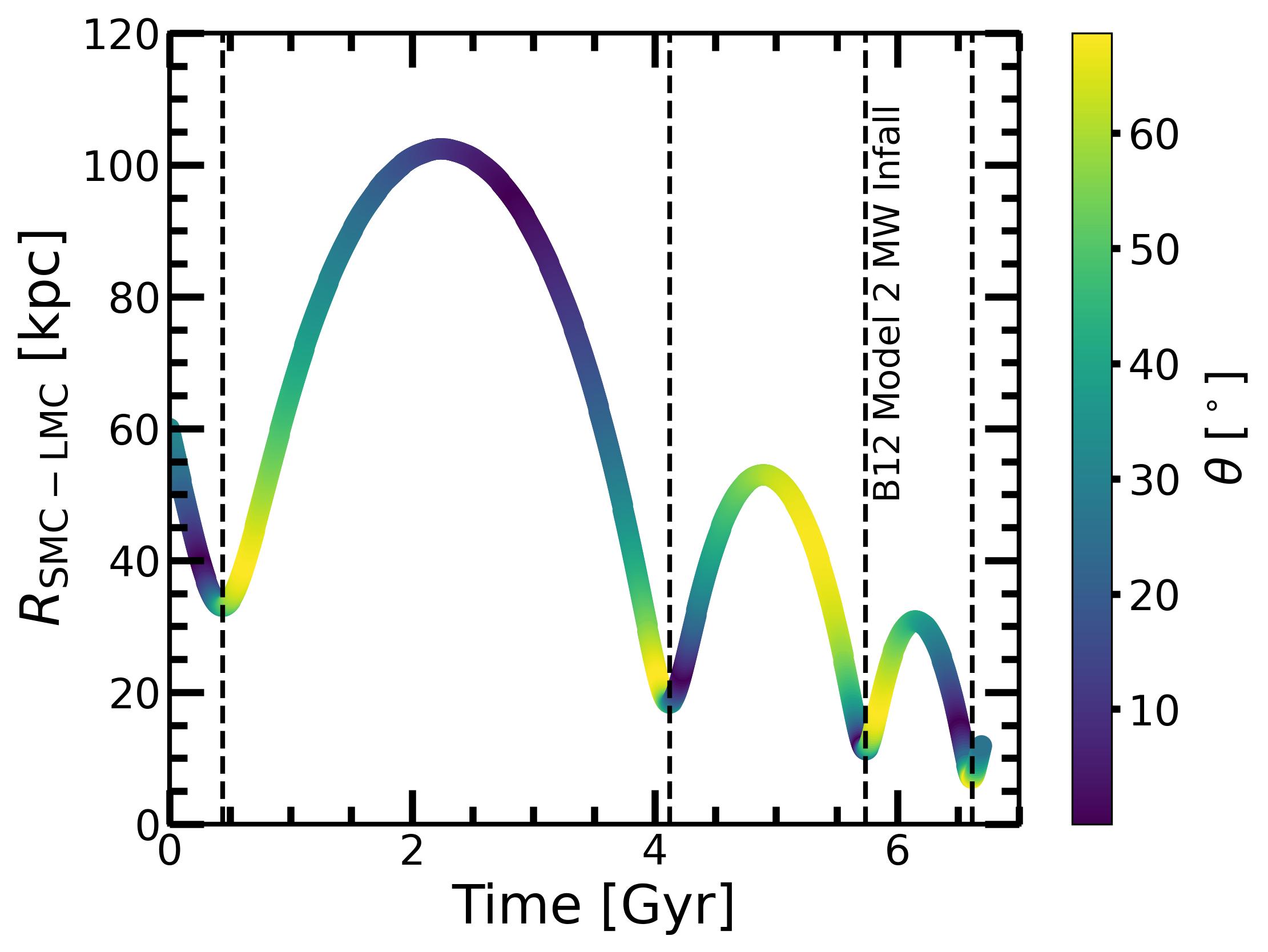}
    \caption{The simulated LMC-SMC orbit. The distance between the LMC and the SMC ($R_{\rm{SMC-LMC}}$) is plotted as a function of simulation time. The pericentric epochs are marked by the black vertical dashed lines. The Clouds' Milky Way infall epoch, as determined by B12, approximately corresponds to the $3^{\rm rd}$ LMC-SMC pericenter. The colormap represents the angle ($\theta$) between the SMC-LMC position vector ($\vect{R_{\rm{SMC-LMC}}}$) and the LMC disk plane. $\theta = 0$ indicates the epochs at which the SMC crosses the LMC's disk plane. Both $R_{\rm{SMC-LMC}}$ and $\theta$ influence the SMC's torques on the LMC's disk.}
    \label{fig:lmc_smc_orbit}
\end{figure}

\begin{figure*}
    \centering
    \includegraphics[width=\textwidth]{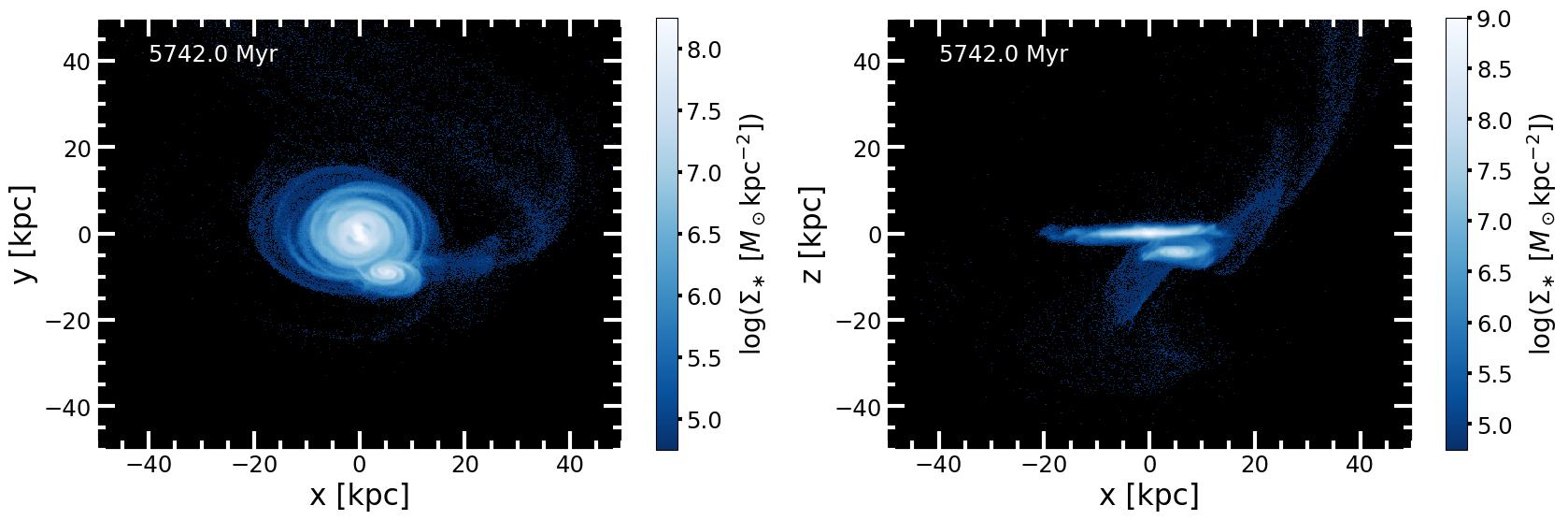}
    \caption{The stellar distribution in the fiducial LMC-SMC simulation at the epoch corresponding to the Clouds' MW infall in B12 ($t \approx 5.7$ Gyr). {\em Left panel} ({\em Right panel}) shows the face-on (edge-on) projections in the LMC disk cartesian frame. The color scale represents the stellar surface density.  The stellar distribution of the LMC and SMC gets significantly perturbed due to their mutual interactions. In particular, the LMC's disk exhibits in-plane and vertical morphological perturbations, and the SMC's disk gets disrupted to form the tidal tails. F26 showed that significant SMC-induced dipole and quadrupole LMC halo distortions also exist at this epoch. Our goal is to disentangle the influence of these halo distortions from the SMC's direct influence in driving the LMC disk perturbations. The full movie of this N-body simulation can be downloaded from \url{https://github.com/himanshrathore/rathore2026a/blob/main/lmsmc_movie.mp4}. This movie is licensed under the GPL-3.0 License, and is 18 MB in size and 56 seconds in duration.}
    \label{fig:lmcsmcsim}
\end{figure*}

The simulated LMC-SMC orbit is shown in Figure \ref{fig:lmc_smc_orbit}. The simulation is stopped at $t\approx 6.7$ Gyr, which roughly corresponds to the present day in the B12 Model 2 simulation. The SMC completes 4 pericentric passages about the LMC until the present day. Note that B12 introduced the Clouds into a static MW potential at $t \approx 5.7$ Gyr. This epoch also corresponds to the $3^{\rm rd}$ pericenter of the SMC about the LMC. Figure \ref{fig:lmcsmcsim} shows the simulated stellar distribution of the Clouds at this epoch. The mutual LMC-SMC interactions results in in-plane and vertical morphological perturbations in the LMC's disk and the tidal disruption of the SMC's disk. F26 showed that the SMC-induced dipole and quadrupole distortions in the LMC's halo are also significant at the shown epoch, and our goal is to disentangle the LMC halo's influence from the SMC's influence in driving perturbations of the LMC's disk.

We do not include the MW in this study as we want to isolate the SMC's impact on the LMC rather than reproducing the present day state of the LMC disk. As the Clouds are on their first infall \citep[e.g.][]{Besla2007}, our simulated orbit reflects the actual LMC-SMC orbit except for the last $\sim 1$ Gyr, where the MW's influence on the LMC-SMC binary can be important. As such, our simulations serve as a critical initial condition that explains the state of the LMC-SMC system upon accretion into the MW. 

The orbit in Figure \ref{fig:lmc_smc_orbit} is colored by the angle ($\theta$) between the SMC-LMC position vector ($\vect{R_{\rm SMC-LMC}}$) and the LMC's disk plane. $\theta = 0$ determines the times at which the SMC crosses the LMC's disk. These disk crossing times will be important for interpreting the instantaneous torques applied by the SMC on the LMC's disk (section \ref{sec:torques}). Simulation snapshots are extracted at a time cadence of $4.85$ Myr, which yields $1381$ snapshots in total.

\subsection{Basis Function Expansion} \label{sec:BFEs}
We represent the distorted density and potential field of the LMC's DM halo and stellar disk in the fiducial LMC-SMC N-body simulation using orthogonal BFEs (see equations 2.212 - 2.214 in \citealt{BT2008}). Traditional BFE treatments \citep[e.g.][]{CluttonBrock73, Kalnajs1977, Fridman1984, Earn1996} estimate the density and potential fields through a superposition of analytic eigenfunctions of the Laplacian. However, most families of analytical eigenfunctions, like the Bessel functions, do not resemble the radial density profiles of typical galactic halos or disks. As such, many eigenfunctions will be required just to represent the equilibrium halo and disk profile, and even more will be required to represent any halo and disk perturbations, making the BFE process computationally very expensive \citep{Weinberg1998, Weinberg1999}.

In this work, we use the \texttt{EXP} code\footnote{https://github.com/EXP-code} \citep{Petersen2025} for computing the BFEs. \texttt{EXP} circumvents the above mentioned limitation by constructing a lowest-order radial basis function that matches a desired halo or disk initial profile. \texttt{EXP} numerically computes such a basis function by solving the Poisson equation expressed in a Sturm-Liouville form \citep{Pryce1991, Press1993}. Any higher-order basis functions will then capture perturbations to the equilibrium galaxy model. Such a numerical approach reduces the computational cost significantly, but the basis functions are no longer analytical and instead need to be tabulated. This tabulation adds to the up-front calculation time with EXP, but once tabulated, is at least as fast as the analytic basis functions.

\subsubsection{DM Halo BFEs} \label{sec:halo_BFEs}
The DM halo BFEs of the fiducial LMC-SMC simulation were already computed by F26 and are used directly in this study to study the time evolution of the DM halo of the LMC. We provide a brief summary below and refer the reader to F26 for more details.

The time-dependent LMC halo density and potential fields were estimated as:
\begin{equation}
    \tilde{\rho}_{\rm halo} (r, \theta, \phi, t) = \sum_{n = 0}^{n_{\rm max}} \sum_{l = 0}^{l_{max}}\sum_{m = -l}^{l} C_{nlm}\rho_n(r)Y_l^m(\theta, \phi)
\end{equation}
\begin{equation}
    \tilde{\Phi}_{\rm halo} (r, \theta, \phi, t) = \sum_{n = 0}^{n_{\rm max}} \sum_{l = 0}^{l_{max}}\sum_{m = -l}^{l} C_{nlm}(t)\psi_n(r)Y_l^m(\theta, \phi)
\end{equation}
\noindent where $(r, \theta, \phi)$ represents a spherical coordinate system where the origin is the LMC's DM density center, the polar axis coincides with $\vect{J_{\rm LMC}}$, and the direction $\phi = 0$ coincides with the $x$-axis of the LMC disk cartesian frame. The index $n$ captures the radial order of the basis, and the indices $(l, m)$ capture the harmonic orders. The harmonic basis functions are chosen to be the spherical harmonics ($Y_l^m(\theta, \phi)$). $C_{nlm}$ are the time-dependent amplitudes of the basis functions, computed by summing the contribution of each particle in the simulation given their location. Hereafter, we will refer to halo basis functions with $l = 1$ as the dipole, and $l = 2$ as the quadrupole. 

Following the methodology of F26, we also perform a BFE of the DM distribution of the LMC-only control simulation. F26 also presented the BFE's of the SMC's halo, which they used to compute the SMC's bound mass at any instant of time in the simulation. Later, we will use this BFE-derived bound mass to calculate the SMC's torques on the LMC's disk. 

\subsubsection{LMC Stellar Disk BFEs} \label{sec:disk_bfe}

In this sub-section, we describe our approach to construct a cylindrical disk basis for representing the disturbed, time-dependent morphology of the LMC's stellar disk in our LMC-SMC N-body simulation. The LMC disk cartesian frame will be used to represent the basis and the corresponding coefficients.

To construct the cylindrical basis, \texttt{EXP} first computes a very high-order spherical basis (like the one used for halo BFEs). The order of this parent spherical basis is denoted by $l_{\rm fid, max}, n_{\rm fid, max}$, where $l_{\rm fid, max}$ is chosen to be $\geq 36$. Given this high-order spherical basis indexed by $l_{\rm fid}, n_{\rm fid}$, \texttt{EXP} computes a new tabulated meridional basis using a superposition of the spherical basis functions. This new basis optimally represents the target density profile of the equilibrium disk in a least-squares sense (see \citealt{Petersen2022}). This basis construction ensures bi-orthogonality of the new meridional basis. The rank of the new basis may then be chosen as the desired number of disk expansion functions.

The resulting cylindrical disk basis functions are characterized by two indices --- $(m, n)$, and the corresponding basis coefficient is denoted by $C_{mn} (t)$. The index $m$ denotes the order of the basis along the azimuthal coordinate. For a given $m$ value, the index $n$ denotes the order of the basis along the radial and vertical coordinate simultaneously, as each basis function is a 2-D meridional plane multiplied by $e^{im\phi}$. The complex valued coefficients capture the overall phase of the basis function with respect to the $\phi = 0$ axis.

A key challenge to constructing a suitable basis is that the LMC disk becomes significantly perturbed in the outer regions ($R > 6$ kpc). The basis needs to be flexible enough to capture such perturbations, particularly in the low density areas of the LMC disk that are either in the outskirts or more than a scale height away from the disk mid-plane. In our usage of specifying a target exponential disk density function (eq. \ref{eq:disk_dens_profile}),
the flexibility of the cylindrical disk basis depends on the following parameters in \texttt{EXP}: (i) the order of the parent spherical basis ($l_{\rm fid, max}, n_{\rm fid, max}$) chosen to construct the disk basis; and (ii) the adopted scale radius ($a_{\rm cyl}$) and scale height ($h_{\rm cyl}$) for the lowest order disk basis function. Most \texttt{EXP} implementations of cylindrical disk basis \citep[e.g.][]{Johnson2023, Hunt2026} use $l_{\rm fid, max}$ values in the range of 36 to 72, and set $a_{\rm cyl} = R_d$ and $h_{\rm cyl} = h_d$. In this work, we use a high value of $l_{\rm fid, max} = 128$ to ensure that the disk basis functions are sufficiently flatenned. We use $a_{\rm cyl} = 2R_d$ and $h_{\rm cyl} = 2h_d$ so that the basis has enough functional support\footnote{The support of a function is the subset of its domain where the value of the function is non-zero.} in the disk outskirts and significantly away from the disk mid-plane. Our EXP disk basis parameters are summarized in Table \ref{tab:basis_params}. 

\begin{table}[]
    \centering
    \caption{Parameters of the LMC cylindrical disk basis.}
    \begin{tabular}{c c}
        \hline
        \hline
        Parameter & Value\\
        \hline
        $a_{\rm cyl}$ [kpc] & 3.4\\
        $h_{\rm cyl}$ [kpc] & 0.68\\
        $n_{\rm fid, max}$ & 72\\
        $l_{\rm fid, max}$ & 128\\
        $m_{\rm max}$ & 16\\
        $n_{\rm max}$ & 72\\
        $n_{\rm cyl, odd}$ & 36\\
        \hline
    \end{tabular}
    \tablenotetext{}{\textbf{Note:} $a_{\rm cyl}$: Scale radius of the cylindrical basis. $h_{\rm cyl}$: scale height of the cylindrical basis. $n_{\rm fid, max}$: maximum radial order for the parent spherical basis. $l_{\rm fid, max}$: maximum angular order for the parent spherical basis. $m_{\rm max}$: maximum azimuthal order of the cylindrical basis. $n_{\rm max}$: total number of radial orders in the disk basis. $n_{\rm cyl, odd}$: total number of vertically anti-symmetric disk basis functions. See the main text in section~\ref{sec:disk_bfe} for more details.}
    \label{tab:basis_params}
\end{table}

We truncate the expansion to a maximum radial order of $n_{\rm max} = 72$ and the maximum azimuthal order of $m_{\rm max} = 16$. Of the $72$ radial orders, we choose $n_{\rm cyl, odd} = 36$ of them to be vertically anti-symmetric. Our choice of $n_{\rm cyl, odd}$ ensures that we have enough basis functions to capture disk perturbations that are not symmetric about the disk mid-plane. Note that the \texttt{EXP} indexing is such that the basis functions corresponding to first 36 radial orders ($n = 0 - 35$) are vertically symmetric and the next 36 orders ($n = 36 - 71$) are vertically anti-symmetric. So, the node spacing of the basis, which controls the representation of fine self-gravitating features, is approximately $\sim \frac{2a_{\rm cyl}}{(n_{\rm max}/2)} = 0.1$ kpc at $R = a_{\rm cyl}$. This radial resolution is sufficient to resolve disk features down to twice the softening length. Our choice of azimuthal truncation order implies that we can resolve features that are up to 16-fold periodic in the disk plane --- more than enough to resolve the bar and spiral structures. In practice, the expansion can be truncated to smaller radial and azimuthal orders, as desired, after a higher-order expansion has been constructed. In Appendix A, we show convergence tests to justify that the chosen truncation orders are appropriate.

\begin{figure*}
    \centering
    \includegraphics[width = 0.49\textwidth]{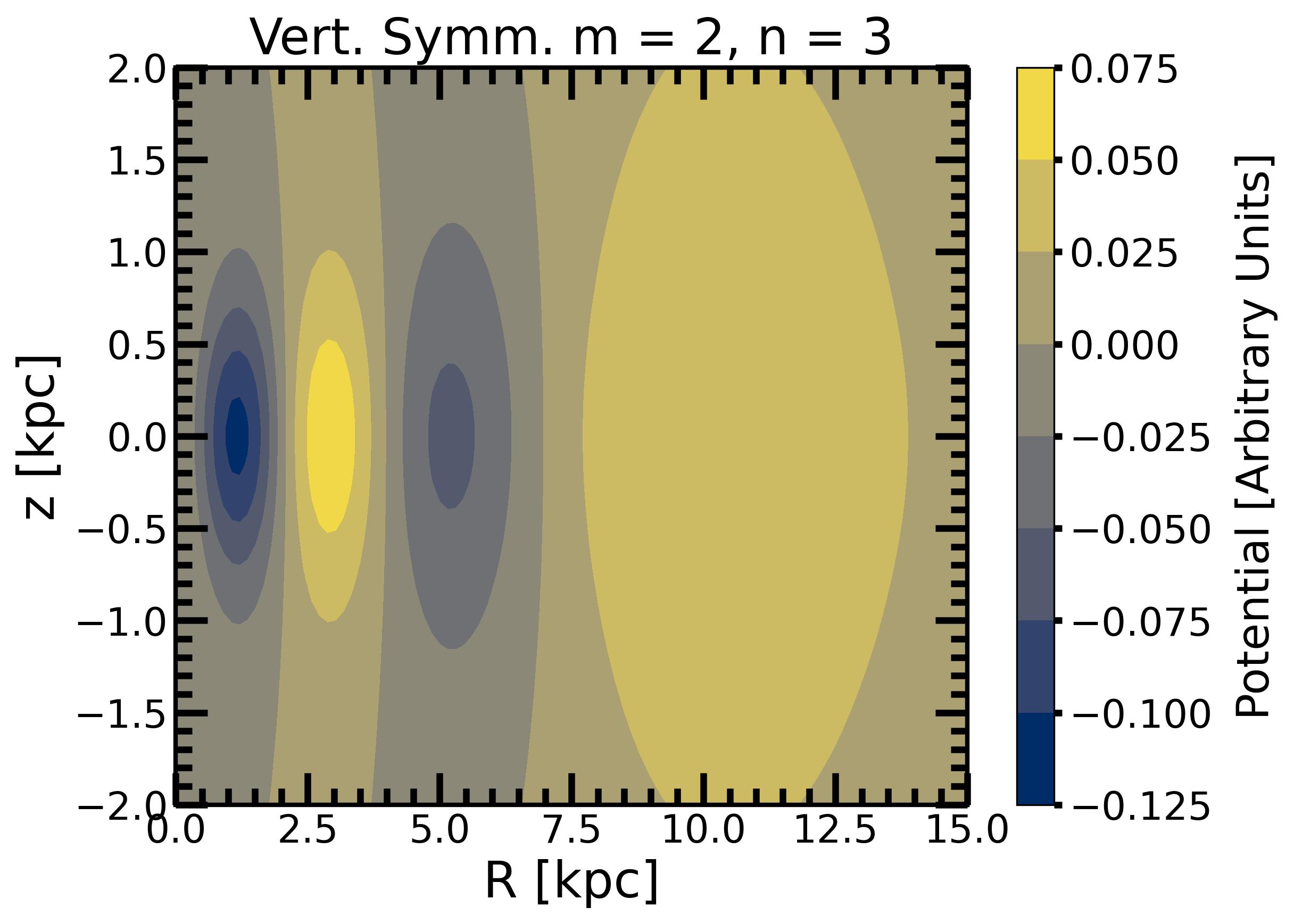}
    \includegraphics[width = 0.49\textwidth]{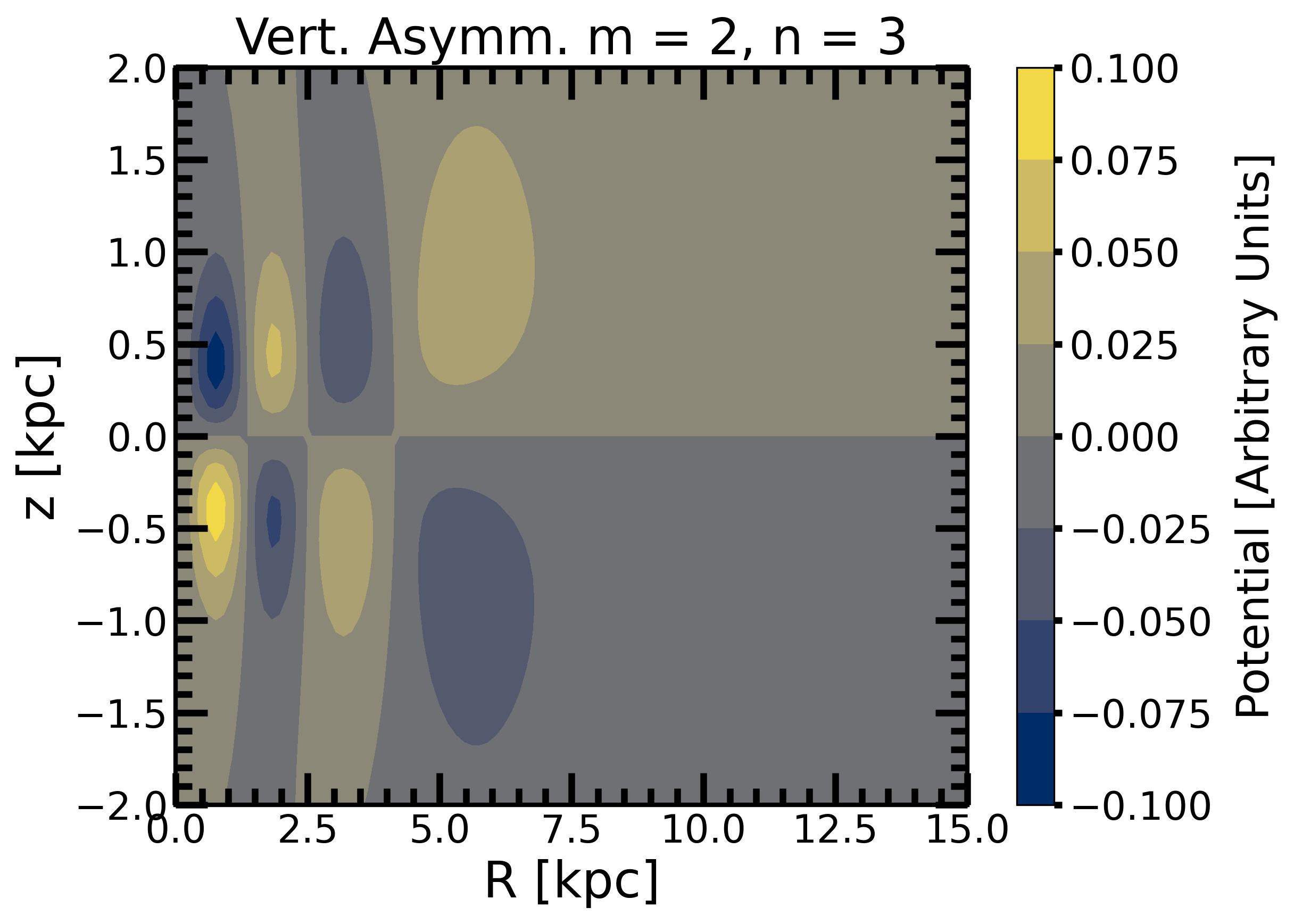}
    \caption{Visualization of the $m = 2,\: n = 3$ disk potential basis functions. {\em Left panel} ({\em Right panel}) shows the vertically symmetric (vertically anti-symmetric) basis function. The colored contours represent the values of the potential on a meridional (R -- z) slice of the LMC disk frame. The radial order ($n$) determines the number of zero crossings (nodes) of the potential basis function along the R-axis. The azimuthal order ($m$) determines the symmetry ($m$-fold) of the potential basis function along the azimuth direction (perpendicular to the plane of the paper).}
    \label{fig:basis_viz}
\end{figure*}

Figure \ref{fig:basis_viz} gives a visualization of one of the potential basis orders in the meridional plane of the disk. Both the vertically symmetric and vertically anti-symmetric basis functions are shown. The computational cost of the basis construction is dictated mainly by the $l_{\rm fid, max}$ parameter, wherein the cost scales as $l_{\rm fid, max}^2$. Our choice of $l_{\rm fid, max} = 128$ requires a RAM of $\approx 25$ GB per core. EXP basis construction is \texttt{OpenMP} parallelized, and our basis generation takes $\approx 2$ hrs when parallelized across 26 cores on a High-Performance Computing (HPC) cluster with AMD Zen2 processors. Once the basis is determined, the coefficients are computed by supplying the simulation star particle data (hereafter referred to as stars) to the \texttt{EXP} basis object. The coefficients are constructed in the LMC disk cartesian frame. Note that choosing the disk density center for centering the disk BFE ensures that the basis coefficients capture any offsets in the mass distribution of the inner disk with respect to the outer disk during the LMC-SMC interactions. This LMC disk frame is slightly different from the reference frame used for computing the halo coefficients. The only difference is that halo coefficients were centered at the halo density center as opposed to the disk density center. We will account for this centering difference while using the basis coefficients to generate the reconstructed potential and density fields. 

The size of the file that encodes the disk coefficients is approximately 30 MB. It is remarkable that a BFE can represent the essential dynamics of a $\approx$TB-sized simulation with a $\sim$ GB basis file and 10's of MBs of coefficients. This information compression capability of BFEs opens a new opportunity space for probing dynamical processes with N-body simulations.

\subsection{Multi-channel Singular Spectral Analysis} \label{sec:mssa_method}
We use the Multichannel Singular Spectral Analysis (mSSA) algorithm to understand the correlations in the time series of the SMC's orbital evolution, SMC-induced LMC halo distortions, and the LMC's disk perturbations (specifically the vertical perturbations). mSSA is a powerful tool to study correlations across several time-dependent dynamical processes, as demonstrated by e.g. \cite{Weinberg2021, Johnson2023, Weinberg2023}. The mSSA algorithm has the following steps.

Time lagged versions of each input time-series (also known as a channel) are constructed and stacked along the rows of a matrix known as the trajectory matrix ($\mathbf{T}$). Singular value decomposition is performed for the covariance matrix of $\mathbf{T}$, and the resulting principal components (PCs) are extracted. Each PC denotes a correlated temporal component across the different channels. Reconstructions of the original channels can be obtained through these PCs (see WP21 for more details). The novelty in mSSA lies in being able to select different subsets (\enquote{groups}) of PCs to perform the reconstruction. As each PC is sensitive to correlations at a particular timescale, reconstructing the original channels using different PC groups enables identifying coupled dynamical processes across different timescales. This mSSA feature allows us to infer at what aspects of the LMC disk's temporal evolution is the most affected by the SMC's influence and the influence of the SMC-induced LMC halo distortions.

For more mathematical details behind this algorithm, we refer the reader to \citet[][hereafter WP21]{Weinberg2021}. In this work, we use the mSSA module supplied with the \texttt{EXP} code.

\section{Results} \label{sec:results}
This section applies the strategy developed in section~\ref{sec:methods} to the isolated LMC-SMC N-body simulation. First, using halo BFEs, we will show that it is possible to distinguish the torques applied by the LMC's distorted halo and the SMC on the LMC's disk (section~\ref{sec:torques}). Then, using disk BFEs, we will quantify the perturbations in the LMC disk that manifest from a combination of the halo's and the SMC's influence (section~\ref{sec:lmc_disk_warps}). Finally, using mSSA, we will find features in the temporal evolution of the LMC's disk that are the most correlated with the LMC halo's influence versus the SMC's influence (section~\ref{sec:mssa}). The above three sub-sections combined constitute our framework for distinguishing the distorted halo's influence from the satellite's influence on the primary's stellar disk.

\subsection{Torques Experienced by the LMC's Disk Leading to Vertical Distortions} \label{sec:torques}

Here, we compute the torques experienced by the LMC's disk that lead to vertical distortions of the disk. These torques arise from the SMC directly, as well as the SMC-induced LMC halo distortions. All torque calculations are performed in the LMC disk cartesian frame (section \ref{sec:sims}).

The torque experienced by any region of the LMC disk will be a 3-D vector in this coordinate frame. Only the in-plane components of the torque vector will contribute to vertical distortions of the disk, which we will analyze in this work. The z-component of the torque vector will impact the angular momentum of the LMC's disk and induce morphological distortions in the disk plane; this will be the subject of future work.

The procedure of \cite{Gomez2016} (hereafter G16) is followed to compute in-plane torques acting on the LMC's disk. First, the LMC's disk is binned in radial annuli of 2 - 4 kpc, 4 - 6 kpc, 6 - 8 kpc, and 8 - 10 kpc. Each annulus is populated with $N_t = 2000$ tracer particles, distributed uniformly along the azimuth. In a radial annulus $i$, let the force experienced by the particle $j$ at a given instant of time $t$ from the LMC's halo and the SMC be $\vect{F^{\rm halo}_{ij}}(t)$ and $\vect{F^{\rm SMC}_{ij}}(t)$ respectively. Then, the total specific torque experienced by that annulus from the halo or SMC is:
\begin{equation}\label{eq:sp_torque}
    \vect{\tau^{\rm halo/SMC}_i} = \frac{\sum_j \vect{R_{ij}} \times \vect{F^{\rm halo/SMC}_{ij}}(t)}{M_i}
\end{equation}
\noindent where $M_i$ is the mass of that annulus. $M_i$ is determined using the density profile of the initial LMC disk (eq. \ref{eq:disk_dens_profile}).

The magnitude of the in-plane specific torque at a given annulus is:
\begin{equation} \label{eq:in_plane_torque}
    \tau^{\rm halo/SMC}_{i, \rm xy} = \sqrt{(\tau^{\rm halo/SMC}_{i, x})^2 + (\tau^{\rm halo/SMC}_{i, \rm y})^2}
\end{equation}

Hereafter, the word \enquote{torque} will refer to the 
magnitude of the {\em in-plane specific} torque.

\subsubsection{The Impact of Torques from the LMC's Distorted Halo on the LMC's Stellar Disk} \label{sec:halo_torques}

Using the BFE representation of the LMC's halo, the \texttt{EXP} computed halo potential ($\tilde{\Phi}_{\rm halo}(\vect{R_{ij}})$) is extracted at the location of the tracer particle $j$ in the annulus $i$. The force on the tracer particle due to the distorted halo is computed by performing a finite difference gradient of the potential:
\begin{equation}
    \vect{F^{\rm halo}_{ij}} = - M_{ij} \sum_k \frac{\tilde{\Phi}_{\rm halo}({\vect{R_{ij}}}_k + \epsilon) - \tilde{\Phi}_{\rm halo}({\vect{R_{ij}}}_k - \epsilon)}{2\epsilon} \hat{\vect{e_k}}
\end{equation}
\noindent where the index $k$ runs over the 3 cartesian directions, and $\epsilon = 0.5$ kpc is the half-width of an annulus. The torque on an annulus due to the distorted halo $(\tau^{\rm halo}_{i, \rm xy})$ is computed using eq. (\ref{eq:sp_torque}) and (\ref{eq:in_plane_torque}).

\begin{figure*}
    \centering
    \includegraphics[width=\textwidth]{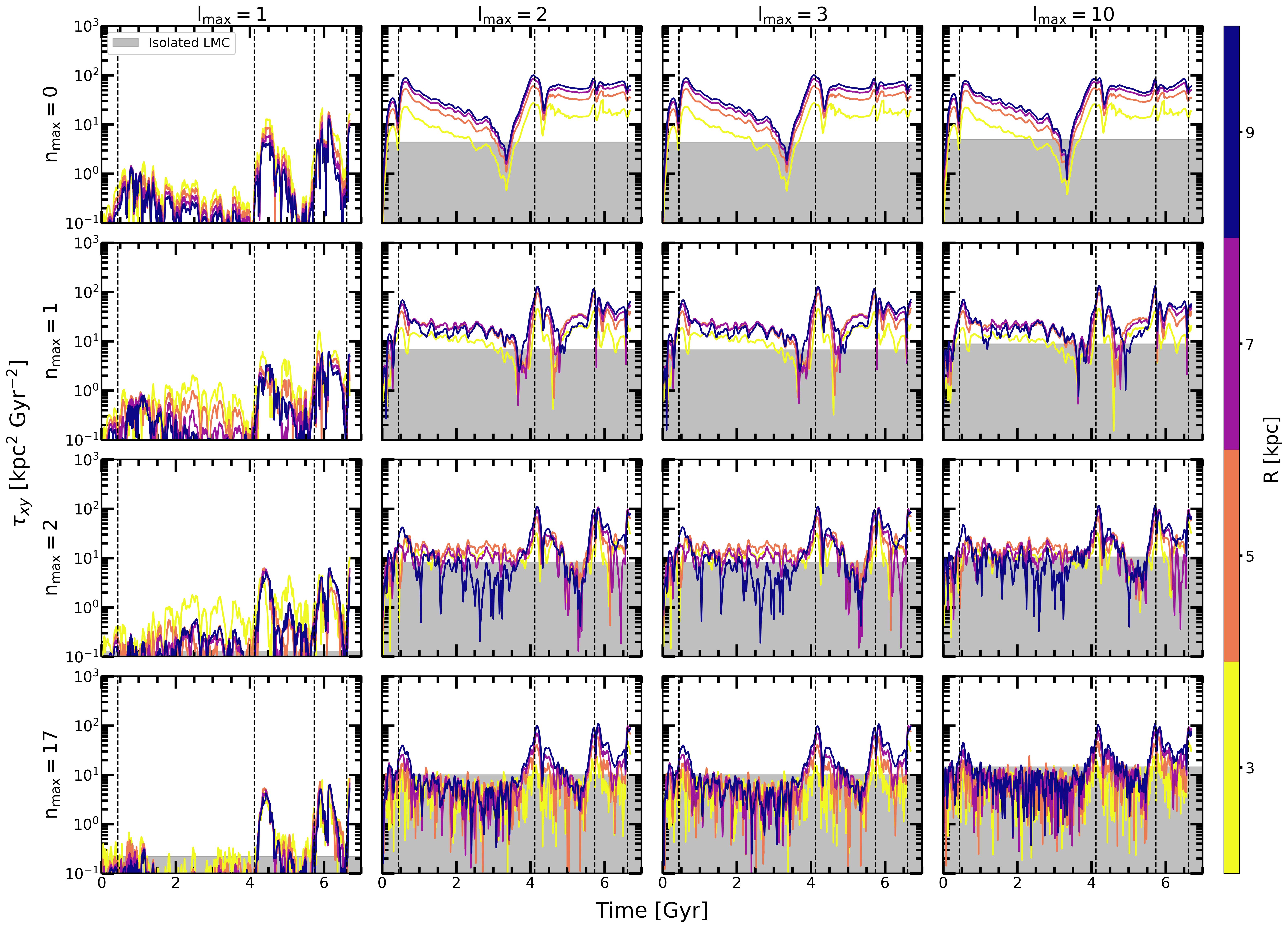}
    \caption{In-plane torques ($\tau_{i, xy}$) experienced by annuli of the LMC's disk from the distorted LMC halo in the LMC-SMC simulation are decomposed into halo modes. Different $l_{\rm max}$ and $n_{\rm max}$ truncation orders chosen for the LMC halo BFE are shown along the columns and rows, respectively. The torques on different radial bins in the LMC disk (2 kpc width; depicted by the color map) are plotted as a function of time. The vertical dashed lines denote the LMC-SMC pericentric epochs. The grey shaded band denotes the $95$ percentile range of halo torques across time and across all radial bins in the isolated LMC simulation (these values are smaller than the y-axis lower limit for $l_{\rm max} = 1$, $n_{\rm max} = 0, 1, 2$). At a given $l_{\rm max}$, torques exerted by modes $n > 1$ in the LMC-SMC simulation fall within the range of torques in the LMC-only simulation, implying that $n > 1$ halo torques are currently not distinguishable from particle noise. At a given $n_{\rm max}$, increasing $l$ beyond $2$ does not change the torques. Hence, the torques applied by the distorted LMC halo on the disk can be reliably distinguished from noise by truncation orders of $l_{\rm max} = 2$, $n_{\rm max} = 1$.}
    \label{fig:nmax_lmax_grid}
\end{figure*}

In Figure \ref{fig:nmax_lmax_grid}, the halo torques are computed by truncating the BFE to different $(l_{\rm max}, n_{\rm max})$ orders. Also shown is the $95$ percentile range ($< 14.5$ kpc$^2$ Gyr$^{-2}$) of the halo torques across time and across radial bins for the LMC-only control simulation, which serves as an estimated noise floor. Adding additional $l$ orders beyond $l_{\rm max} = 2$ does not significantly affect the $\tau^{\rm halo}_{i, \rm xy}$ values. Further, the value of torques applied by orders beyond $n > 1$ are comparable to the noise floor imposed by the simulation's mass resolution. As such, with the current resolution, it cannot be determined if the torques applied by modes $n > 1$ in the LMC-SMC simulation are physically real, or if they are caused by particle noise. We do not consider the torques exerted by $n > 1$ halo modes in further analysis due to our inability to interpret them. For a truncation order of $(l_{\rm max} = 2, n_{\rm max} = 1)$, the maximum value of $\tau^{\rm halo}_{i, \rm xy}$ across time and across LMC annuli is more than an order of magnitude above the noise level. As such, we can confidently say that torques applied by $n \leq 1$ halo modes can be distinguished from the particle noise. For $l_{\rm max} = 1$, the value of torques in the control simulation are below $10^{-1}$ kpc$^2$ Gyr$^{-2}$.

\begin{figure}
    \centering
    \includegraphics[width=\columnwidth]{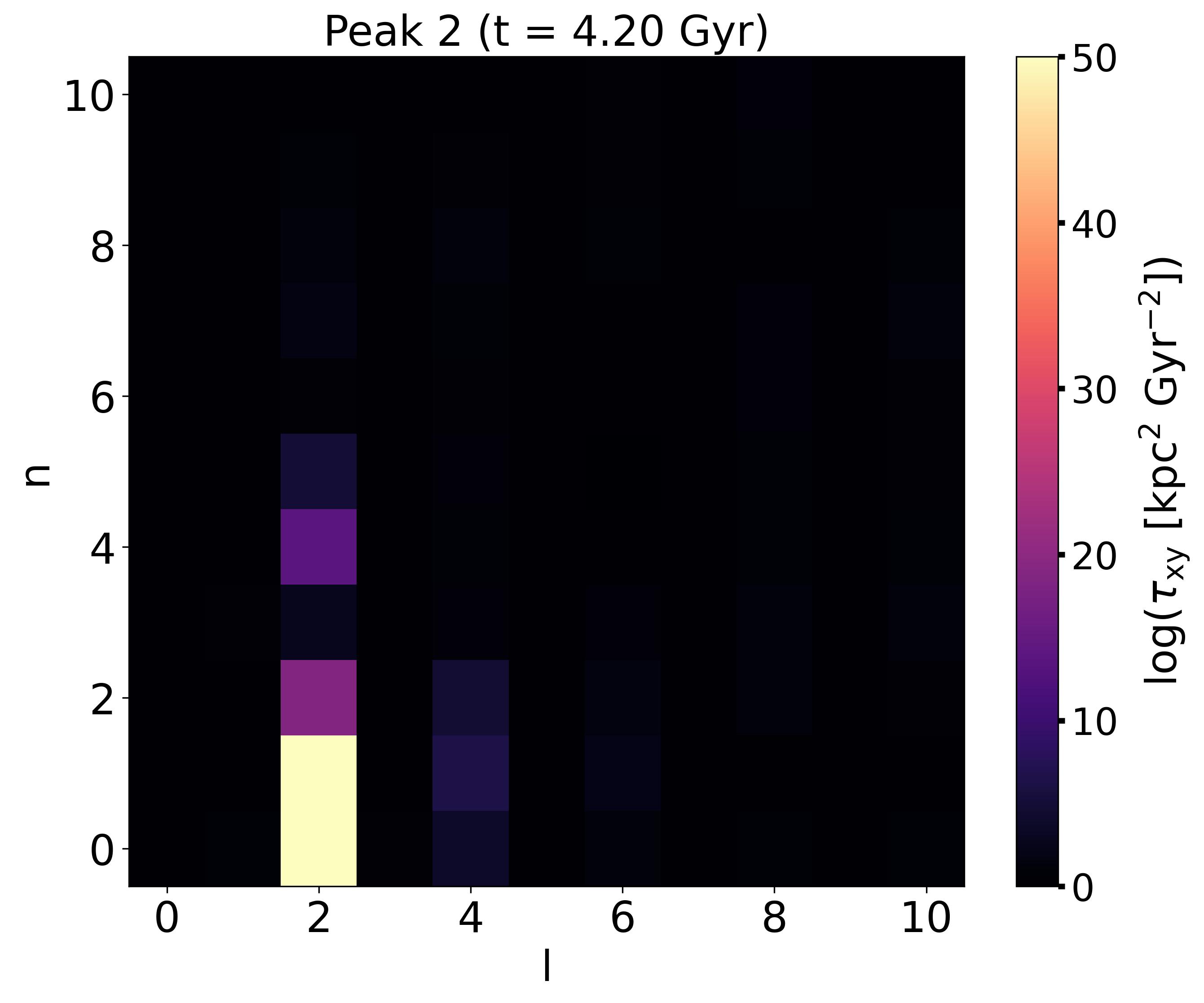}
    \caption{Contribution of the individual halo modes to vertically warping the LMC's disk via torques from the distorted LMC halo. The epoch chosen ($t = 4.20$ Gyr) corresponds to the peak of the $l_{\rm max} = 2$, $n_{\rm max} = 1$ halo torque (see Figure \ref{fig:nmax_lmax_grid}), just after the 2$^{\rm nd}$ LMC-SMC pericenter. The color scale denotes the torque experienced by the 8 - 10 kpc radial bin of the LMC disk. The quadrupole ($l = 2$) mode of the LMC's halo) dominates the LMC halo torques acting on the LMC's disk. This torque is strongest for $n = 0, 1$, consistent with Figure \ref{fig:nmax_lmax_grid}.}
    \label{fig:n_l_mesh}
\end{figure}

To further quantify the torque contributions of different halo orders, the halo torque experienced by the outermost LMC annuli (8 - 10 kpc) is computed for individual $(n, l)$ modes. Figure \ref{fig:n_l_mesh} shows these contributions for the epoch where the $(l_{\rm max} = 2, n_{max} = 1)$ torque is maximized ($t = 4.20$ Gyr), just after the 2$^{\rm nd}$ LMC-SMC pericenter (Figure \ref{fig:lmc_smc_orbit}). Note that this epoch of local maximum happens $\sim 100$ Myr after the pericenter. Consistent with the findings of Figure \ref{fig:nmax_lmax_grid}, $l = 2$ and $n = 0, 1$ orders are the dominant contributors to the halo torque.

These findings are consistent with the analytical works of \cite{Weinberg1998, Vesperini2000}, who showed that the $l = 1$ mode or the halo dipole cannot exert in-plane torques on the disk that are responsible for vertical perturbations, and that the $l = 2$ mode is the lowest $l$-order capable of exerting in-plane torques. The potential corresponding to the $l = 1$ term is proportional to the radial coordinate, leading to a spatially constant force field that does not exert any net torques on an annulus in the disk plane. Moreover, in the LMC-SMC simulations, the $l > 2$ orders have a smaller contribution to the BFE by a factor of $\sim 5$ (F26) in gravitational power, and hence they do not exert significant torques. 

\subsubsection{The Impact of the SMC's Direct In-Plane Torques on the LMC's Disk}

The gravitational force applied by the SMC on a tracer particle $j$ in the annulus $i$ is:
\begin{equation}
    \vect{F^{\rm SMC}_{ij}}(t)  = \frac{GM_{\rm SMC} (t)M_{ij}}{|\vect{R_{\rm SMC}}(t) - \vect{R_{ij}}|^3} [\vect{R_{\rm SMC}}(t) - \vect{R_{ij}}] 
\end{equation}
\noindent where $M_{ij} = M_i/N_t$ is the mass of the tracer particle, and $\vect{R_{ij}}$ is the position vector of the tracer particle. 

$M_{\rm SMC}(t)$ is defined as the SMC's bound mass at an instant of time in the simulation. $M_{\rm SMC}$(t) was calculated by F26 using BFEs of the SMC in the LMC-SMC N-body simulation (see their Figure 9). The SMC's halo becomes tidally disrupted throughout the LMC-SMC interactions, where F26 showed that by each of the $4$ pericenters, the SMC has lost $15\%, 52\%, 72\%, 86\%$ of its initial halo mass due to tidal disruption. By only considering the bound mass, torques from the unbound SMC halo particles will be missed. However, these halo particles are expected to become out of phase with the SMC's orbit after a dynamical time in the LMC halo \citep[e.g.][]{Johnston1996, Tremaine1999, Necib2026}, and consequently, these particles are not expected to exert significant net torques on the disk. Using Figure 9 of F26, the mass loss rate of the SMC is estimated to be:
\begin{equation}
    \dot{M}_{\rm SMC} \approx 1.7 \times 10^9 \:\: {\rm M}_{\odot} \: {\rm Gyr}^{-1}
\end{equation}
Hence, the mass of disrupted SMC halo particles that are still in phase with the SMC's orbit at a given moment in time is roughly given by $\Delta M_{\rm phase} = \dot{M}_{\rm SMC} \times t_{\rm dyn, halo}$. The LMC halo dynamical time $t_{\rm dyn, halo}$ at the largest LMC-SMC pericentric distance of $\approx 35$ kpc is $\approx 0.5$ Gyr, which yields an upper bound of $\Delta M_{\rm phase} = 8.5 \times 10^8$ M$_{\odot}$ --- still an order of magnitude smaller than the SMC's bound mass at the end of the simulation. As such, our assumption that most of the SMC's torques arise from its bound halo is reasonable. The torque on a given annulus due to the SMC $(\tau^{\rm SMC}_{i, \rm xy})$ is computed using eq. (\ref{eq:sp_torque}) and (\ref{eq:in_plane_torque}). 
\\
\\
\begin{figure*}
    \centering
    \includegraphics[width = \textwidth]{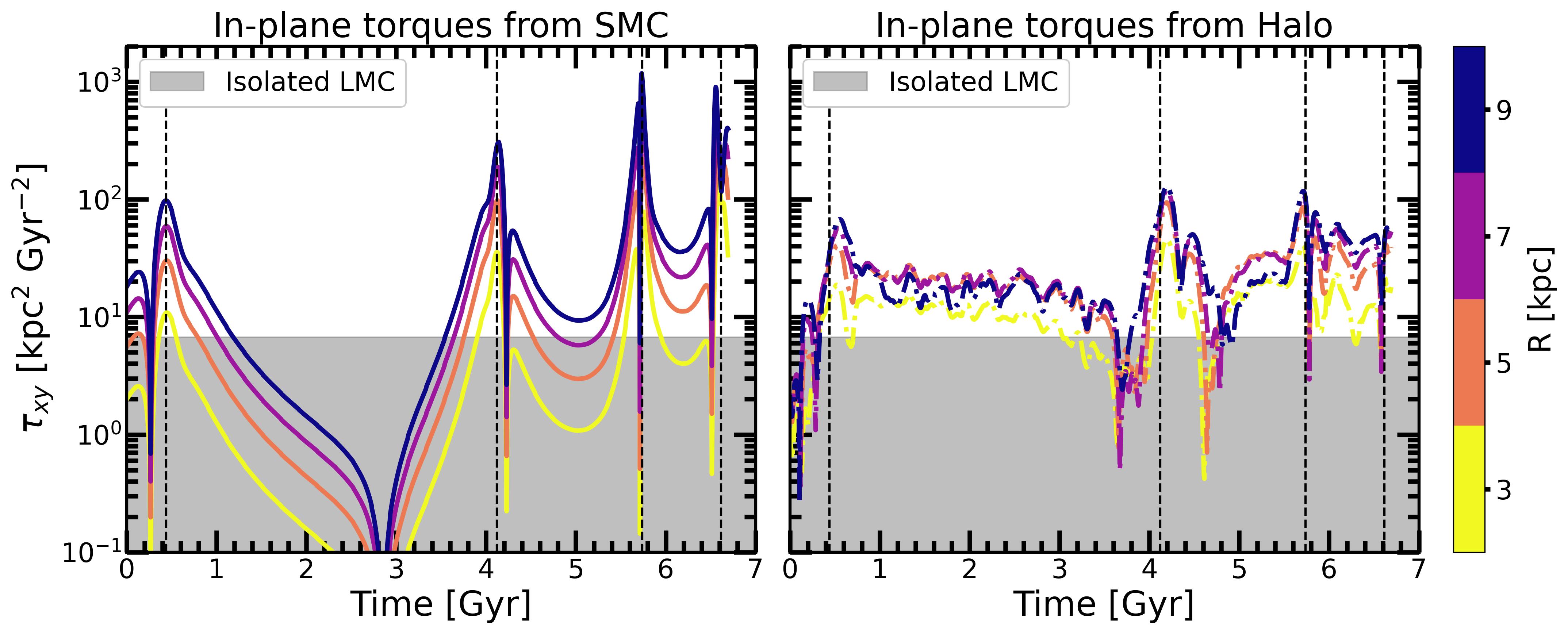}
    \caption{The in-plane torques ($\tau_{i, xy}$) experienced by the LMC's disk directly from the SMC ({\em left panel}) and the LMC's distorted halo ({\em right panel}) are plotted as a function of time throughout the LMC-SMC simulation. The dashed vertical lines correspond to the LMC-SMC pericenters. The color map denotes different radial bins (2 kpc width, as in Figure \ref{fig:nmax_lmax_grid}) in the LMC disk. {\em Left panel:} the SMC's direct torques ($\tau^{\rm SMC}_{i, xy}$) peak at the LMC-SMC pericenters for all LMC radial bins, but are maximized at large LMC radius. $\tau^{\rm SMC}_{i, xy}$ increases as the LMC-SMC pericentric distance reduces from $\approx 35$ kpc to $\approx 7.5$ kpc. {\em Right panel:} The halo torques ($\tau^{\rm halo}_{i, xy}$) are computed using the LMC halo's $n_{\rm max} = 1$, $l_{\rm max} = 2$ BFE representation (Figure \ref{fig:nmax_lmax_grid}). The grey shaded band denotes the $95$ percentile range of the $n_{\rm max} = 1, l_{\rm max} = 2$ halo torque across time and across all radial bins for the LMC-only simulation. $\tau^{\rm halo}_{i, xy}$ values in the LMC-SMC simulation are comparable to $\tau^{\rm SMC}_{i, xy}$ for the first two pericenters, but are almost an order of magnitude smaller than $\tau^{\rm SMC}_{i, xy}$ for the third and fourth pericenter. However, the halo torques are up to an order of magnitude larger than the SMC torques when the SMC is not at pericenter.}
    \label{fig:smc_direct_torques}
\end{figure*}

In Figure \ref{fig:smc_direct_torques}, we perform a comparative study of the torques exerted by the SMC directly ({\em left panel}) and the torques exerted by the LMC halo distortions ({\em right panel}). The latter are computed using a BFE truncation order of $l_{\rm max} = 2, n_{\rm max} = 1$ (Figure \ref{fig:nmax_lmax_grid}).

Both the direct torques and the halo torques peak near the LMC-SMC pericenters, where the former peak almost exactly at the pericenters and the latter are delayed by $\sim 100 - 200$ Myr. Both torques decline after the pericenters, but the SMC's torques decline at least 2 - 3 times faster than the halo torques. The strength of both the torques generally increases with radius in the LMC disk. Note that the sudden dip in the strength of the SMC's torques in the vicinity of a pericenter is due to the SMC crossing the LMC's disk plane ($\theta = 0$ in Figure \ref{fig:lmc_smc_orbit}). 

The SMC's torques attaining a local maximum near the pericenters is a direct consequence of the SMC's force field being proportional to $R_{\rm SMC-LMC}^{-2}$. The halo torques being delayed is likely due to the additional time required for the halo quadrupole response to fully develop (F26). The halo torques being more sustained than the satellite's torques was also noted by \cite{Weinberg1998, Gomez2016, Laporte2018}, and is likely a consequence of the timescale involved in the decay of the quadrupole. For example, using Figure 10 of F26, we can estimate that it takes $\approx 300$ Myr for the halo quadrupole strength to reduce by a factor of 2 after attaining the first peak. But, $R_{\rm SMC-LMC}$ increases by $\approx 2.5$ times over 300 Myr after the first pericenter. Hence, the SMC's direct torques reduce by a factor of $2.5^2 = 6.25$, but the halo torques only reduce by a factor of 2 over the timescale considered. Thus, the SMC's torques reduce much faster than the halo torques after attaining a local maximum.

The SMC's torques at the last two pericenters are almost an order of magnitude stronger than that at the first two pericenters. This evolution in the strength of SMC torques can be attributed to the decay of the LMC-SMC orbit due to dynamical friction, where the pericentric distance decreases from $\approx 35$~kpc at the first pericenter to $\approx 7.5$~kpc at the fourth pericenter. Consequently, at the initial two pericenters ($R_{\rm SMC-LMC} \gtrsim 20$ kpc), the halo torques are comparable in strength to the SMC's torques. At the latter two pericenters ($R_{\rm SMC-LMC} \lesssim 20$ kpc), the SMC's torques are an order of magnitude stronger than the halo torques. However, at all apocenters, the halo torques are stronger than the SMC's torques by at least a factor of $\sim 5$, likely because the halo torques are more sustained after the pericenters. 

The above analysis, based on Figure \ref{fig:smc_direct_torques}, quantifies the instantaneous torques applied to the LMC's disk. However, it is also important to analyze the cumulative effect of the torques on the disk. As shown by e.g. \cite{Mayer2001, Lokas2015, Kazantzidis2017, Semczuk2018}, the vertical structure and kinematics of a disk can significantly evolve even due to weak but sustained tidal fields. Although the SMC's torques are much stronger at the later pericenters in an impulsive way, the halo torques are generally more sustained. As such, it is non-trivial and interesting to understand which of the two torques has a larger cumulative contribution to the disk.

\begin{figure*}
    \centering
    \includegraphics[width=0.49\textwidth]{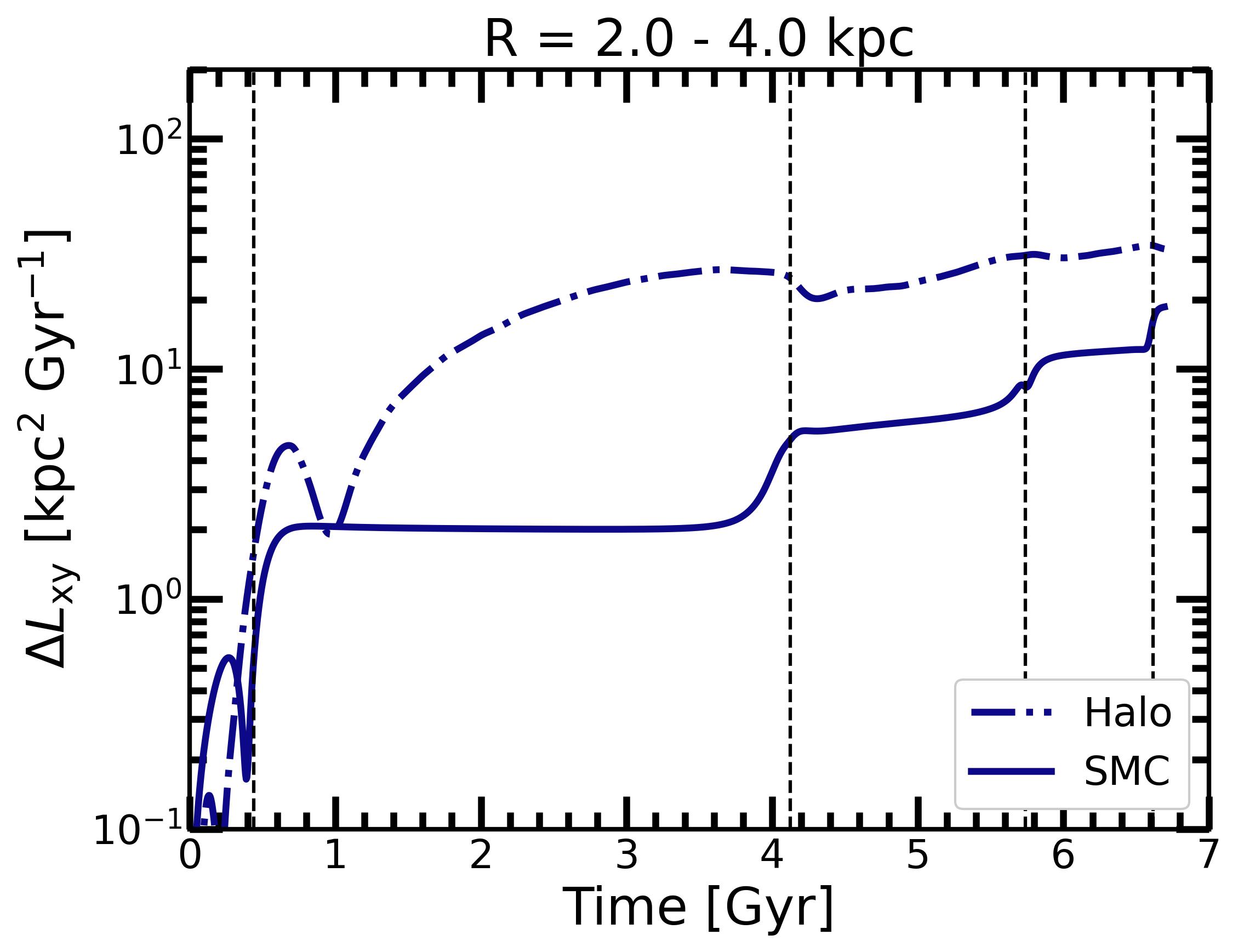}
    \includegraphics[width=0.49\textwidth]{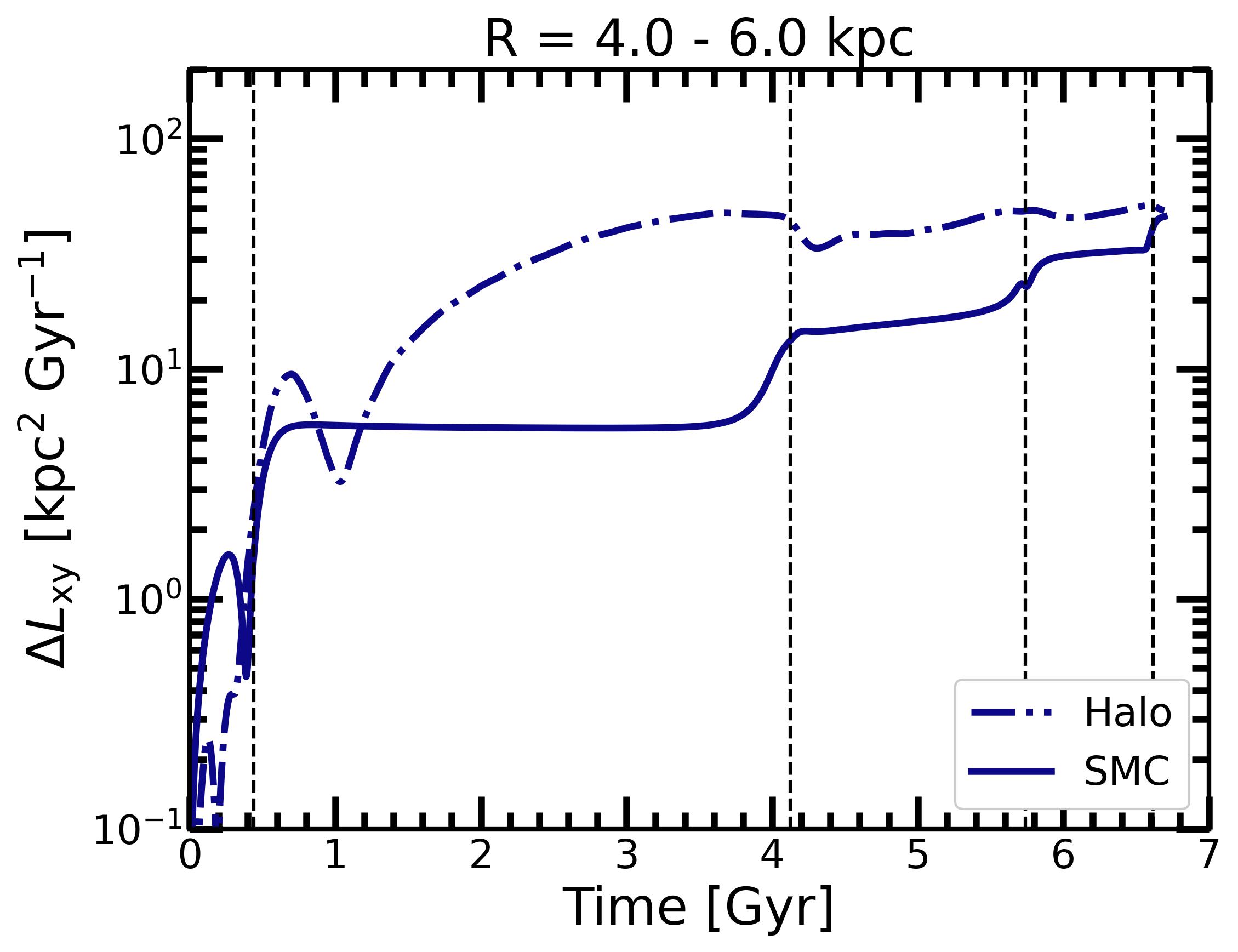}
    \includegraphics[width=0.49\textwidth]{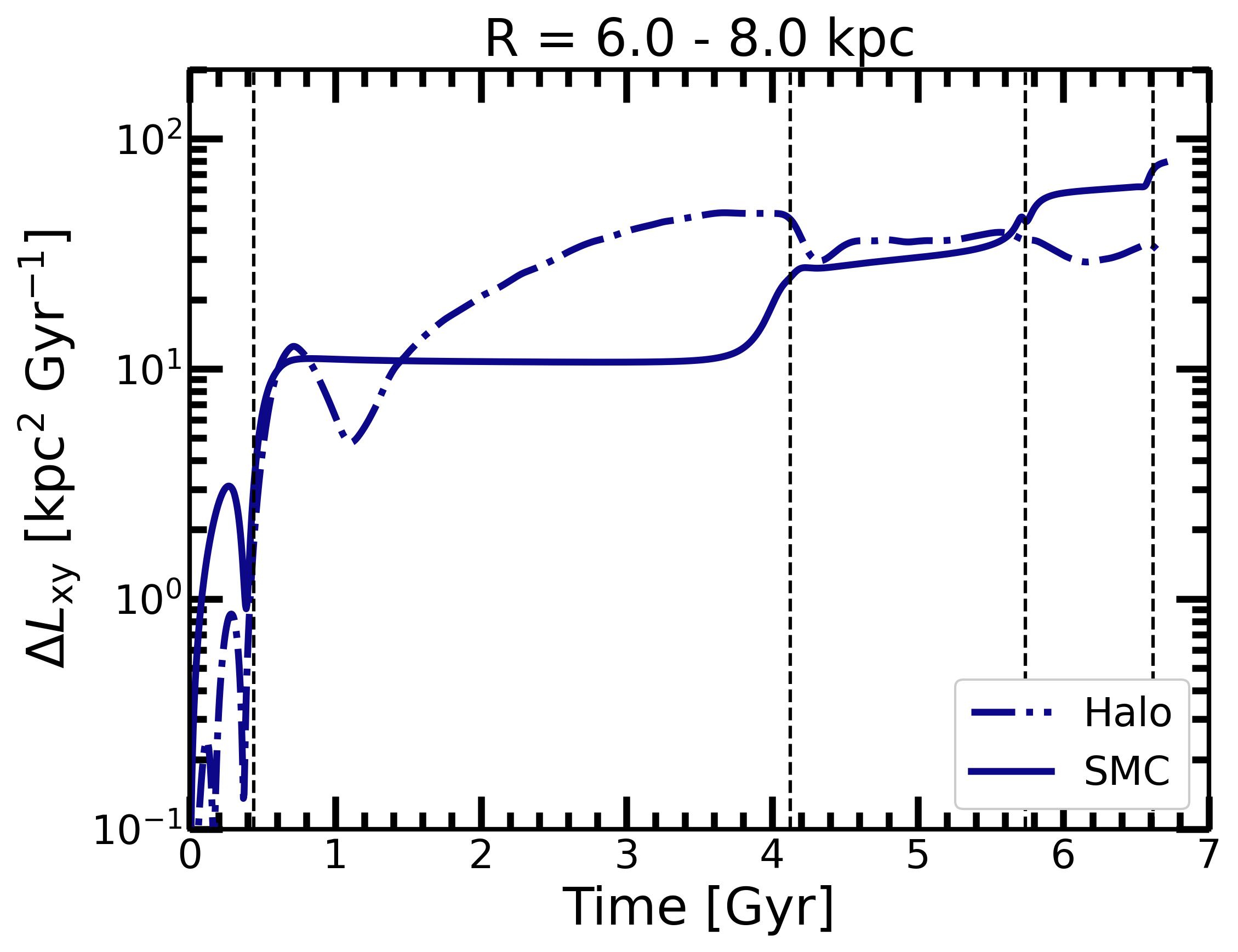}
    \includegraphics[width=0.49\textwidth]{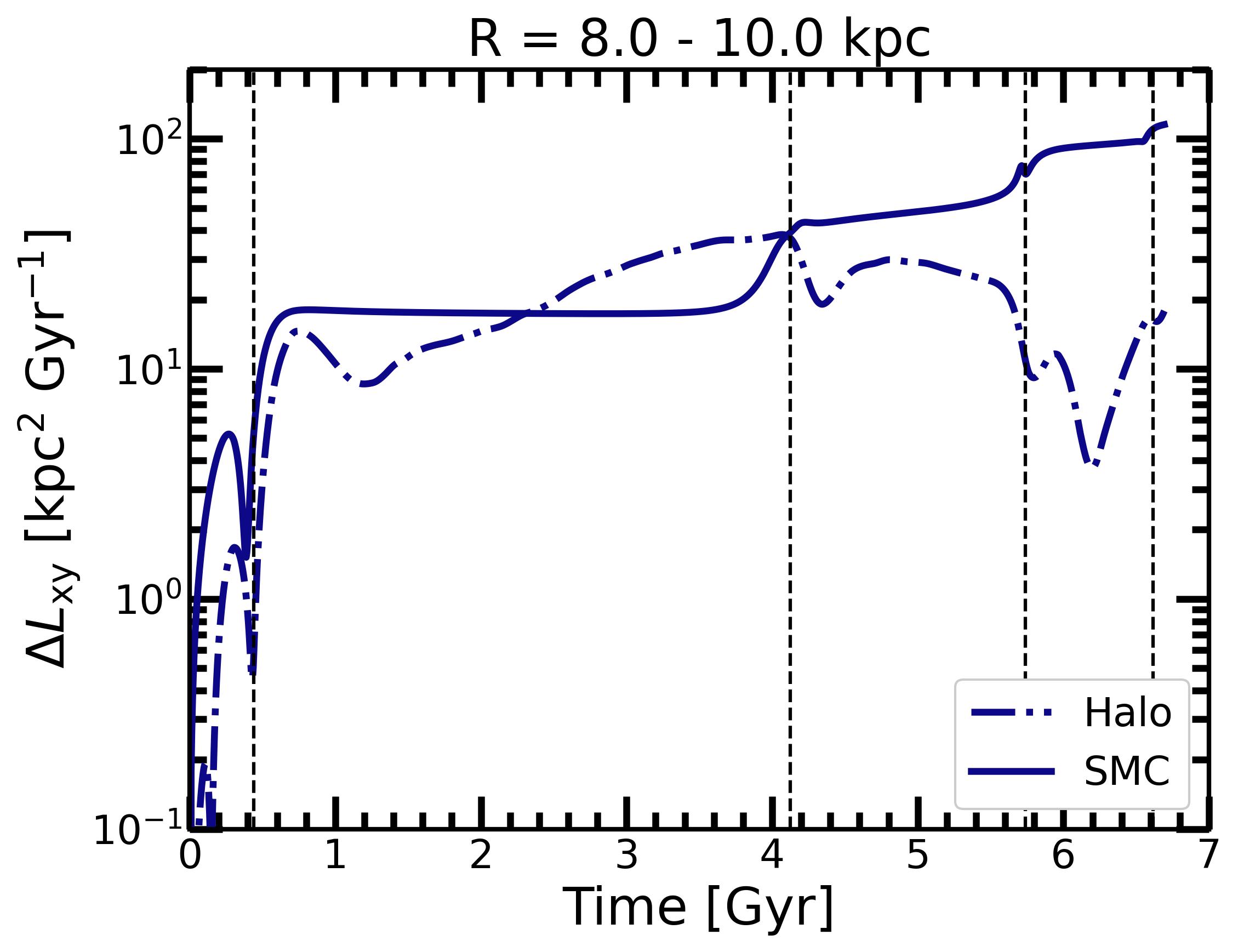}
    \caption{The cumulative change of the in-plane angular momentum in the LMC disk due to the torques from the SMC and the distorted LMC halo, plotted as a function of time throughout the LMC-SMC simulation. The 4 panels correspond to 4 radial bins in the LMC disk. The LMC-SMC pericentric epochs are marked by the vertical dashed lines. For the inner LMC disk (2-6 kpc; top panels), the cumulative effect of the halo torques is stronger than the SMC's torques by up to an order of magnitude almost throughout the simulation, although the difference reduces by the $4^{\rm th}$ pericenter. For the outer LMC disk (6-10 kpc; bottom panels), the cumulative SMC torques dominate after the second or third pericenter. Halo torques are thus potentially dominant in the inner disk of the LMC, depending on the exact orbital history of the SMC.}
    \label{fig:cumulative_torques}
\end{figure*}

The cumulative change of in-plane angular momentum of the $i^{\rm th}$ radial bin ($\Delta L^{\rm halo/SMC}_{i, \rm xy}$, responsible for vertical perturbations) of the LMC disk, resulting from the SMC's direct torques or the LMC halo torques, is defined as:
\begin{equation}
\begin{split}
    \Delta L^{\rm halo/SMC}_{i, \rm xy} (t) = & \\\sqrt{\left(\int_0^t \tau^{\rm halo/SMC}_{i, x} (t^\prime) dt^\prime \right)^2 + \left(\int_0^t \tau^{\rm halo/SMC}_{i, y} (t^\prime) dt^\prime \right)^2}
\end{split}
\end{equation}
\noindent Figure \ref{fig:cumulative_torques} illustrates the resulting cumulative change of in-plane angular momentum from the SMC and the halo as a function of time for the 4 radial bins.

For the inner LMC disk (R $< 6$~kpc), the cumulative contribution of the halo torques is almost consistently larger than that of the SMC's torques, with a maximum difference of almost an order of magnitude occurring between the first two pericenters. The difference between the torques reduces with time, and by the fourth pericenter, the halo's contribution is at most a factor of 2 larger than the SMC's contribution. The SMC contributes significantly more in the outer LMC disk (R $> 6$~kpc) as compared to the inner disk. In particular, by the fourth pericenter, the SMC's contribution is up to an order of magnitude larger than the halo's contribution.

The halo's relatively stronger contribution in the LMC's inner disk can be understood as follows. \cite{Weinberg1998} suggested that the halo's response to a satellite includes significant distortions in the inner regions of the halo where the disk resides. From Figure 4 of F26, it is evident that the $l = 2$ quadrupole includes significant density perturbations (up to $\sim 15\%$ of the spherically averaged field) in the inner LMC halo ($R < 20$ kpc). Because of this de-localization of density perturbations, the halo is able to torque the inner disk more efficiently than the satellite itself.

Using the example of an LMC-SMC-like binary orbit, we have shown that in-plane torques resulting from satellite-induced distortions to the primary galaxy's halo can be as, or more, important than the in-plane torques from the satellite. In particular, halo torques can dominate in the inner stellar disk, depending on the orbital history of the binary. To the best of our knowledge, this work is the first to have rigorously decomposed the torque contributions from orbiting satellites, including their distortions to the host halo, in controlled high-resolution N-body simulations. Importantly, the results of these numerical experiments confirm the theoretical arguments put forward by previous works. There is now motivation to believe that the halo torques on the disk are distinguishable from the satellite torques in LMC-SMC-like systems. So, the next step is to understand how these torques manifest in the stellar structure of the disk, and whether these torques can be distinguished methodically using the disk perturbations.

\subsection{Characterizing Morphological Distortions of the LMC's Stellar Disk using BFEs} \label{sec:lmc_disk_warps}

The goal of this sub-section is to characterize the temporal evolution of the LMC disk morphological perturbations using the disk BFE developed in section \ref{sec:disk_bfe}. As a first step, it needs to be ensured that the disk BFE is reproducing the key morphological features like the simulated LMC's bar, spirals, BP/X structure and warps. Then, we will focus on the temporal evolution of the vertical asymmetries/warps as encoded by the BFE coefficients.

\subsubsection{BFE Reconstructions of the Simulated LMC Disk} \label{sec:disk_bfe_recon}

\begin{figure*}
    \centering
    \includegraphics[width=\textwidth]{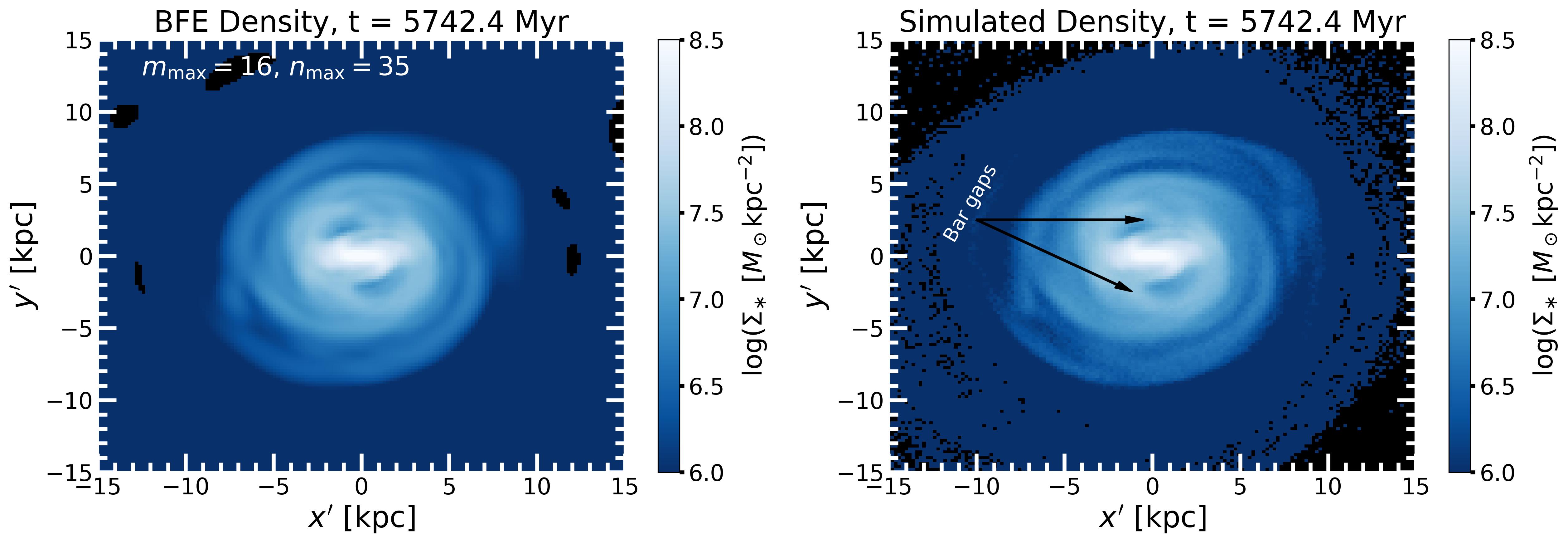}
    \includegraphics[width=\textwidth]{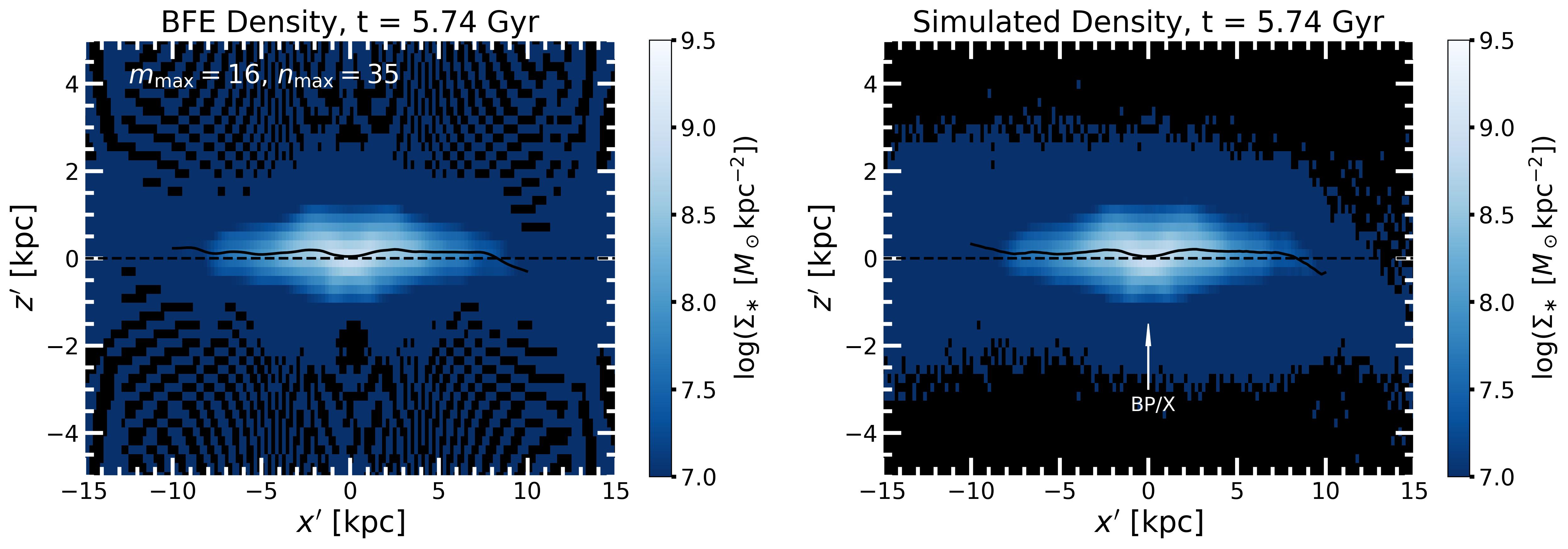}
    \includegraphics[width = \textwidth]{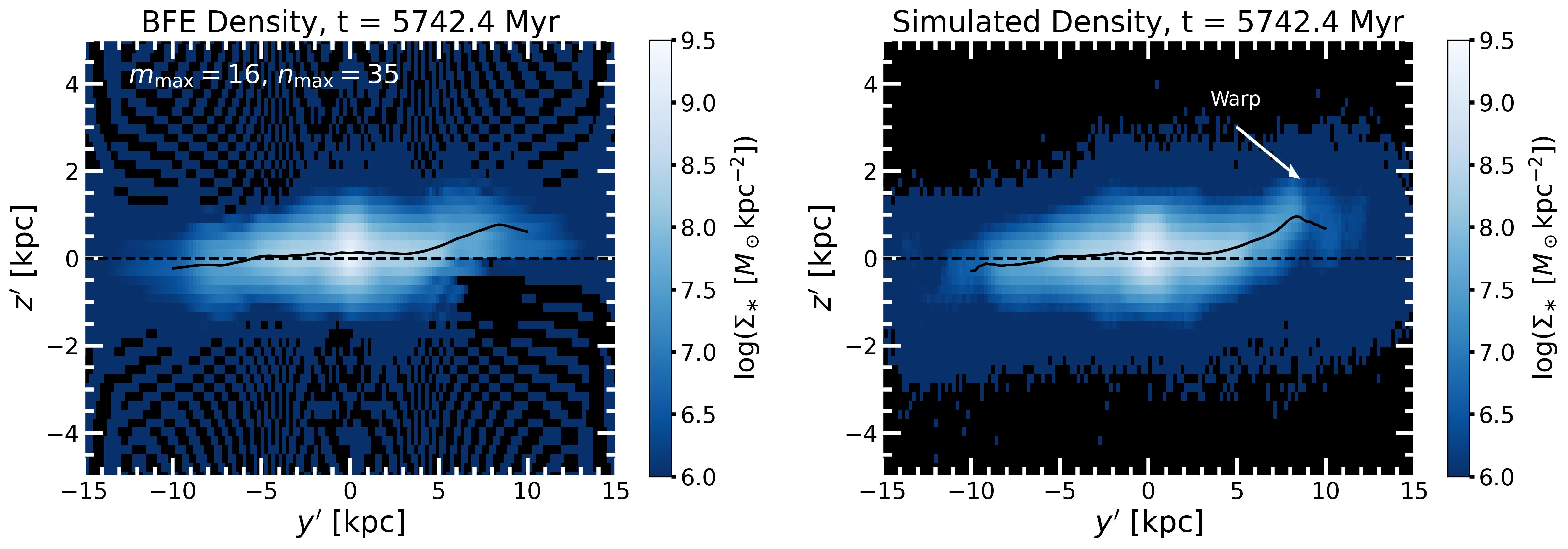}
    \caption{The BFE reconstruction of the disturbed LMC disk in the LMC-SMC simulation. The {\em top, middle and bottom rows} shows the face-on ($x^\prime - y^\prime$) and the two edge-on ($x^\prime - z^\prime$, $y^\prime - z^\prime$) projections respectively in the LMC bar cartesian frame at the third LMC-SMC pericenter/MW Infall epoch (t = $5.74$ Gyr, see Figure \ref{fig:lmc_smc_orbit}). The {\em left column} represents the BFE reconstructed stellar density field of the simulated disk ({\em right column}), where the color scale represents the stellar surface density. In the edge-on projections, the mean $z$ coordinate of stars is denoted by the black solid curve, and the overall disk plane is denoted by the horizontal dashed line at $z=0$. In the face-on projection, the BFE reconstruction captures several features of the LMC disk, like: the central bar; spiral arms at various radii; and the gaps along the bar minor axis. In the edge-on projections, the BFE reconstruction captures features like the X-shaped bar ({\em middle panel}) and the warping of the disk ({\em bottom panel}). The BFE coefficients of the simulated LMC disk reasonably capture both the in-plane and vertical morphological distortions. This figure shows that the LMC disk is expected to be warped by the SMC's influence (direct and halo distortions) prior to the system's infall into the MW. A movie illustrating how the BFE captures finer disk structures as more expansion orders are added can be found at \url{https://github.com/himanshrathore/rathore2026a/blob/main/bfe_reconstruction_movie.mp4} (1.6 MB, GPL-3.0 License, 11 seconds duration).}
    \label{fig:bfe_reconstruction_eg}
\end{figure*}

Figure \ref{fig:bfe_reconstruction_eg} shows an example comparison between the BFE reconstructed disk and the simulated LMC disk at the third LMC-SMC pericenter ($t = 5.74$ Gyr), which also corresponds to the MW infall epoch in the B12 timeline (Figure \ref{fig:lmc_smc_orbit}). The simulated and BFE-reconstructed density fields ($\rho_{\rm disk}(x^\prime, y^\prime, z^\prime)$) are sampled on a 3-D grid in the LMC bar cartesian frame (section \ref{sec:sims}) --- $x^\prime, y^\prime, z^\prime \in (-15, 15) \: {\rm kpc} \times (-15, 15) \: {\rm kpc} \times (-5, 5) \: {\rm kpc}$ with a grid size of $0.2 \: {\rm kpc} \times 0.2 \: {\rm kpc} \times 0.2 \: {\rm kpc}$. The face-on surface density projections ($\Sigma(x^\prime, y^\prime)$) are obtained by:
\begin{equation}
    \Sigma_{\rm disk}(x^\prime, y^\prime) = \int_{z^\prime} \rho_{\rm disk}(x^\prime, y^\prime, z^{\prime\prime}) dz^{\prime\prime}
\end{equation}
\noindent and the two edge-on surface density projections are obtained by:
\begin{equation} \label{eq:edge_on_sigma}
    \Sigma_{\rm disk}(x^\prime, z^\prime) = \int_{y^\prime} \rho_{\rm disk}(x^\prime, y^{\prime\prime}, z^\prime) dy^{\prime\prime}
\end{equation}
\noindent and,
\begin{equation} \label{eq:edge_on_sigma}
    \Sigma_{\rm disk}(y^\prime, z^\prime) = \int_{x^\prime} \rho_{\rm disk}(x^{\prime\prime}, y^{\prime}, z^\prime) dx^{\prime\prime}
\end{equation}
\noindent where the integrals are evaluated discretely on the grid using the python package \texttt{scipy.integrate.simpson}. For the two edge-on projections, we also show the mean vertical deviations, which are computed as follows:
\begin{equation} \label{eq:mean_z}
    \bar{z}_x (x^\prime) = \frac{\int_{z^\prime} z^{\prime\prime} \Sigma(x^\prime, z^{\prime\prime}) dz^\prime}{\int_{z^\prime} \Sigma(x^\prime, z^{\prime\prime})dz^{\prime\prime}} 
\end{equation}
\noindent and,
\begin{equation} \label{eq:mean_z}
    \bar{z}_y (y^\prime) = \frac{\int_{z^\prime} z^{\prime\prime} \Sigma(y^\prime, z^{\prime\prime}) dz^\prime}{\int_{z^\prime} \Sigma(y^\prime, z^{\prime\prime})dz^{\prime\prime}} 
\end{equation}

From the face-on projections, the BFE reasonably captures several in-plane morphological perturbations in the disk, like --- the central bar, spiral features, and under-densities along the bar minor axis. In the edge-on projections, the BFE captures features like --- the BP/X shape in the bar, vertical deviations, and warps in the disk. In particular, the warp at $x \approx 10$ kpc with a mean vertical deviation of $\sim 1$ kpc is represented in the BFEs, even though the exact shape of the warp may not be captured due to the limited support of the basis functions away from the disk mid-plane. 

\begin{figure*}
    \centering
    \includegraphics[width=0.49\textwidth]{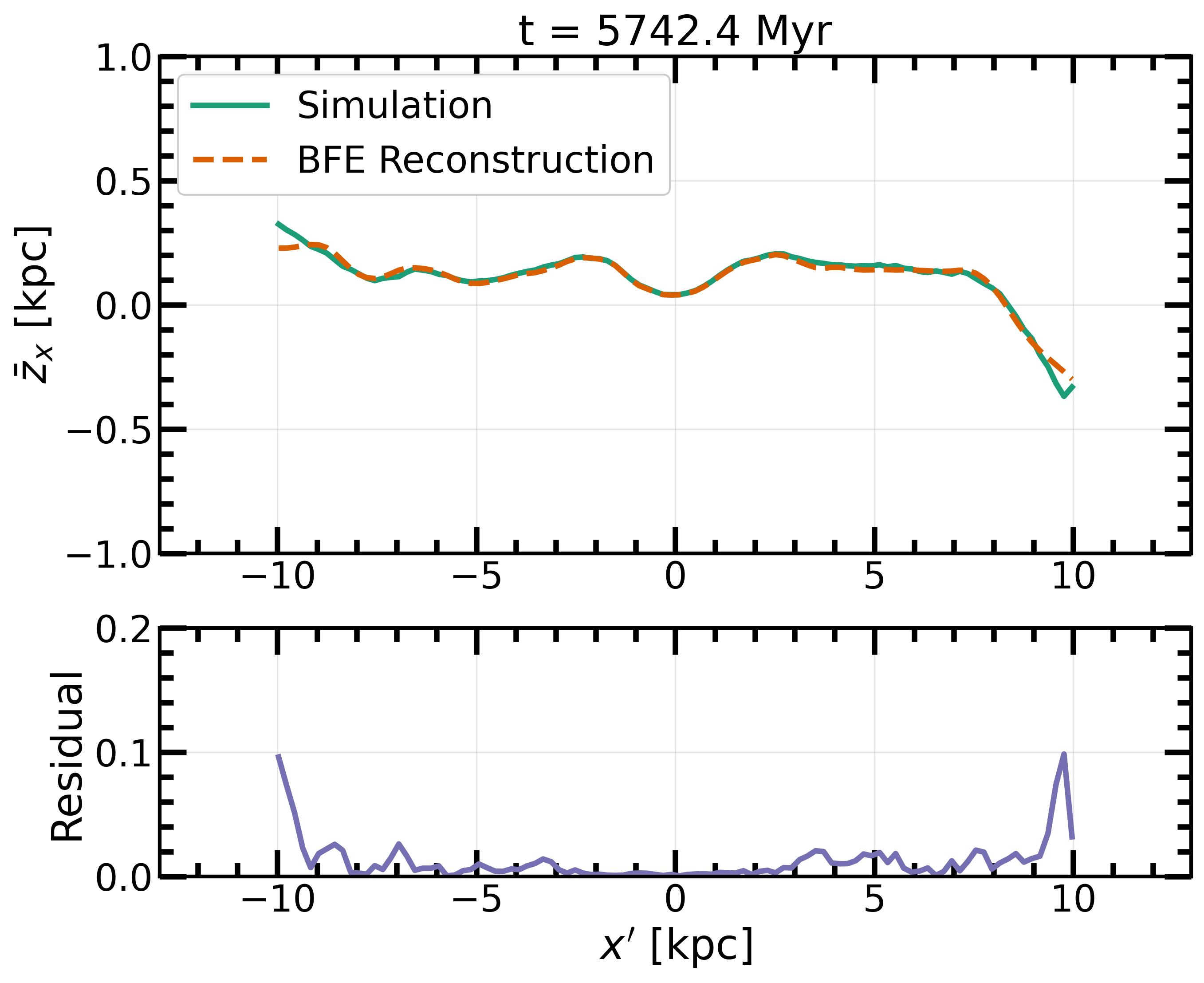}
    \includegraphics[width=0.49\textwidth]{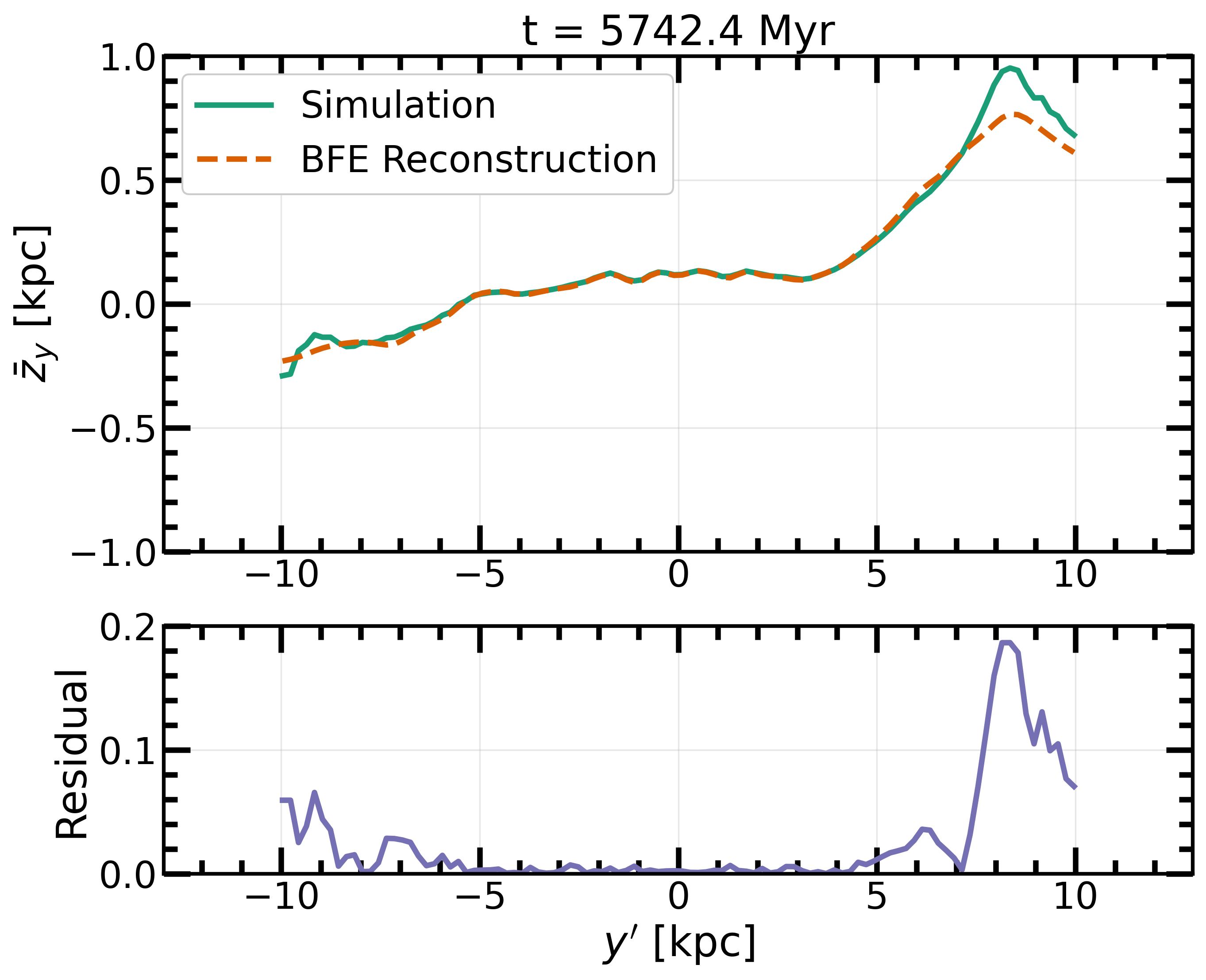}
    \caption{The mean vertical deviations ($\bar{z}_x$ and $\bar{z}_y$, see Figure \ref{fig:bfe_reconstruction_eg} and text for definition) are quantified and compared for both the BFE-reconstructed and simulated LMC disk at the $3^{\rm rd}$ LMC-SMC pericentric/MW infall epoch. The residual is the difference between the $\bar{z}$ values of the reconstruction and the simulation. The BFE and simulation agree within 0.2 kpc, which is comparable to the observational uncertainties on the LMC's vertical perturbations. As such, the BFE adequately captures the vertical distortions for the bulk of the LMC's disk. This figure also illustrates that the LMC disk is expected to have been significantly warped ($\bar{z} \sim 1$ kpc) {\it prior} to the Clouds' MW infall.}
    \label{fig:mean_z}
\end{figure*}

To further quantify the agreement in the vertical structure of the BFE reconstruction and the simulation, in Figure \ref{fig:mean_z} we plot the $\bar{z}_x$ and $\bar{z}_y$ values for the chosen snapshot. The $\bar{z}$ values agree within 0.2 kpc (as indicated by the residuals), which is comparable to the observational uncertainties \citep[e.g.][]{Ferreira2025}. In Appendix A, a formal evaluation of the quality of the disk BFE reconstruction is performed using a statistical measure.

The epoch shown in Figure \ref{fig:bfe_reconstruction_eg} corresponds to the MW infall epoch in B12. This figure implies that the observed LMC disk should already have been significantly distorted at the onset of infall. In section \ref{sec:obs}, we will further discuss how the dynamical state of the LMC at infall affects our interpretation of the observed LMC disk today. Now that the LMC disk BFE has been validated, next we use these BFEs to understand the evolution of the LMC's disturbed morphology during the LMC-SMC interactions.

\subsubsection{Time-dependent Morphology of the LMC's Disk} \label{sec:disk_morphology}
\begin{figure*}
    \centering
    \includegraphics[width=0.49\textwidth]{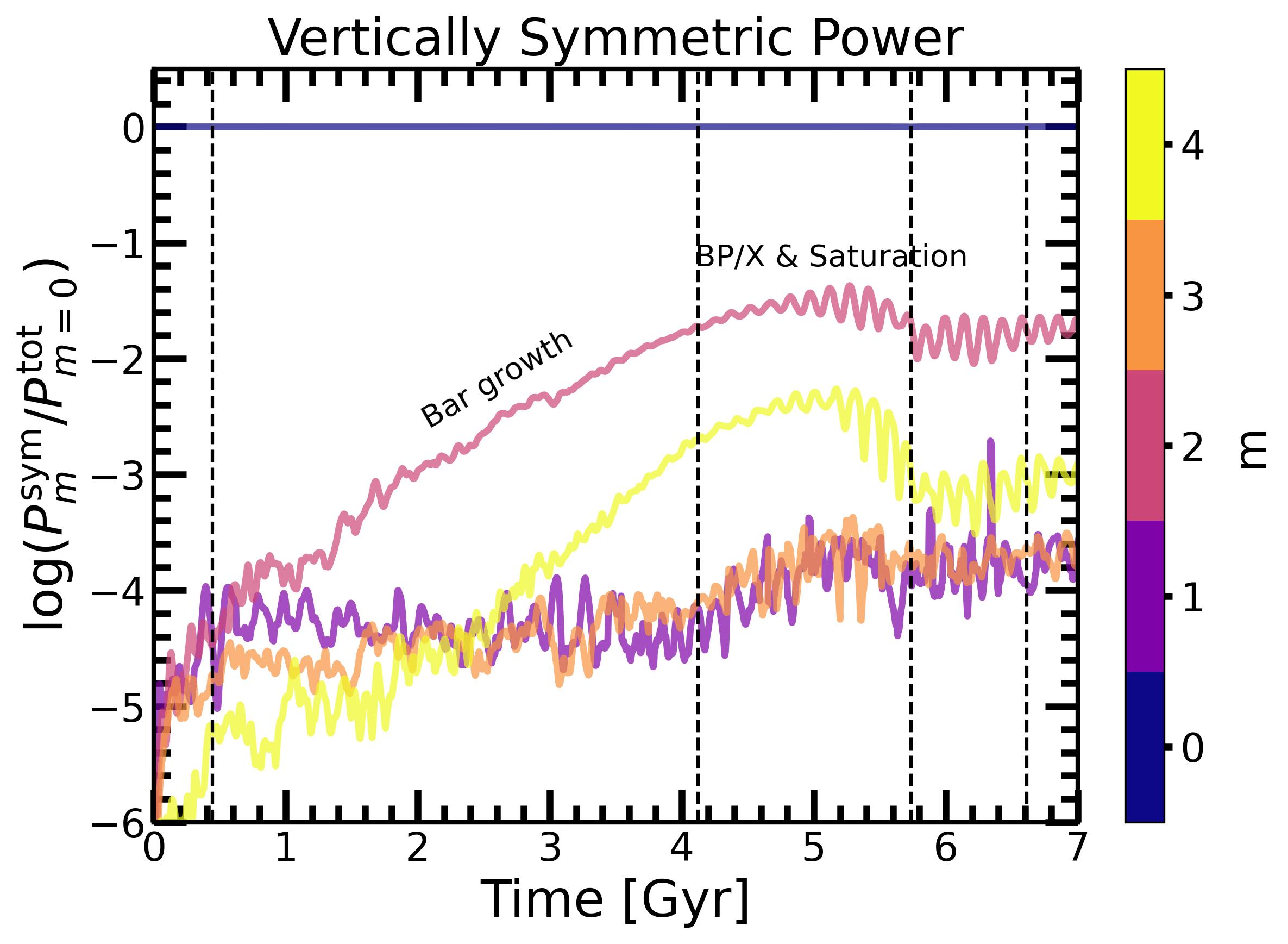}
    \includegraphics[width=0.49\textwidth]{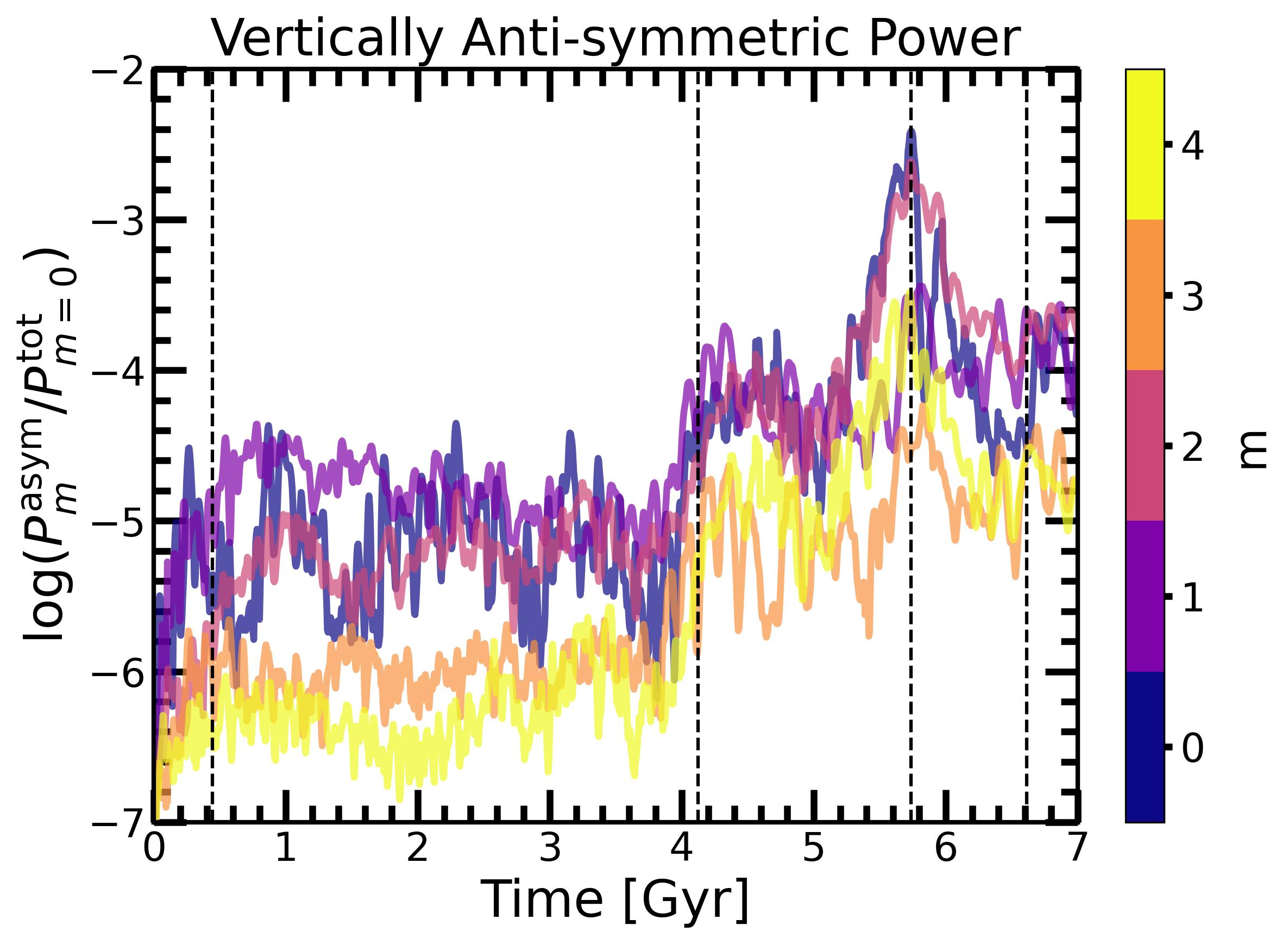}
    \caption{Coefficient power as a function of time for the BFE decomposition of the simulated LMC disk in the LMC-SMC simulation. {\em Left panel} ({\em Right panel}) shows the power across the vertically symmetric (anti-symmetric) basis functions, normalized by the total disk monopole ($m = 0$) power. The LMC-SMC pericentric epochs are marked by the vertical dashed lines. The colorbar depicts different $m$ values. Most of the total $m = 0$ power is contributed by the vertically symmetric basis functions. {\em Left panel}: $m = 2$ and $m = 4$ are the strongest perturbations in the disk. These modes capture the exponential growth of the bar for $t = 0 - 5$ Gyr, followed by a BP/X formation at $t \approx 5$ Gyr and the saturation of the bar instability. {\em Right panel:} The power for all $m$ values show enhancements near the LMC-SMC pericentric epochs. Hence, a combination of the SMC's and the LMC halo's influence drive significant vertically anti-symmetric perturbations in the LMC disk. However, the rapid power increase in the even-m vertically anti-symmetric coefficients between $t = 4-6$ Gyr is primarily due to the bending of the bar during BP/X formation. The BFE coefficients successfully capture the key features in the dynamical evolution of the LMC's disk.}
    \label{fig:coeff_power}
\end{figure*}

Through the BFE reconstruction of the simulated LMC's stellar disk, the time-dependent dynamics of the LMC's disk has now been encoded in the form of coefficient time series. We analyze these disk basis coefficients. Following \cite{Petersen2019}, the coefficient power is defined as (for $m \geq 1$):
\begin{equation}
    P^{\rm sym}_m(t) = \sum_{n = 0}^{35} |C_{mn}(t)|^2
\end{equation}
\noindent and,
\begin{equation}
    P^{\rm asym}_m(t) = \sum_{n = 36}^{71} |C_{mn}(t)|^2
\end{equation}
\noindent where the superscripts \enquote{sym} and \enquote{asym} denote the power in the vertically symmetric and vertically anti-symmetric basis functions, respectively. For $m = 0$, we have:
\begin{equation}
    P^{\rm tot}_{m = 0} (t) = \sum_{n = 0}^{71} |C_{0n}(t)|^2
\end{equation}
Note that the vertically symmetric $m = 0$ coefficient has the highest power, as the lowest order basis function is designed to match the unperturbed disk. In Figure \ref{fig:coeff_power}, we show the coefficient power as a function of time (normalized by $P^{\rm tot}_{m = 0}$) for both the vertically symmetric and vertically anti-symmetric basis functions.

Among the vertically symmetric basis functions, the coefficient with the second-highest power is $m = 2$, followed by $m = 4$. The $m = 2$ coefficient captures perturbations that are bi-symmetric along the azimuth, which includes the bar and two-armed spirals. We have checked that most of the $m = 2$ power is contributed by the $m = 2, n = 0$ basis function, which captures disk perturbations at the the largest ($\sim R_d$) radial scales --- essentially the bar. As such, $m = 2$ mainly corresponds to the LMC's bar, and the quantity $P^{\rm sym}_{m = 2}/P^{\rm tot}_{m = 0}$ is the bar strength (WP21). The bar strength increases exponentially from $< 10^{-5}$ to $\approx 10^{-1.5}$ over a period of $\approx 5$ Gyr. Such an exponential growth of the bar instability has been well studied with linear perturbation theory \citep[e.g.][]{Kalnajs1977, Jalali2007, Sellwood2014}. When the growth saturates at $t \approx 5$ Gyr, a BP/X bulge forms along with a slight reduction in the bar strength. The small amplitude oscillations in the bar strength after $t \approx 5$ Gyr are a consequence of stars trapped in the bar resonances \citep[e.g.][]{Chiba2021, Hamilton2023}. The coefficient with the next highest power is $m = 4$, whose evolution mirrors the evolution of the bar strength.

Among the vertically anti-symmetric coefficients of the simulated LMC disk, $m = 0, 1, 2$ have a comparable power throughout, and $m = 3, 4$ are about an order of magnitude weaker. Generally, all vertically anti-symmetric $m$ orders exhibit local enhancements near LMC-SMC pericenters, indicating that the LMC-SMC interactions --- either through the SMC's direct influence or the influence of the LMC's halo --- drive significant vertical asymmetries in the LMC disk. Between $t = 4 - 6$ Gyr, there is a sharp increase (almost 2 orders of magnitude) in power within the even-m vertically anti-symmetric coefficients. While the SMC’s pericentric passage at $t \approx 6$ Gyr contributes to this trend, the primary driver is likely bar bending. As the bar begins to bend, it creates a strong vertically anti-symmetric perturbation that is also azimuthally bi-symmetric. 

\begin{figure*}
    \centering
    \includegraphics[width=0.337\textwidth]{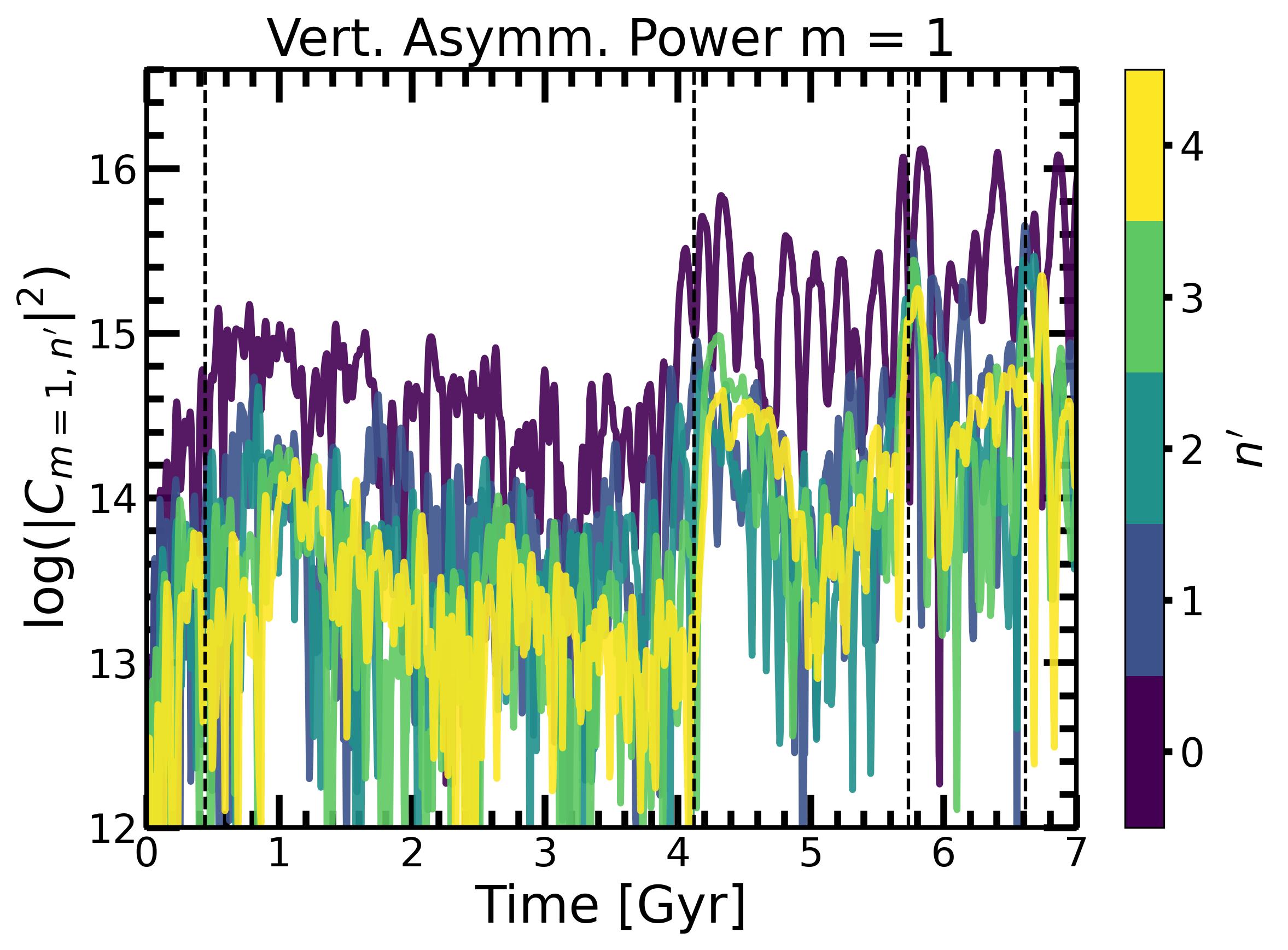}
    \includegraphics[width=0.653\textwidth]{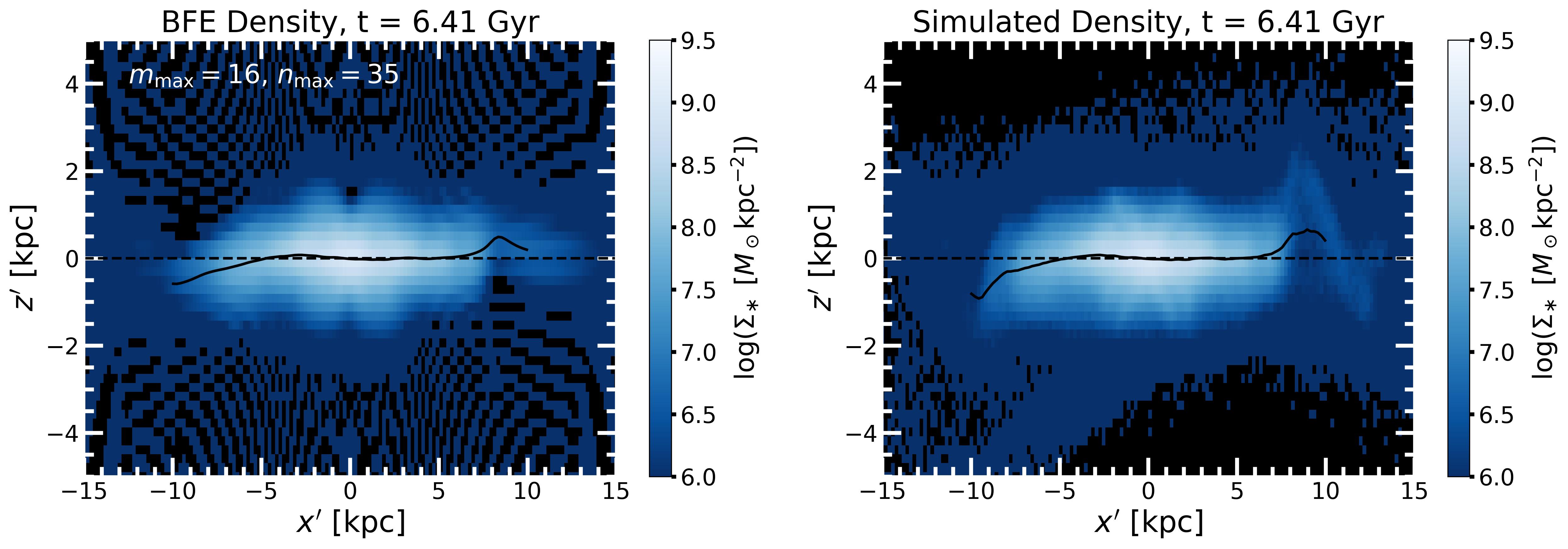}
    \caption{The vertically anti-symmetric $m = 1$ perturbation in the simulated LMC disk of the LMC-SMC simulation. {\em Left panel:} Contribution of individual $n^\prime$ values (depicted by the colorbar) to the $m = 1$ perturbation as a function of time. The vertical dashed lines mark the LMC-SMC pericenter. $n^\prime = 0, 1$ are the largest contributors to the $m = 1$ perturbation, suggesting that the LMC's vertical asymmetries possess the most power at the largest radial scales. {\em Middle} and {\em Right} panels show the edge-on ($x^\prime - z^\prime$) projection of the BFE reconstructed and simulated density fields at $t = 6.41$ Gyr. This time epoch corresponds to the global maximum of $m = 1$ vertically antisymmetric power (see Figure \ref{fig:coeff_power} {\em right panel}). The mean vertical deviation is also shown with the black solid line. A significant warp between $x^\prime = -10$ kpc and $x^\prime = -7$ with a mean vertical deviation of $\approx 1$ kpc is evident. This snapshot represents the point in the simulation where the LMC disk exhibits maximum vertical perturbations.}
    \label{fig:coeff_asymm_m1}
\end{figure*}

To investigate LMC disk warps, we focus on odd-$m$ vertically anti-symmetric coefficients. In particular, we analyze the $m = 1$ mode, which represents the strongest signal among these odd coefficients. In Figure \ref{fig:coeff_asymm_m1} {\em left panel}, we perform a breakdown of $P^{\rm asym}_{m = 1}$ in n-orders. Here, we define the index $n^\prime = n - 36$, so the lowest order vertically anti-symmetric basis function corresponds to $n^\prime = 0$. The lower $n^\prime$ orders have the largest amplitude, and the amplitude decreases as $n^\prime$ increases. All of the $n^\prime$ orders show amplitude enhancements near the LMC-SMC pericenters.

In Figure \ref{fig:coeff_asymm_m1} {\em middle and right panels}, we show the edge-on ($x^\prime - z^\prime$) surface density projections (eq. \ref{eq:edge_on_sigma}) of the BFE reconstructed and the simulated LMC disk respectively, at the epoch where $P^{\rm asym}_{m = 1}$ achieves a global maximum ($t = 6.41$ Gyr). The mean vertical deviation ($\bar{z}_x$) is also shown. A significant warp between $x^\prime = -10$ kpc to $x^\prime = -7$ kpc is evident, with a maximum mean vertical deviation of $\approx 1$ kpc. We highlight that the above epoch represents the point in our simulation where the LMC disk exhibits its peak vertical perturbations. In section \ref{sec:obs}, we compare this simulated maximum perturbation with those observed in the actual LMC disk, demonstrating how this comparison provides insights into the LMC-SMC interaction history.

The disk BFE framework developed in this section paves the way for understanding the various modes that get excited in a disk due to repeated satellite encounters, and how these modes couple to the secular evolution of the disk. We will study these couplings in more detail in future work. Next, in section \ref{sec:mssa}, we identify different components of the temporal evolution of the LMC's vertical perturbations that are influenced by the LMC's distorted halo and by the SMC directly.

\subsection{Understanding Dynamical Correlations with mSSA} \label{sec:mssa}

The disk BFE's presented in the previous section encodes the LMC's vertical perturbations into a time-series of coefficients. Here, we seek to identify correlations between the temporal evolution of the LMC-SMC orbit, LMC's halo distortions and the LMC disk's vertical perturbations. As described in section \ref{sec:mssa_method}, mSSA allows us to identify correlated temporal components across different time series. These components then allow us to infer coupled dynamics. As we will see, the PCs identified by mSSA can be fairly non-trivial to interpret. In this work, we provide a proof-of-concept of mSSA's capabilities by considering the simplest possible time series that can encode the LMC-SMC orbit, the evolution of the LMC's halo distortions, and the evolution of the LMC disk's vertical asymmetries.

\begin{enumerate}
\item For the LMC-SMC orbit, the chosen timeseries is the scalar
distance between the LMC and SMC ($R_{\rm SMC-LMC}$) as a function of time. 
\item As shown in section \ref{sec:halo_torques}, most of the halo torques on the disk arise from the $l = 2$ quadrupole distortion. So, for quantifying the time evolution of the halo distortions, we compute the power in the halo quadrupole ($P^{\rm halo}_{l = 2} (t)$) across radial orders $n = 0, 1$ and all angular orders (following F26 and \citealt{Arora2025}):
\begin{equation}
    P^{\rm halo}_{l = 2} = \sum_{n = 0, 1} \sum_{m = -2}^{2} |C_{n,l = 2, m}|^2
\end{equation}
\noindent where $C_{n, l = 2, m}$ is the halo quadrupole coefficient.
\item As shown in section \ref{sec:disk_morphology}, the $m = 1$ mode is the dominant vertically-antisymmetric mode in the disk. So, to quantify the disk vertical perturbations, we choose the time-evolving power in the $m = 1$ vertically anti-symmetric basis function ($P^{\rm asym}_{m = 1} (t)$). 
\end{enumerate}

\begin{figure*}
    \centering
    \includegraphics[width=0.32\textwidth]{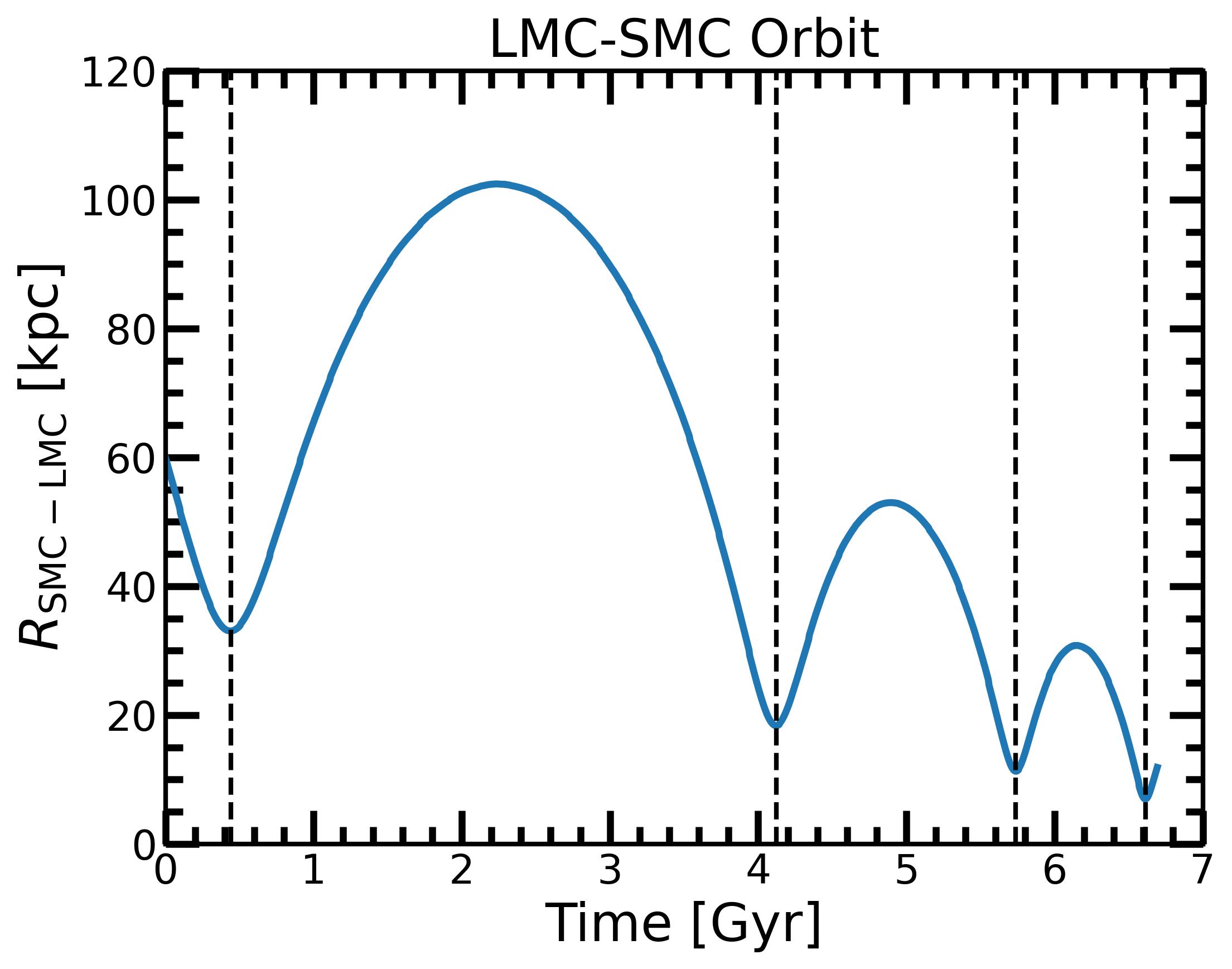}
    \includegraphics[width = 0.32\textwidth]{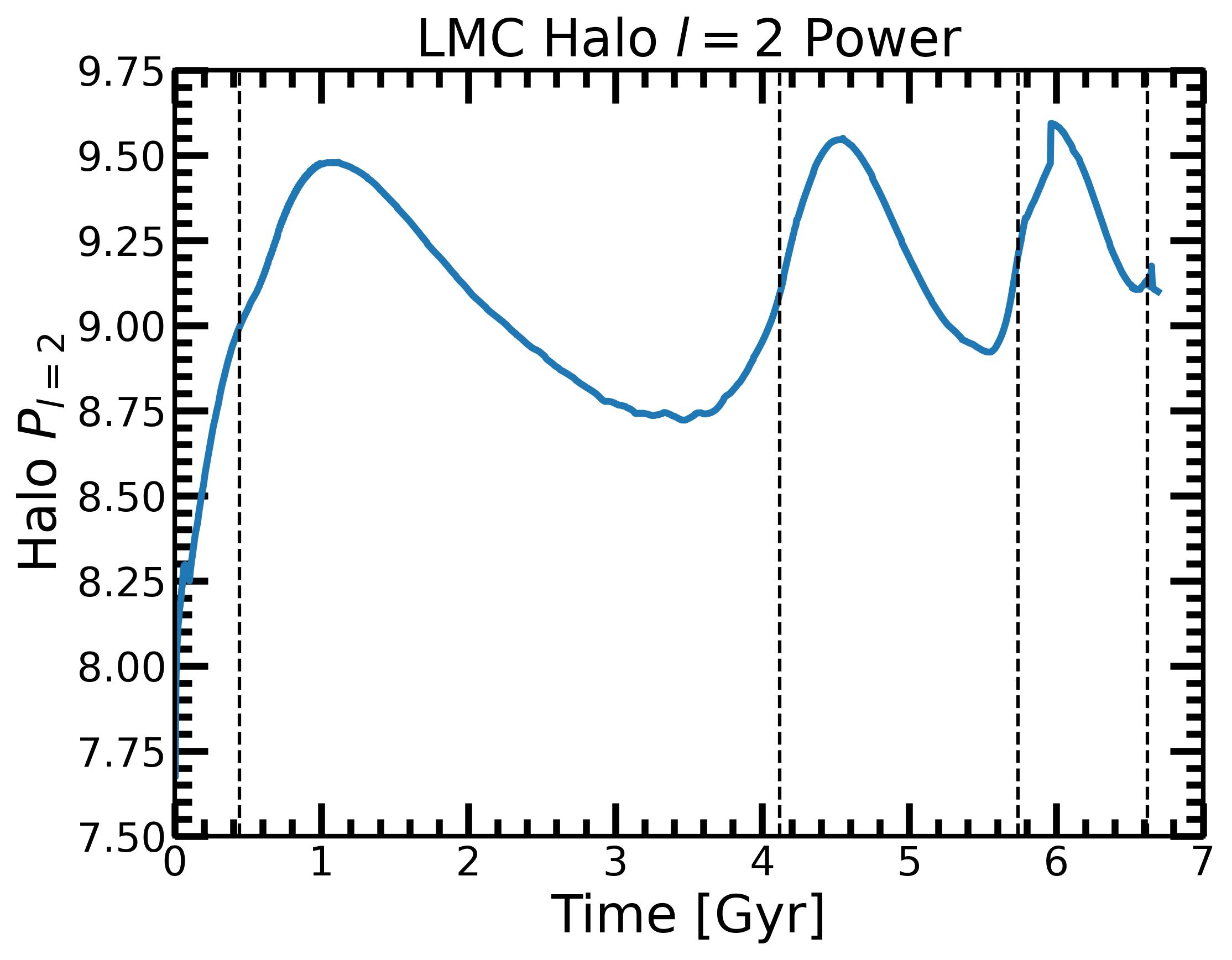}
    \includegraphics[width = 0.32\textwidth]{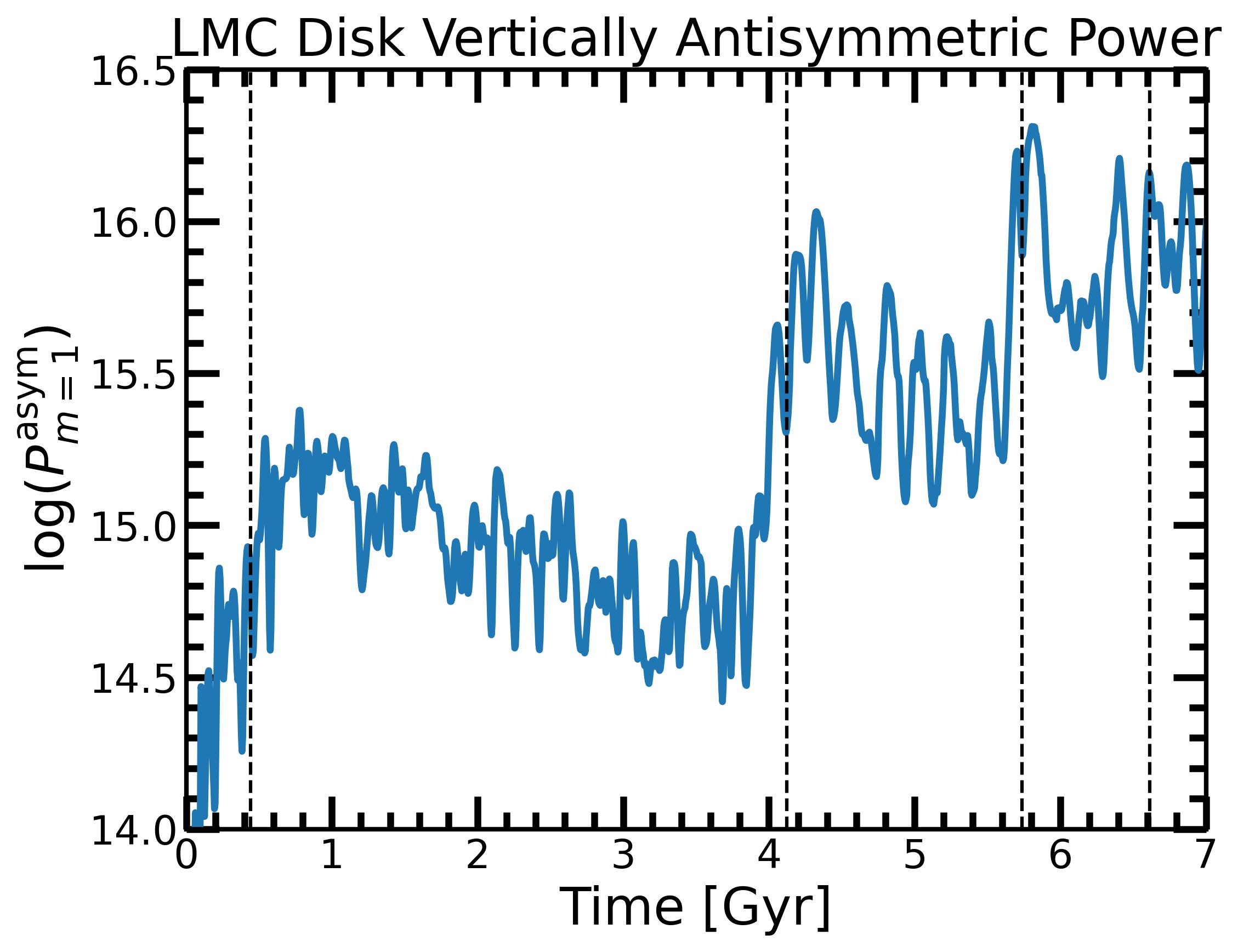}
    \caption{The three time series used for understanding the correlations between the simulated LMC-SMC orbit, LMC halo quadrupole distortion, and the LMC disk vertical asymmetries. The vertical dashed lines mark the LMC-SMC pericenters. {\em Left panel:} similar to Figure \ref{fig:lmc_smc_orbit}, repeated here for the reader's convenience. {\em Middle panel:} The LMC halo quadrupole ($l = 2$) power as a function of time. {\em Right panel:} The total power in the LMC disk's $m = 1$ vertically antisymmetric basis functions, which encodes the warps. The correlations between these time series' will be characterized using mSSA.}
    \label{fig:mssa_time_series}
\end{figure*}

The above three time series are shown in Figure \ref{fig:mssa_time_series}. Note that here we have preferred to use $R_{\rm SMC-LMC} (t)$ as the first mSSA channel as opposed to the SMC's direct torque mainly due to two reasons: (i) the orbital separation is a direct observable that can be obtained through backward integration of the Clouds' velocity vectors; and (ii) the torque changes rapidly during the SMC's crossings of the LMC disk (Figure \ref{fig:lmc_smc_orbit}), which can introduce spurious PCs in the mSSA decomposition.

To further simplify the interpretation of the mSSA PCs, we correlate pairs of time series involving the LMC disk: $R_{\rm SMC-LMC} (t)$ with $P^{\rm asym}_{m = 1} (t)$ (\enquote{Orbit $+$ Disk}), and then $P^{\rm halo}_{l = 2}(t)$ with $P^{\rm asym}_{m = 1} (t)$ (\enquote{Halo $+$ Disk}). Note that the above time series represent different physical quantities and that the numerical values spanned by each of them are also quite different, as evident in Figure \ref{fig:mssa_time_series}. These varying value ranges could cause the PCs to be biased by one of the time series. To avoid this bias, mSSA de-trends each channel by subtracting the mean and normalizing by the variance. This ensures that the channels span a similar range of values and each channel gets an equal weight in the trajectory matrix, allowing the channels to effectively mix with each other. This de-trending is fundamentally similar to the commonly applied feature normalization technique in Machine Learning.

The key parameter controlling the sensitivity of mSSA to different timescales is the length of the time-lagged copies of the individual channels, also known as the window length ($L$). Mathematically, we require $L \leq N/2$, where $N$ is the length of the channels (WP21). The larger (smaller) the value of $L$, the coarser (finer) the correlations in time would be identified across the input channels. Some experimentation is required to determine the appropriate value of $L$ that is sensitive to a desired timescale. 

Since LMC-SMC interactions are the driver for both the LMC halo distortions and the LMC disk vertical asymmetries, the timescales associated with the LMC-SMC orbit will be imprinted on both the halo and disk evolution. As such, $L$ needs to be small enough that the PC's with significant eigenvalues are sensitive to the smallest orbital period in the LMC-SMC orbit ($\approx 800$~Myr). This sensitivity is not achieved if we choose $L = N/2$. On the other hand, if $L$ is too small ($\leq N/10$), the dominant PCs contain contributions from the high frequency oscillations occurring at timescales $< 100$ Myr in the time series of $P^{\rm asym}_{m = 1}$. We find $L = N/4$ to be a good choice for the window length, as this value captures the relevant timescales in the LMC-SMC orbit while being insensitive to the high frequency oscillations in $P^{\rm asym}_{m = 1}$.

\subsubsection{Orbit $+$ Disk mSSA Analysis}

\begin{figure*}
    \centering
    \includegraphics[width=0.49\textwidth]{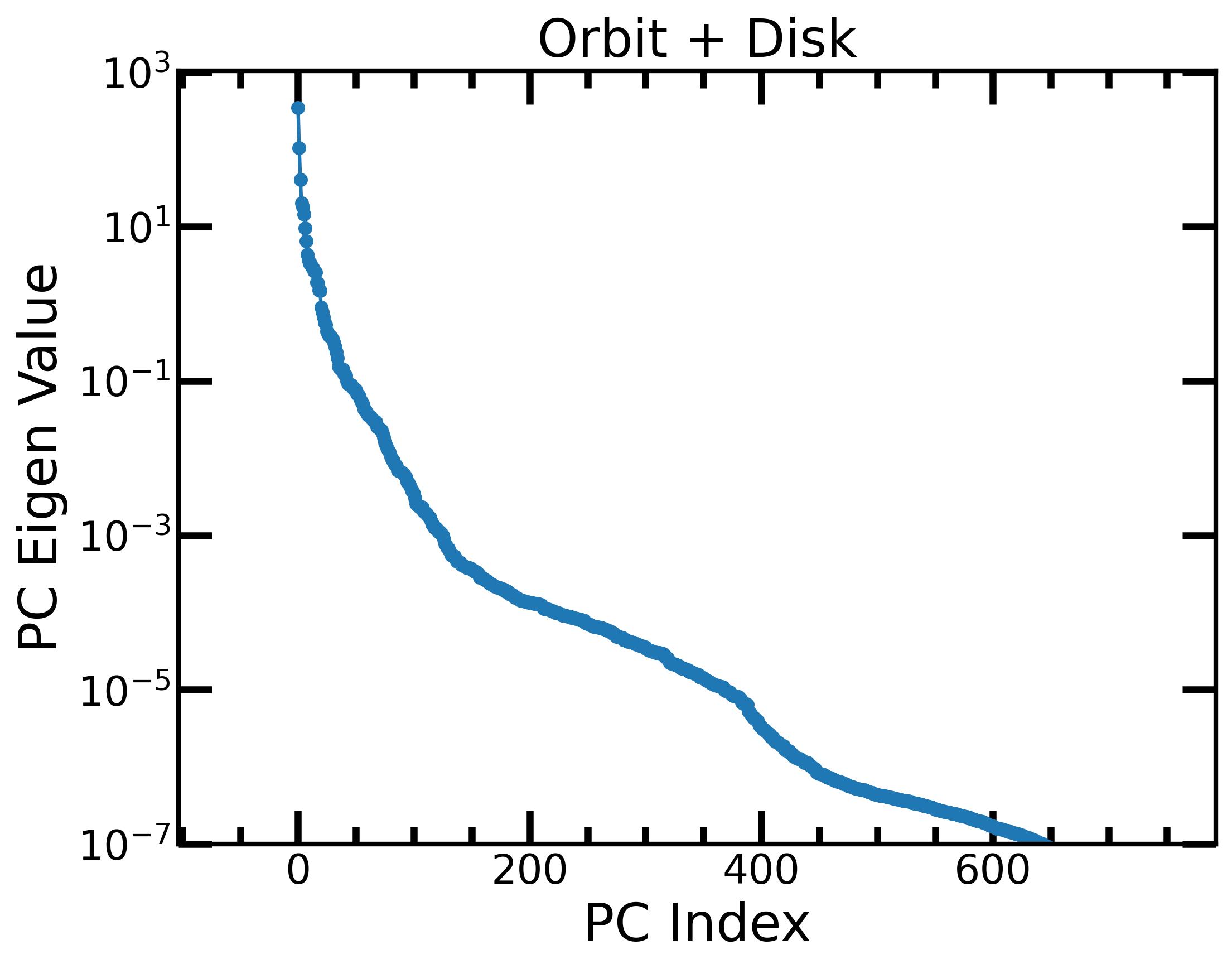}
    \includegraphics[width=0.49\textwidth]{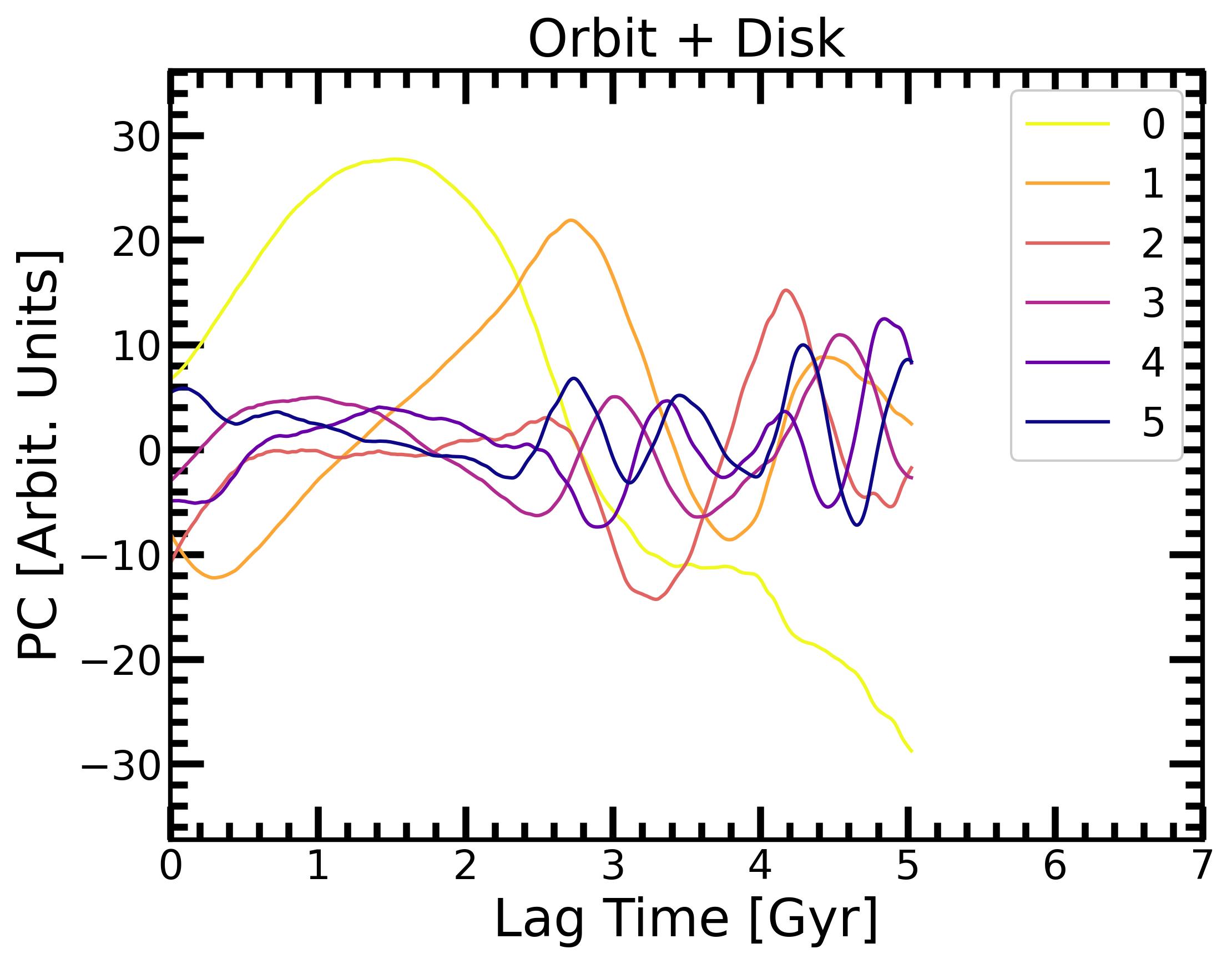}
    \includegraphics[width=0.49\textwidth]{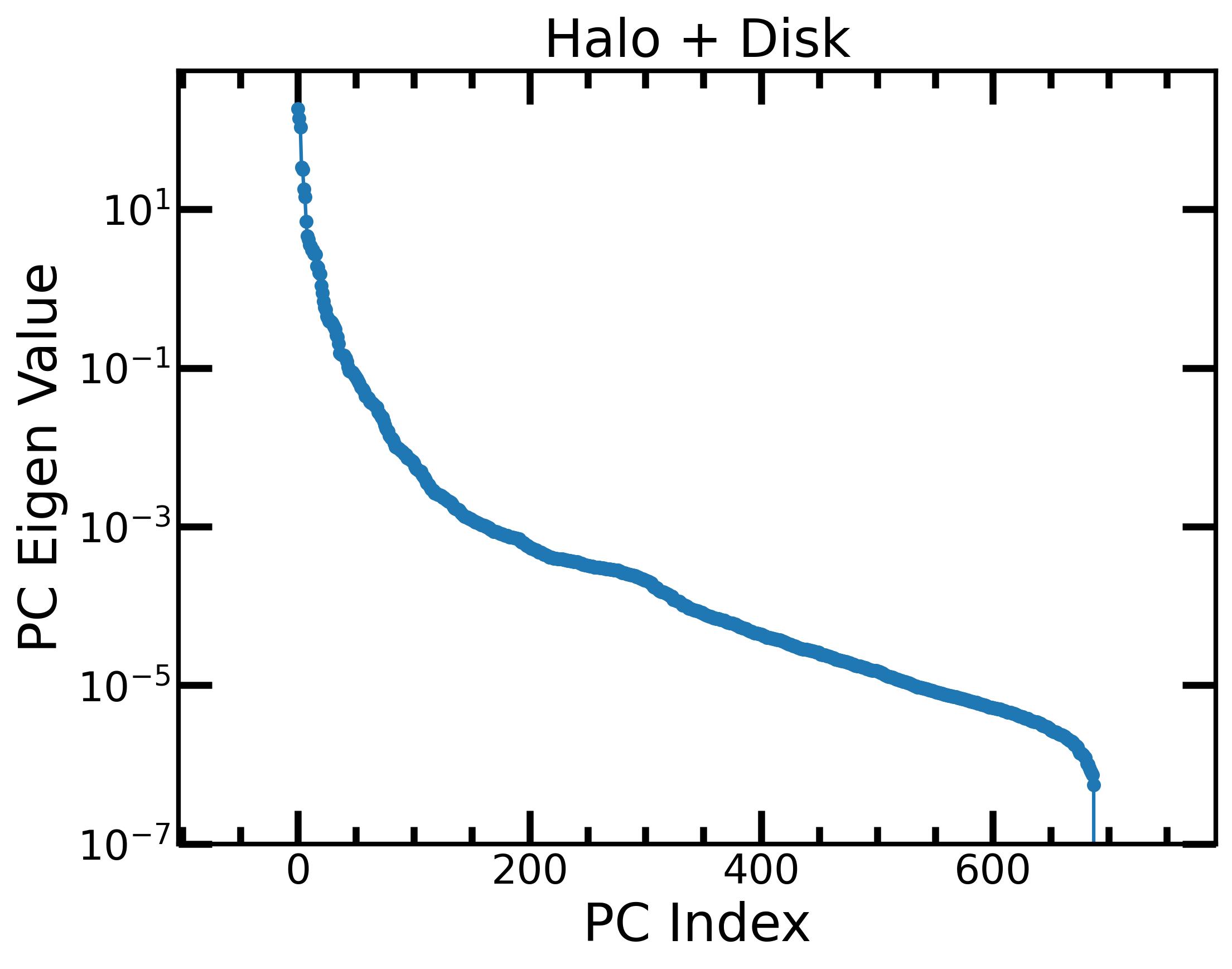}
    \includegraphics[width = 0.49\textwidth]{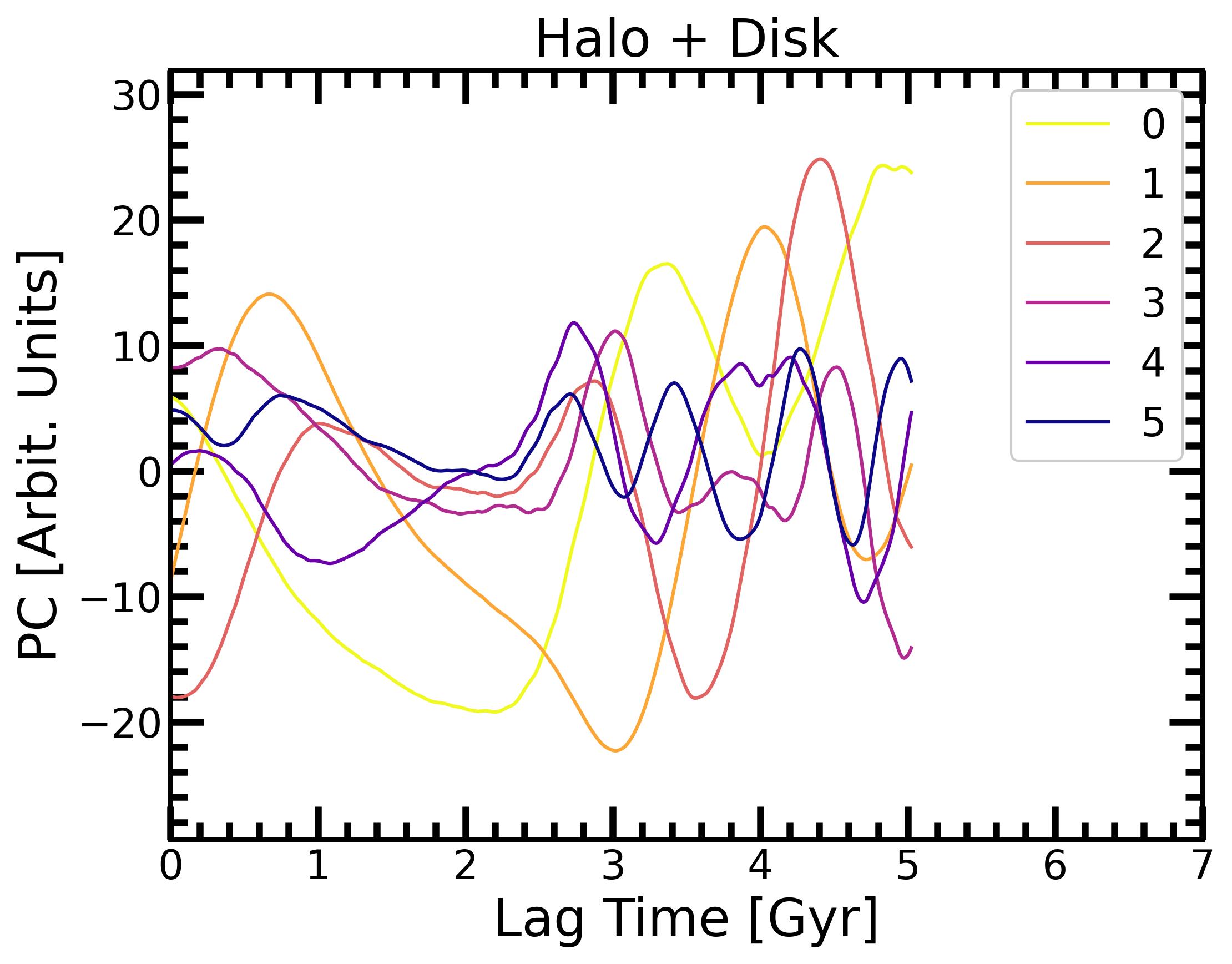}
    \caption{Eigenvalues and Principal Components (PCs) of the two mSSA analyses considered. The mSSA window length (1.67 Gyr) is chosen to be a quarter of the time series length (6.69 Gyr). The {\em top row} depicts the analysis with the LMC-SMC orbit (Figure \ref{fig:mssa_time_series} {\em left panel}) and the disk $m = 1$ vertically antisymmetric power (Figure \ref{fig:mssa_time_series} {\em right panel}) as the two mSSA time series channels. The {\em bottom row} depicts the analysis with the halo quadrupole power (Figure \ref{fig:mssa_time_series} {\em right panel}) instead of the LMC-SMC orbit. The {\em left column} shows the run of the PC eigenvalues. The {\em right column} shows the lag time dependence of the PC's corresponding to the top 6 eigenvalues. The eigenvalues drop sharply, indicating that the original time series' used is dominated by only a few PCs. The PC's with the largest amplitude and the least fluctuation as a function of lag-time (like PC 0 in the {\em top row}) encode the variation of the original time series over the longest time scales (i.e., over the full duration). The PC's that vary quasi-periodically with lag time likely encode the oscillatory component of the original time series. The smaller the amplitude of a quasi-periodic PC, the finer the feature it encodes in the original time series. Hence, mSSA breaks down the original time series' into components encoding different temporal behaviors, thereby allowing us to infer which temporal components of the LMC-SMC orbit, halo quadrupole, and the disk warps are correlated to each other.}
    \label{fig:evals_pcs}
\end{figure*}

Figure \ref{fig:evals_pcs} {\em top panel} shows the PCs and their corresponding eigenvalues for the Orbit $+$ Disk analysis. The eigenvalues ({\em top left panel}) drop sharply, indicating that most of the information in the time series is contained in a few PCs with the largest eigenvalues. In the \texttt{EXP} convention, the PCs are labeled in descending order of eigenvalues. So, the PC with index \enquote{0} (in short PC [0]) has the largest eigenvalue, followed by PC [1] and so on. In the {\em top right} panel, each PC vector is expressed as a time series in the space of lag times. Only the top 6 PCs are shown. The PC [0] has the largest amplitude, and it does not oscillate with a lag time. As such, the PC [0] encodes a correlated non-oscillatory temporal component in the evolution of $R_{\rm SMC-LMC} (t)$ and $P^{\rm asym}_{m = 1} (t)$. 

\begin{figure*}
    \centering
    \includegraphics[width=0.32\textwidth]{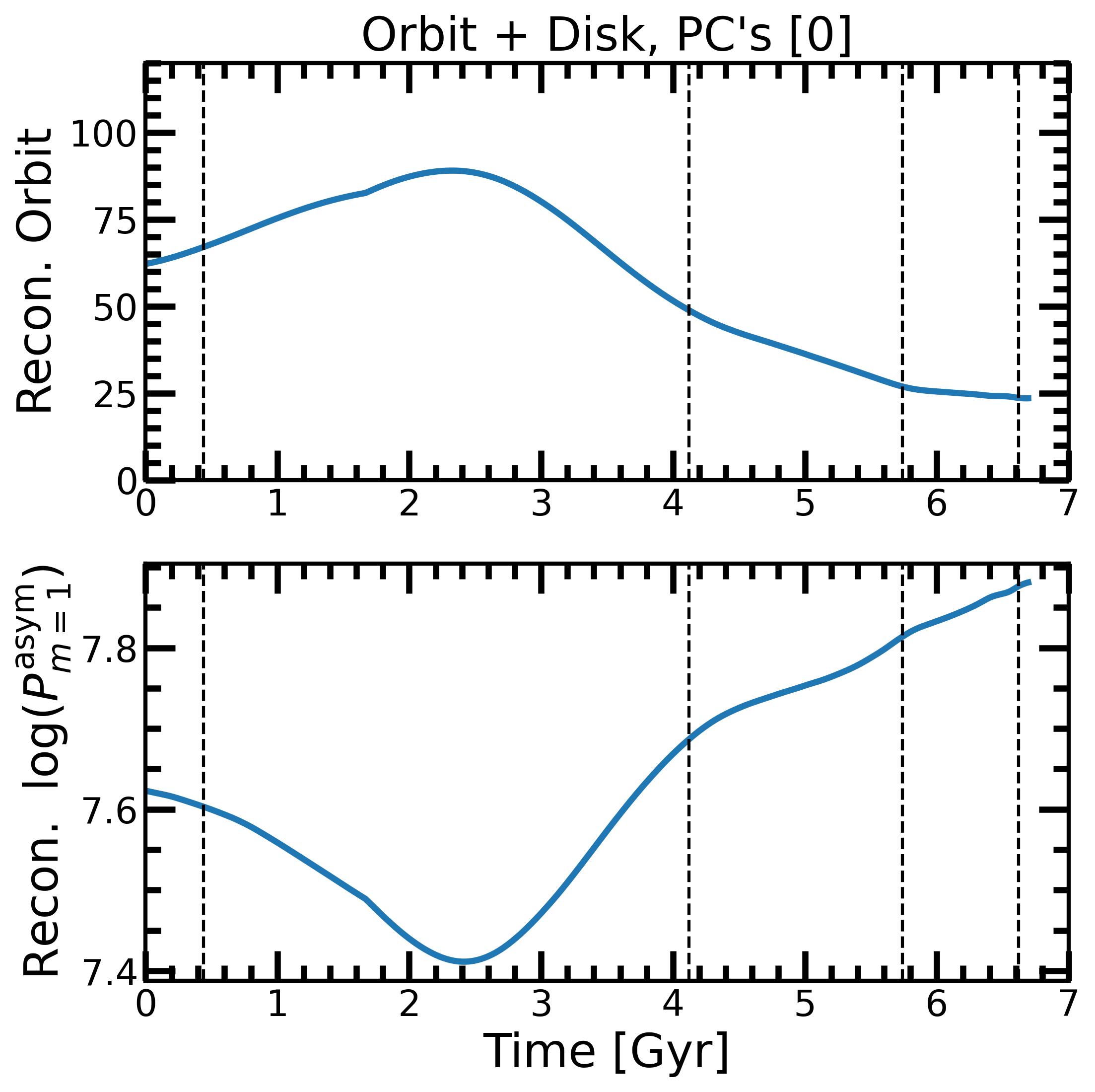}
    \includegraphics[width = 0.32\textwidth]{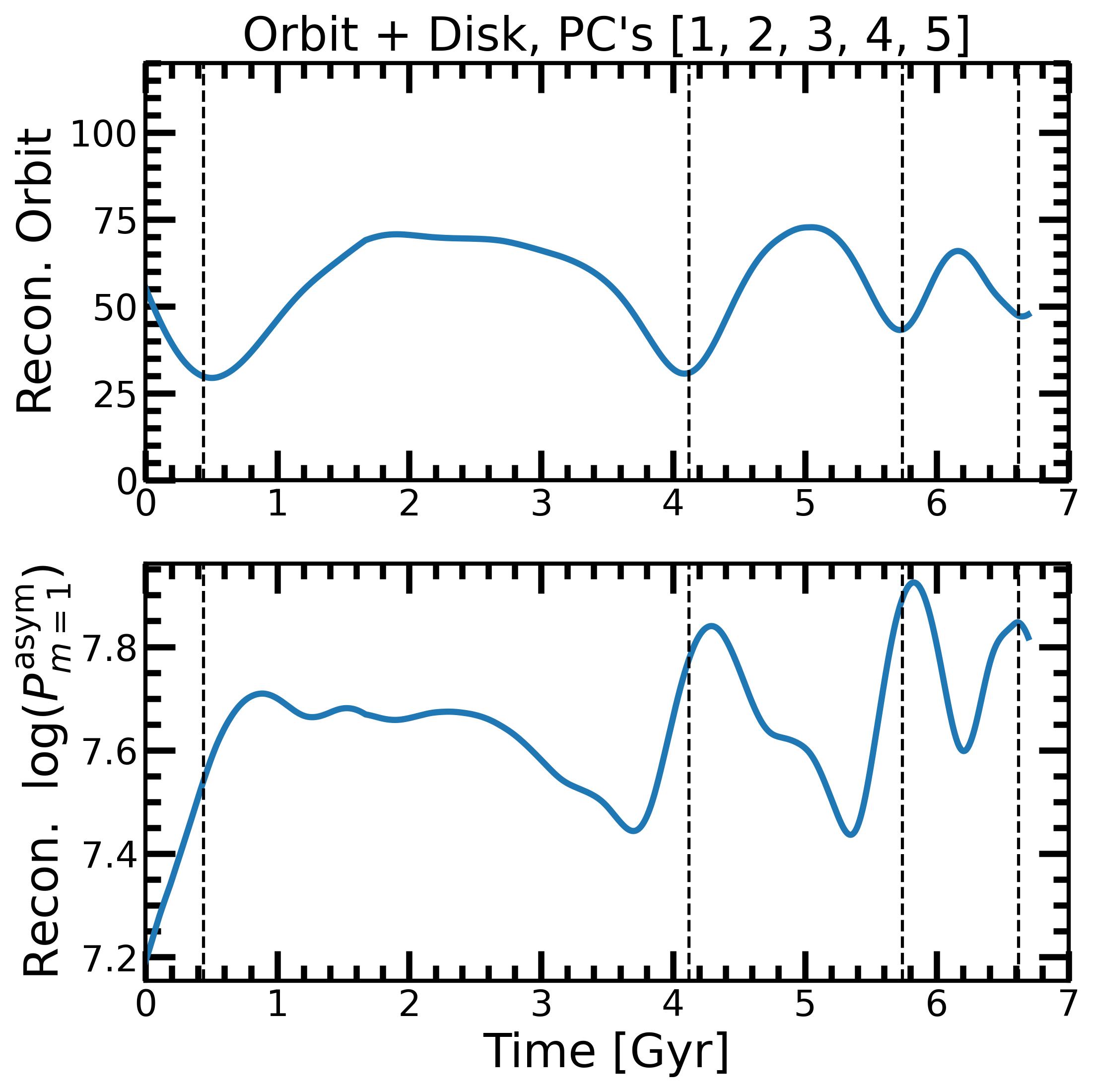}
    \includegraphics[width = 0.32\textwidth]{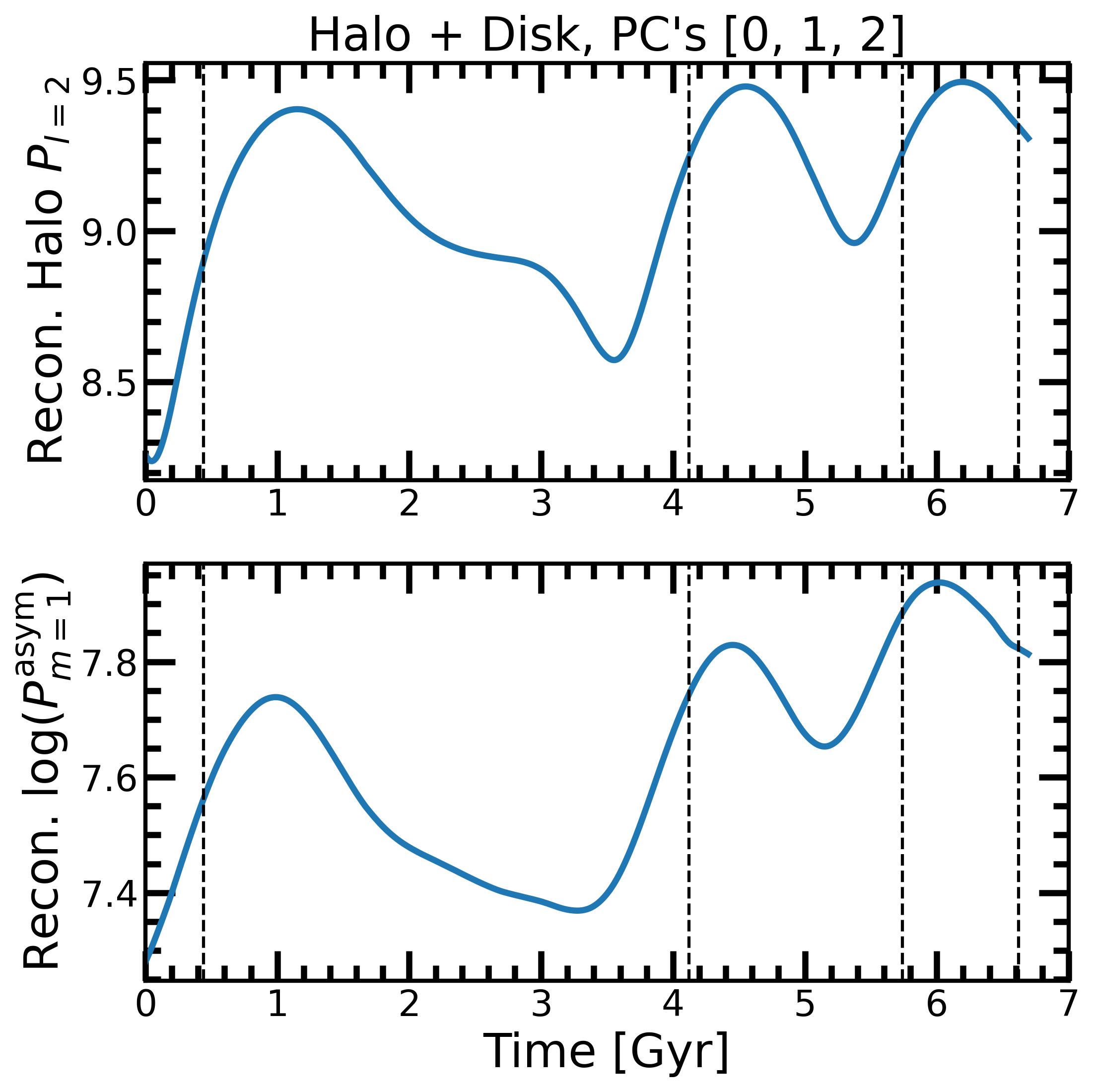}
    \caption{Reconstructions of the LMC-SMC orbit, LMC's halo quadrupole power, and the LMC's disk $m = 1$ vertically antisymmetric power using different PC groups selected from Figure \ref{fig:evals_pcs}. The {\em left and middle} columns depict the reconstructions involving the LMC-SMC orbit (Figure \ref{fig:mssa_time_series} {\em left panel}) and the disk $m = 1$ power (Figure \ref{fig:mssa_time_series} {\em middle panel}) as the two mSSA channels. The {\em right column} depicts the reconstructions involving the halo quadrupole (Figure \ref{fig:mssa_time_series} {\em middle panel}) instead of the LMC-SMC orbit. The {\em bottom row} denotes the reconstructed disk $m = 1$ time series for each case. {\em Left column:} the reconstruction using PC [0] captures the long-range behavior, suggesting that as the LMC-SMC orbit becomes more compact, the disk $m = 1$ power increases. {\em Middle column:} the reconstruction using PC's [1, 2, 3, 4, 5] captures the oscillatory behavior, suggesting that the oscillations in the disk $m = 1$ power are coupled to the LMC-SMC pericenters, albeit with a time delay. {Right panel:} the reconstruction using PC's [0, 1, 2] suggests that the disk $m = 1$ power responds roughly in-phase to the halo quadrupole power, and the time delay between the LMC-SMC pericenters and the peaks of the disk $m = 1$ power is likely in part due to the delayed peak of the halo quadrupole. This analysis suggests that both the LMC-SMC orbit and the LMC's halo quadrupole shape the temporal evolution of the LMC's disk warps.}
    \label{fig:mssa_reconstructions}
\end{figure*}

The {\em left column} of Figure \ref{fig:mssa_reconstructions} shows the reconstruction of the $R_{\rm SMC-LMC} (t)$ and $P^{\rm asym}_{m = 1} (t)$ using only PC [0]. The orbit reconstruction tracks the overall growth and decay of the distance between the LMC and SMC. Because the period of the first orbit likely exceeds the timescale probed by PC [0], the component captures the initial rise ($t = 0–2$ Gyr) and subsequent decline in $R_{\rm SMC-LMC}$. However, since the periods of the latter two orbits are likely shorter than what PC [0] probes, the reconstruction cannot resolve the peri-apo-peri oscillations in $R_{\rm SMC-LMC}$; instead, it isolates the overall orbital decay resulting from dynamical friction. The reconstruction of $P^{\rm asym}_{m = 1}$ using the same PC[0] reveals an initial decline that mirrors the rise in the reconstructed $R_{\rm SMC-LMC}$. Subsequently, the power increases steadily, mirroring the long-term orbital decay. Consistent with the logic applied to the orbit reconstruction, the timescale probed by PC [0] is large enough that the reconstruction isolates the broad trend in $P^{\rm asym}_{m = 1} (t)$ rather than resolving the oscillations.

Through PC [0], we identify a correlated temporal component common to both the LMC-SMC orbit and the disk $m = 1$ vertical asymmetries. This component illustrates an inverse average relationship between the two: as the LMC-SMC distance grows, the vertically asymmetric disk power decays, and as the orbit becomes more compact, the disk power increases. Therefore, the steady increase in $P^{\rm asym}_{m = 1}$ over the time span of the simulation is likely driven by the overall decay of the LMC-SMC orbit due to dynamical friction. Using mSSA allows for a clear isolation of this physical connection.

The other PCs [1--5] are all oscillatory in nature. The phases of these PCs are also different at any given lag time. Oscillatory PCs generally represent quasi-periodic signals in the original channels. Indeed, both the LMC-SMC orbit and the disk power show oscillations with time. Figure \ref{fig:mssa_reconstructions} {\em middle panel} shows the reconstructions of $P^{\rm asym}_{m = 1} (t)$ and $R_{\rm SMC-LMC} (t)$ using PCs [1--5]. The reconstructed orbit clearly shows the oscillations in $R_{\rm SMC-LMC} (t)$, and consistently recovers the LMC-SMC pericenters. Note that if we naively superpose the orbit reconstructions with PC [0] and PC's [1--5], the overall shape of $R_{\rm SMC-LMC} (t)$ would be reasonably captured. 

The reconstruction of the disk power captures the overall oscillations in the original time series of $P^{\rm asym}_{m = 1} (t)$, and shows the power enhancements near the LMC-SMC pericenters. The reconstruction also shows a power suppression when the SMC is at an apocenter. As such, mSSA illustrates that the LMC's disk response in the vertical direction is coupled to the periodic increase and decrease in the LMC-SMC distance. However, several aspects of the the disk's temporal evolution are still not clear, like the cause of the: $\sim 100 - 200$ Myr delay between the peaks of $P^{\rm asym}_{m = 1}$ and the LMC-SMC pericenters; and the sawtooth-like shape of the oscillations in the reconstructed $P^{\rm asym}_{m = 1}$ after $t = 4$ Gyr as opposed to sinusoidal-like oscillations in the reconstructed $R_{\rm SMC-LMC}(t)$. Later, we shall see that these temporal features are explainable by accounting for the halo.

Note that the $P^{\rm asym}_{m = 1}$ reconstruction is also free from the high-frequency ($< 100$ Myr time period) fluctuations in the original time series, thereby giving a much cleaner view of the overall oscillations in the disk power evolution. As such, mSSA can also be used as a powerful tool to selectively remove features of a time series that are dynamically not relevant and retain features that are.

\subsubsection{Halo $+$ Disk mSSA Analysis}

The {\em bottom row} of Figure \ref{fig:evals_pcs} shows the PCs and their corresponding eigenvalues for the Halo $+$ Disk analysis. Similar to the Orbit $+$ Disk analysis, the eigenvalues display a sharp drop-off ({\em bottom left panel}), suggesting that only the top few PCs are important contributors to the original time series. The PCs [0--5] are expressed as a time series in lag time space in the {\em bottom right panel}. All of the PC's are oscillatory in nature. PCs [0, 1, 2] have similar amplitudes of oscillations, but have different phases at any given lag time. PCs [3--5] have significantly smaller amplitudes of oscillation as compared to PCs [0--2]. Hence, it is likely that PCs [0--2] encode a dominant correlated quasi-periodic signal across the input halo and disk time series. In Figure \ref{fig:mssa_reconstructions} {\rm right panel}, we reconstruct $P^{\rm halo}_{l = 2}(t)$ and the $P^{\rm asym}_{m = 1}$ disk power using the PCs [0--2].

The halo quadrupole time series is reasonably captured by the reconstruction. In particular, the peaks of $P^{\rm halo}_{l = 2}$ and their time delay with respect to the LMC-SMC pericenters, as well as the shape of the halo power decay after a peak, are well represented in the reconstruction. The reconstruction of $P^{\rm asym}_{m=1} (t)$ qualitatively mirrors that of $P^{\rm halo}_{l = 2}(t)$, exhibiting consistent peak delays following LMC-SMC pericenters and similar sawtooth-like power decay profiles. This close alignment suggests that the dominant correlated components of both time series evolve synchronously, indicating a significant dynamical link between the halo quadrupole and the development of disk vertical asymmetries.
\\
\\
\begin{table*}
\centering
\caption{Dynamical correlations between the LMC-SMC orbit, LMC's halo distortions and LMC disk vertical perturbations.}
\begin{tabular}{c c c}
\hline
\hline
 mSSA Analysis   & PC Groups &  Dynamical Coupling \\
 \hline
 \multirow{2}{6em}{Orbit $+$ Disk} & [0] & Orbital decay leads to gradual increase of disk vertical asymmetries\\
 & [1, 2, 3, 4, 5] & Enhanced disk vertical response near the LMC-SMC pericenters\\
 \hline
 Halo $+$ Disk & [0, 1, 2] & $\sim 200$ Myr delay in peak disk vertical response after a pericenter\\
 \hline
 \end{tabular}
 \tablenotetext{}{{\textbf Note:} {\em Left column:} specifies the original time-series used for the mSSA analysis (see Figure \ref{fig:mssa_time_series}). {\em Middle column:} PCs used for the reconstruction of the original time-series (see Figure \ref{fig:evals_pcs}). {\em Right Column:} the dynamical correlations inferred using the mSSA reconstructions (see Figure \ref{fig:mssa_reconstructions}).}
 \label{tab:mssa_table}
\end{table*}

Table \ref{tab:mssa_table} provides a summary of the dominated correlated PCs in the Orbit $+$ Disk and the Halo $+$ Disk mSSA analysis, and the information contained in those PCs that is suggestive of dynamical couplings between the LMC-SMC orbit, LMC halo distortions and the LMC disk perturbations. Our mSSA experiments suggest that the LMC disk vertical asymmetries are driven by both the direct gravitational influence of the SMC and the influence of the SMC-induced LMC halo distortions. While these processes --- the SMC-LMC orbit, halo quadrupole, and disk vertical power --- are naturally coupled by the SMC’s orbital evolution, the mSSA PCs allow us to characterize their individual contributions to some extent. Notably, the time lag between the orbital pericenters and peak disk power, as well as the subsequent decay of disk power, appears to be primarily governed by the halo's response rather than the direct impact of the satellite. These $\sim 200$'s of Myr time delays are particularly significant during later times in the LMC-SMC orbital evolution, as the increase in orbital frequency results in these delays becoming a significant fraction of the orbital time period (for instance, the last orbit of the SMC has a time period of $\approx 800$ Myr). In the next section, we will explore potential consequences of these findings for the evolution of the observed LMC-SMC system.

The mSSA application we have shown here is only a surface-level illustration of the true power of this algorithm. Given the framework we have developed, there is now an opportunity to use mSSA directly on the time series of individual complex valued coefficients of the disk and halo as opposed to summary quantities like the coefficient power. Further, the individual cartesian components of the $\vect{R_{\rm SMC-LMC}}(t)$ can be utilized instead of just the LMC-SMC distance. Using the coefficients directly would enable the spatial reconstruction of the potential and density fields of the disk and halo. As such, these more sophisticated analysis will enable identifying even more dynamical couplings, like: the relation between the orbital geometry, spatial orientation of the halo quadrupole, and phases of the disk excitations; and the role of secular perturbations like the bar in governing the evolution of the excitations. We will pursue a more complex mSSA analysis in future work. Finally, just like any other machine learning algorithm, mSSA results must always be validated against physical intuition as there can be a possibility of misidentifying linked features across multiple time-series channels or spurious systematic features in the reconstructions.

\section{Discussion} \label{sec:discussion}

In section \ref{sec:results}, we presented a generic framework that enables: (i) quantification of the torques applied on the primary's disk by the satellite and the satellite-induced primary halo distortions; (ii) understanding how the disk vertical perturbations develop in response to these torques; (iii) dis-entangling the contributions of the halo and the satellite in the temporal evolution of the disk perturbations. Next, we discuss: (i) insights into the observed LMC's pre-MW-infall and present day properties  (section \ref{sec:obs}); (ii) implications for other interacting galaxy systems in the local universe (section \ref{sec:other_galaxies}); (iii) how our BFE + mSSA framework enables the use of Clouds as a new DM testbed (section \ref{sec:dm_physics}); and (iv) further limitations of our work and scope for the future (section \ref{sec:future}).

\subsection{Implications for Observations} \label{sec:obs}

Due to the absence of the MW in our simulation setup, our simulated LMC-SMC orbital evolution after the $3^{\rm rd}$ pericenter (which corresponds to the Clouds' MW infall time, see Figure \ref{fig:lmc_smc_orbit}), does not capture the evolution of the LMC-SMC binary to the present day. The Clouds are expected to infall to the MW at $t \approx 5.7$ Gyr, where the MW tidal field will alter the SMC's orbital trajectory with respect to the LMC.

There is a growing body of evidence that the SMC's most recent ($100-200$ Myr ago) pericenter about the LMC was likely a direct collision, with an impact parameter of $\sim 2$ kpc \citep[e.g.][]{Besla2016, Choi2018a, Choi2022, Rathore2025b, Rathore2026, Dhanush2024, Arranz2025b}. Our simulation does not capture this collision. 

The goal of this study is to establish a generalizable framework that quantifies and connects prominent distortions in the DM halo and stellar disk, demonstrating how the impact of the DM halo can be disentangled from the direct tidal field of the satellite. This framework will be applied to upcoming simulations of the LMC-SMC-MW system (H. Rathore et al. {\em in prep}) to directly connect to present day observations of the LMC's stellar disk. In addition, this study affords insight into the evolution of the LMC's disk {\it prior} to the Clouds' infall into the MW. Figure \ref{fig:bfe_reconstruction_eg} indicates that the LMC disk should have had significant vertical asymmetries (mean vertical deviation of up to 1 kpc) at MW infall due to the previous LMC-SMC interactions. This pre-infall state of the disk needs to be accounted for while interpreting the observed LMC's disk perturbations today. In particular, this work indicates that during the recent direct collision between the Clouds, the SMC must have impacted an LMC disk that was {\em already} significantly warped. Figure \ref{fig:bfe_reconstruction_eg} provides a suitable initial condition to evaluate how much the recent SMC collision builds on to the pre-collision distortions of the LMC disk.

For example, as shown in Figure \ref{fig:coeff_asymm_m1}, the maximum mean vertical deviation of the outer (R $\approx 10$ kpc) LMC disk obtained in our simulations is $\lesssim 1$ kpc, which happens near the $4^{\rm th}$ pericenter. The observed LMC disk is found to have a prominent U-shaped warp \cite{Saroon2022, Oden2025, Garver2026} with a mean vertical deviation of at least 2 kpc at $R \approx 7-10$ kpc. This difference indicates that a wide LMC-SMC orbit (closest approach $\geq 7$ kpc) does not lead to large enough vertical perturbations in the LMC's disk. Interestingly, such a wide orbit is the mean result of analytical LMC-SMC orbit integrations \citep[e.g.][]{Zivick2018}. Our findings, therefore, indicate that the analytically derived LMC-SMC orbit is not consistent with the vertical deviations in the outer LMC disk, and a much closer LMC-SMC encounter, like a direct collision discussed before, is needed.

Figure \ref{fig:bfe_reconstruction_eg} also shows a BP/X morphology around the LMC's bar. Generally, in simulations, dynamical processes that contribute to the development of the BP/X feature depend on the assumed initial conditions, like the initial disk and halo profiles. The LMC initial conditions adopted in our simulations are motivated by the properties of the observed LMC. As such, a more detailed investigation of the observed bar is thus motivated, in order to identify features indicative of BP/X-like orbits \citep[e.g.][]{Tahmasebzadeh2024b}. However, the observed LMC bar has been challenging to understand, as it shows several strange properties like offset from the disk center, tilt with respect to the disk plane, and lack of gas inflows \citep{deVFreeman72, Choi2018a, Stavely-Smith2003}. These strange properties have even led early works to question whether the LMC's bar is actually a dynamical galactic bar \citep[e.g.][]{Zhao2000, Zaritsky2004}. Recently, \cite{Rathore2025a} showed that the observed LMC bar’s strength and length are consistent with bar-galaxy co-evolution, despite the strange properties. Subsequently, \cite{Rathore2025b, Arranz2025b} showed that the bar's strange properties are a result of the LMC-SMC interactions. As such, the LMC's bar is an important testbed for both standard bar theory and how bars are affected by external perturbations \citep[e.g.][]{Berentzen2003, Purcell2011, Zhou2026}. Excitingly, precision observations from surveys like Gaia DR3 \citep{Luri2021} and VMC \citep{Cioni2011} have started to enable a detailed characterization of the {\em internal} structure and kinematics of the LMC's bar \citep{Niederhofer2022, Arranz2024b, Scholch2025}. In future work, we plan to investigate the LMC bar's constituent orbits using the BFE approaches we have developed.

The mSSA reconstructions (Figure \ref{fig:mssa_reconstructions} {\em right panel}) show that the disk's peak response to the halo quadrupole lags behind the pericenter by roughly $100 - 200$ Myr. Significantly, the LMC-SMC collision occurred at a similar time ago, suggesting that the correlated Disk $+$ Halo response (see also Table \ref{tab:mssa_table}) could be at its maximum today. Moreover, the simulated orbit (Figures \ref{fig:lmc_smc_orbit}) and torque calculations (Figure \ref{fig:smc_direct_torques}) reveal that at a separation of $R_{\rm SMC-LMC} \approx 20$ kpc (t $\approx 6.4$ Gyr), which corresponds to the observed LMC-SMC separation at present day, the halo torques are comparable to or larger than the SMC's direct torques across all radii in the LMC disk. Consequently, the halo may play an important role in perturbing the LMC disk at present day. However, this interpretation remains tentative, as the precise dynamical response of the LMC halo to a direct SMC-LMC collision is not yet fully understood.

Finally, the LMC's halo is expected to be distorted by the MW as well \citep{GC2021, Lilleengen2023, Vasiliev2024}, which is not accounted for in our simulations. The strength of the MW-induced quadrupole in the LMC halo can be an order of magnitude larger than the SMC-induced quadrupole (F26). This effect would make the LMC halo torques even stronger. Moreover, the SMC's influence on the LMC's halo is also likely to be amplified if the LMC's halo is already distorted by the MW's tides \citep{GC2019, Dillamore2026}. Hence, in cases where the MW's influence is significant, our finding that the LMC's halo is a significant perturbing agent to the LMC's disk at present day becomes even more important.

In the future, we will apply our BFE $+$ mSSA framework to observationally tailored high-resolution simulations of the LMC-SMC-MW interaction history (\texttt{MEGHA} simulations, H. Rathore et al. {\em in prep}), where all three galaxies will be represented with live DM halos and stellar disks. The \texttt{MEGHA} simulations will enable a direct comparison with the observed LMC's disk, consequently allowing us to differentiate between observed disk perturbations that are caused by the SMC-induced DM halo distortions versus the SMC directly.

\subsection{Implications for Other Interacting Systems} \label{sec:other_galaxies}

Our work highlights that mutual interactions between pairs of dwarf galaxies can significantly affect the morphology of the primary's disk through a combination of direct torques and halo torques, without requiring the tidal field of a massive MW-like neighbor. Many isolated $1:10$ mass ratio pairs of galaxies have been identified in the local universe \cite{Zaritsky1993, Zaritsky1997, Pearson2016, Pearson2018, Besla2018}, with a significant fraction of these systems showing disturbances in the primary's disk (see \cite{Phookun1992, Wilcots2004} for LMC-like morphologies). The number of such galaxies with well-quantified morphologies is expected to further increase with deeper photometric surveys undertaken with DECam --- like DELVE \citep{Drlica-Wagner2021} and MERIAN \citep{Danieli2025}, Rubin-LSST, and eventually with Roman.

F26 argued that dipole and quadrupole distortions in the halos of interacting galaxies should be common (see also \citealt{Arora2025, Darragh-Ford2025}). This generality of halo distortions motivates using our BFE framework with tailored simulations of more and more interacting systems in the local universe, with the goal of understanding how the DM distortions affect the stellar distribution of these galaxies. In particular, we have shown that the distorted halo can still apply significant torques on the disk even when the satellite is far away ($> 20$ kpc) (see also arguments put forth by \citealt{Vesperini2000, Gomez2016, Laporte2018}). As such, significant primary disk distortions will be expected even for widely separated interacting pairs, due to the existence of their DM halos.

Our findings are also important for interpreting the morphologies of satellite galaxies around other hosts, including M31 \citep{Savino2025} and MW-like satellite systems identified using surveys like the SAGA survey \citep{Mao2024}. We have shown that the SMC can significantly distort the LMC's disk even prior to the Clouds becoming satellites of the MW. As such, the origin of disturbed morphologies of satellite galaxies (e.g., M33 --- \citealt{Corbelli2024}) needs to be revisited. In particular, one needs to ask whether these satellite galaxies could have experienced significant morphological processing through interactions with other satellites, or through interactions within the parent group that they were a part of prior to infall \citep[e.g.][]{Patel2026}.

\subsection{The Clouds as Testbeds for DM Particle Physics} \label{sec:dm_physics}

We have shown that the torques on the LMC's disk arise from both the SMC-induced LMC halo distortions and the SMC directly. The strength of torques applied by both of these agents is sensitive to the DM properties of the Clouds as well as the nature of DM particle interactions. As such, the morphological perturbations in the LMC's disk are a unique probe of DM theory.

In section \ref{sec:torques}, we argued that the SMC's direct torques at any instant of time depend on the SMC's bound mass. F26 showed that even prior to the Clouds' MW infall, the SMC loses almost $\frac{2}{3}^{\rm rd}$ of its initial halo mass due to the LMC's tides. The SMC's mass loss will be sensitive to its DM profile \citep[e.g.][]{Nadler2020, Du2024}. As such, the strength of the SMC's direct torques and consequently their effect on the LMC's disk will be a function of the SMC's DM profile, which in turn is sensitive to the underlying DM particle interactions. In particular, DM self-interactions are expected to make the profile cored \citep{Weinberg2015}, which is more susceptible to tidal disruption \citep{Zeng2022, Zhang2024} as compared to a cusped NFW \citep{NFW1997} or Hernquist-like profile expected in CDM. For this reason, we predict that the SMC-induced morphological perturbations in the LMC's disk will be weaker if DM were significantly self-interacting.

The SMC-induced halo distortions are also expected to be sensitive to DM physics. In particular, \cite{Foote2023} showed that the structure of the LMC's dynamical friction wake in the MW's halo is different in Fuzzy Dark Matter models because of interference between the de-Broglie wavelengths of individual DM particles (see also \citealt{Lancaster2020}). The difference is expected to be more pronounced in the LMC-SMC system as the SMC-LMC relative orbital speed is lower than the LMC-MW relative speed, resulting in a relatively stronger density contrast of the dynamical friction wakes (F26). \cite{Glennon2024} showed that the wake can also be different in Ultra-Light self-interacting DM models. In particular, attractive DM self-interactions can enhance the wake strength, whereas repulsive self-interactions can suppress the wake. The effects on the wake would consequently affect the torques applied by the LMC's distorted halo on the disk, thereby affecting the morphological perturbations in the disk. 

The observational characterization of the Clouds' disequilibrium morphology and kinematics is expected to become increasingly precise, thanks to existing surveys like Gaia \citep{Luri2021}, DECam-SMASH \citep{Nidever2017} and SDSS \citep{Nidever2026}, as well as upcoming surveys with Roman, Rubin-LSST and Stage-V Spectroscopic Experiments. This level of observational precision requires accurate and sophisticated modeling frameworks to appropriately infer DM physics. Through the BFE+mSSA framework we have developed using a CDM simulation of the Clouds, there is now an opportunity to understand how the morphology and internal kinematics of the Clouds will be affected in alternative DM models. We will carry out these explorations in future work.

\subsection{Other Limitations and Future Scope} \label{sec:future}
In this section, further limitations and caveats whose resolution is beyond the scope of the current study are listed, but we mention how we plan to address them in future works:

\begin{itemize}
    \item The halo BFE's that we use from F26 do not resolve the inner $1$ kpc of the LMC's halo. That is why we have not analyzed the torques on the inner 2 kpc of the LMC's disk. In future works, we will construct a halo basis that enables the computation of halo torques near the LMC's center, particularly on the bar. 
    \item As evident from Figure \ref{fig:bfe_reconstruction_eg}, our disk BFE's reasonably captures a bulk of the simulated LMC disk's vertical structure. However, some discrepancies remain in the outer LMC disk ($R \approx 10$ kpc), where the BFE underestimates the mean vertical deviation by $\sim 0.2$ kpc and does not exactly capture the shape of the warp. Resolving these discrepancies would require an even more flexible disk basis, and hence an increase in the value of the \texttt{EXP} $l_{\rm fid, max}$ parameter beyond 128.
    \item Even though our disk BFEs are fully self-consistent, and therefore account for self-gravity of the LMC's disk, we have not carefully quantified the restoring force the disk would apply once it gets warped. This restoring force would affect the amplitude of vertical perturbations for a given strength of external torques; outer regions of a disk distort more easily due to weaker self-gravity as compared to inner regions \citep{Jog2000, Ghosh2022, Varela-Lavin2023, Tavangar2026}. To understand the role of these restoring forces, a controlled numerical experiment is required wherein the disk is represented as a collection of non-gravitating tracer particles in our time-dependent SMC $+$ distorted halo BFE potential, and the result of this experiment needs to be compared with our fully live N-body simulation. Such tracer particle experiments are feasible by combining the \texttt{EXP} code with integrators like \texttt{Gala} \citep{Price-Whelan2017}, \texttt{Galpy} \citep{Bovy2015} or \texttt{AGAMA} \citep{Vasiliev2019}.
    \item Similar controlled experiments like the one mentioned above are required to predict exactly which warps in the observed LMC's disk primarily originate from the halo torques versus the warps which originate from the SMC's torques. These experiments would include simulating the LMC-SMC system in a rigid LMC halo, and evolving the LMC's disk in our time-dependent BFE halo potential without the SMC.
    \item Finally, our simulations do not include hydrodynamics. In the future, we will apply our BFE framework to hydrodynamic simulations of the LMC-SMC-MW interaction history to assess the impact of processes like gas dynamics, radiative cooling, and stellar feedback on the evolution of the LMC's disk perturbations.
\end{itemize}

\section{Conclusion}
The mutual LMC-SMC interactions over the last $\sim6$~Gyr have left the LMC's DM halo significantly perturbed, including dipole and quadrupole distortions in the halo (F26). We aim to build a framework that can quantify perturbations in the LMC's stellar disk that arise from torques owing to SMC-induced halo distortions versus direct torques owing to the SMC itself. In particular, the LMC stellar disk is observed with pronounced stellar warps at different radii, offering a potential novel test of DM theory. In this work, we present a framework to: (i) quantify the relative influence of the LMC's distorted halo torques and the SMC's torques on the LMC's disk; (ii) characterize how the halo and SMC torques manifest in the LMC disk perturbations; and (iii) characterize the LMC halo and the SMC contributions in the temporal evolution of the LMC disk perturbations.

In this work, we construct basis function expansions (BFEs) of the LMC's halo and stellar disk (Figure \ref{fig:basis_viz}) in a high-resolution N-body simulation of an isolated LMC-SMC-like galaxy interaction. Both the LMC and SMC are initialized with live DM halos and stellar disks, and the Clouds interact for a period of $\sim$6 Gyr on an eccentric, decaying orbit (Figure \ref{fig:lmc_smc_orbit}). The MW is not included in this study as our goal is to understand how the mutual LMC-SMC interactions shape the LMC's stellar disk. As such, this study also presents a new understanding of the initial conditions of the LMC-SMC system, and the dynamical state of the LMC's stellar disk, prior to MW infall. 

In the first component of the framework, the relative significance of the LMC's halo torques and the SMC's direct torques were established. Warps in the inner LMC disk are found to be the most promising probe of the perturbed structure of the LMC's DM distribution. Additional findings are:   

\begin{itemize}
    \item {\em LMC halo torques act on the LMC stellar disk primarily through 
    the quadrupole distortion caused by the SMC:} See Figures \ref{fig:nmax_lmax_grid} and \ref{fig:n_l_mesh}. The LMC halo quadrupole includes the SMC's dynamical friction wake and the over/under densities in the inner halo ($R < 20$ kpc, F26).
    \item {\em The relative strength of the torques from the LMC halo and the SMC on the LMC disk is time and radius dependent:} both the halo and SMC torques peak near SMC orbital pericenters (Figure \ref{fig:smc_direct_torques}). At the earlier two pericenters (LMC-SMC separation $>20$~kpc), the halo torques are comparable to the SMC's torques. At the latter two pericenters (LMC-SMC separation $<20$~kpc), the SMC's torques are stronger by up to an order of magnitude. However, after each pericenter, the SMC's torques decay $\sim 2-3$ times more rapidly than the halo torques. As such, the halo torques are sustained for longer times than the SMC's torques throughout the SMC's orbit. In particular, at all apocenters, the halo torques on the inner LMC disk (R$<6$~kpc) are stronger than the SMC's torques by at least a factor of $\sim5$.
    
    \item {\em Consequently, the time-integrated torques from the halo are stronger than those from the SMC in the inner LMC disk:} The cumulative time integral of the halo torques and the SMC's torques (see Figure \ref{fig:cumulative_torques}) indicate that the halo torques have a larger contribution to the change in the LMC's disk angular momentum for $R<6$~kpc, but the SMC's torques have a larger contribution for $R>6$~kpc. Hence, warps in the inner LMC disk probe the LMC's DM distribution.
\end{itemize}
 
In the second component of the framework, disk BFEs were used to investigate perturbations to the LMC's stellar morphology that arise from the halo's torques and SMC's torques. We demonstrate a disk BFE framework that captures perturbations of a highly disturbed disk, such as that of the LMC (Figure \ref{fig:bfe_reconstruction_eg} and Figure \ref{fig:mean_z}). 
We find that:
\begin{itemize}
    \item {\em The LMC's disk has a significant global vertical response to the SMC:} we quantify the LMC disk warps through the vertically anti-symmetric basis functions. The power in the vertically anti-symmetric coefficients systematically increases as the LMC-SMC orbit becomes more compact (Figure \ref{fig:coeff_power}), with the coefficients attaining local maxima near the LMC-SMC pericenters. Further, the warps have the most power at the largest radial scales of the LMC disk ($\sim$ scale radius, see Figure \ref{fig:coeff_asymm_m1}).
    \item {\em The LMC's disk is already significantly perturbed before MW Infall:} The simulated LMC's disk develops warps with a mean vertical extent of $\sim 1$~kpc (Figure \ref{fig:mean_z}) as a result of the SMC and halo torques prior to the Clouds' MW infall epoch. This infall state of the LMC disk needs to be accounted for while interpreting the observed LMC's morphology at present day.
\end{itemize}

Finally, in the third component of the framework,  multi-channel singular spectral analysis (mSSA) was applied on the time series of the LMC-SMC orbit, the LMC halo quadrupole power, and the disk vertically anti-symmetric power (Figure \ref{fig:mssa_time_series}). The goal is to identify temporal components of the LMC warp evolution that are directly correlated with the LMC's halo quadrupole. mSSA separated the key temporal components (such as overall growth, decay, and quasi-periodic oscillations) of the above time series into different principal components (PCs, Figure \ref{fig:evals_pcs}). We find that:
\begin{itemize}
    \item {\em mSSA identifies dynamical correlations between the LMC warp evolution and the LMC halo quadrupole:} By reconstructing the original time series from the PCs (Figure \ref{fig:mssa_reconstructions}), we identified correlations between different temporal components of the LMC's vertically anti-symmetric coefficient, the SMC's orbit, and the LMC's distorted halo. This illustrates mSSA's ability to identify couplings across different temporal processes. 
    \item{\em In the observed, present-day SMC-LMC system, the LMC halo quadrupole likely has a significant effect on the LMC's disk.} We find a $100-200$ Myr delay in the peak of the LMC warp amplitude with respect to the LMC-SMC pericenters. This is explained by a delayed peak of the halo quadrupole. The SMC's most recent pericenter approach to the LMC occurred $<200$ Myr ago. Given the expected delay time, we anticipate that the LMC's halo quadrupole is currently acting on the LMC's disk.
    \end{itemize}

We conclude that the presented framework can separate the effects of the LMC's distorted halo torques and the SMC's torques on the LMC's disk. Excitingly, this means that the LMC-SMC system is a promising new testbed for DM particle physics. The strength of the SMC's torques at any moment in time depends on its tidally truncated (bound) DM profile, which in itself is determined by the underlying DM particle interactions. The torques from the LMC's distorted halo depend on the strength of the SMC-induced LMC halo distortions, which also depends on the nature of the DM particle.

In future work, the framework developed in this work will be applied to an observationally fine-tuned CDM N-body simulation of the LMC-SMC-MW interaction history. This framework would enable disentangling the effect of the MW-induced LMC halo distortions on the LMC's disk from the SMC-induced LMC halo distortions, and allow the SMC's torques and the LMC's halo torques to be directly connected to the observed LMC disk perturbations.

\begin{acknowledgments}
HR would like to acknowledge interesting discussions with Kathryn Johnston, Mathieu Renzo, Dennis Zaritsky, Chris Hamilton, Scott Tremaine, Kaisey Mandel, Jason Hunt, Uddipan Banik, Kiyan Tavangar, Michael Petersen and S{\'o}ley Hyman. We would like to thank the anonymous referee and the journal data editor for providing insightful comments that improved the quality as well as clarity of the paper. HR, HRF, GB would like to acknowledge financial support from NASA FINESST 80NSSC24K1469 (FI - H. Rathore, PI - G. Besla), NASA ATP 80NSSC24K1225 (PI - G. Besla) and NSF CAREER AST 1941096 (PI - G. Besla). SVL acknowledges financial support from ANID/Fondo ALMA 2024/31240070. KJD and NGC acknowledge support from the Heising-Simons Foundation grant \# 2022-3927. The Theoretical Astrophysics Program (TAP) at the University of Arizona provided resources to support this work. This work utilized the Puma and ElGato High Performance Computing clusters at the University of Arizona. We thank Ethan Jahn for his assistance with installing and troubleshooting \texttt{EXP} and \texttt{MakeGalaxy} on the Puma cluster, which was made possible through the University of Arizona Research Technologies Collaborative Support program. We have made extensive use of NASA’s Astrophysics Data System and the arXiv preprint service in the preparation of this paper. We respectfully acknowledge the University of Arizona is on the land and territories of Indigenous peoples. Today, Arizona is home to 22 federally recognized tribes, with Tucson being home to the O’odham and the Yaqui. The university strives to build sustainable relationships with sovereign Native Nations and Indigenous communities through education offerings, partnerships, and community service.
\software{\texttt{EXP} \citep{Petersen2025}. \texttt{Gadget-4} \citep{Springel2021}. \texttt{combine\_ics} \citep{combine_ics}. \texttt{snapAnalysis} \citep{snapAnalysis}. Python, and its packages like \texttt{numpy} \citep{vanderWalt2011, Harris2020}, \texttt{scipy} \citep{Virtanen2020} and \texttt{matplotlib} \citep{Hunter2007}. \texttt{Jupyter} \citep{Kluyver2016, Granger2021}.}
\end{acknowledgments}

\begin{contribution}
HR is responsible for designing and running the N-body simulations, performing the torque calculations, constructing the disk BFEs, and doing the mSSA analysis. HR also wrote and submitted the manuscript. HRF performed the halo BFEs, provided significant help in the construction of the disk BFEs and in the mSSA analysis, and edited the manuscript. GB supervised HR and HRF, obtained funding, provided guidance on the analysis, and edited the manuscript. SVL provided significant help with the calculation of torques, and edited the manuscript. NGC provided guidance on the torque and mSSA analysis, and edited the manuscript. MDW is a primary developer of the EXP code, and provided significant help with using the code and interpretation of the results, and edited the manuscript. FAG, KJD and CFPL provided significant guidance on the analysis and interpretation of the results, and edited the manuscript.
\end{contribution}

\appendix

\section{LMC Disk BFE Validation}

\begin{figure}
    \centering
    \includegraphics[width=\columnwidth]{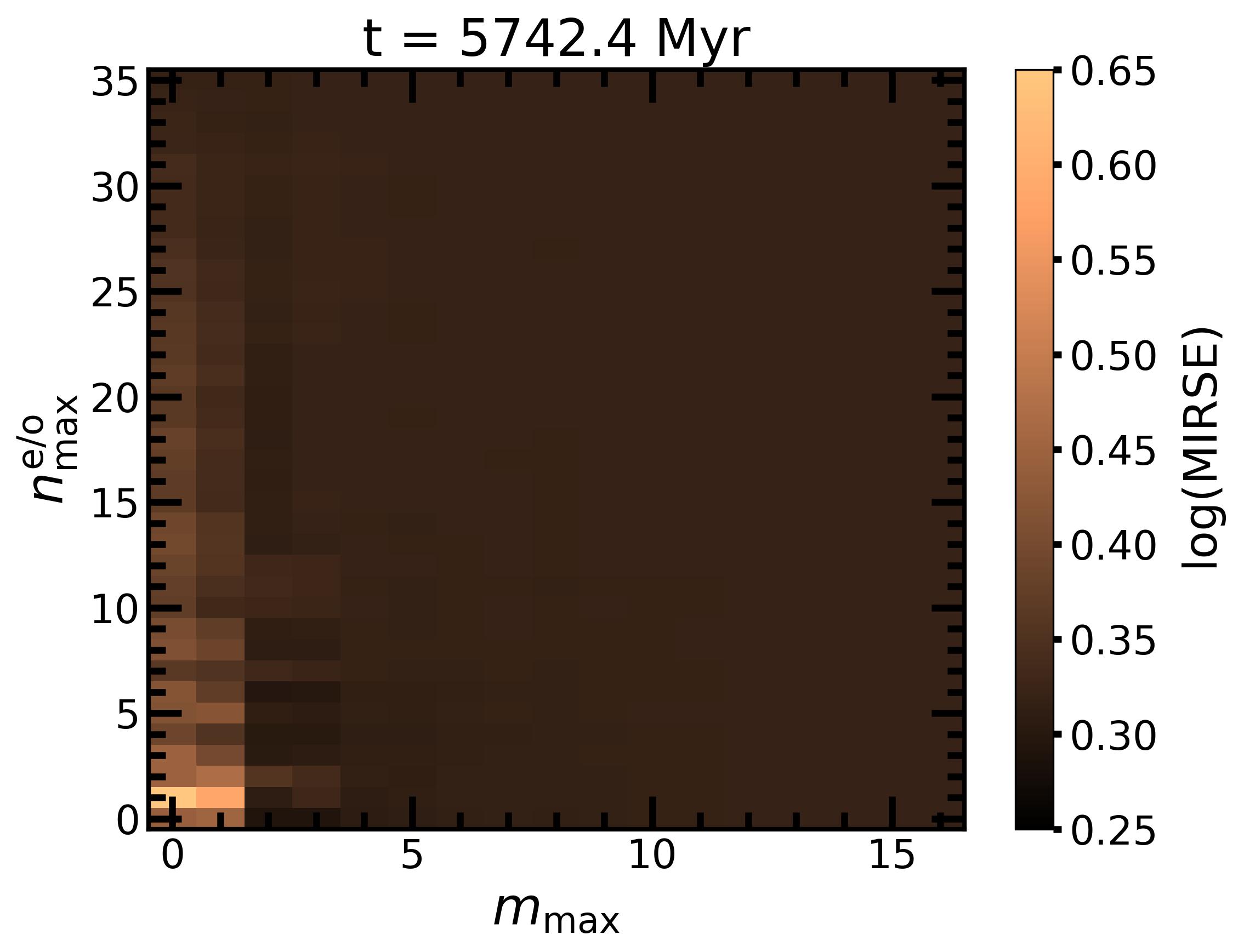}
    \caption{The Mean Integrated Relative Square Error (MIRSE) of the simulated LMC disk's BFE reconstruction at the third LMC-SMC pericenter/MW Infall Epoch (t $=$ 5.74 Gyr, see Figure~\ref{fig:bfe_reconstruction_eg}). The colorbar represents the MIRSE value obtained after truncating the disk coefficient series up to a given ($m_{\rm max}, n^{\rm e/o}_{\rm max}$) truncation order. The worst MIRSE is obtained for $m_{\rm max}, n^{\rm e/o}_{\rm max}$ close to 0 as the basis is not able to capture the small-scale density perturbations in the disk. $m_{\rm max} > 5$ and $n^{\rm e/o}_{\rm max} > 25$ is required for obtaining convergence in the MIRSE of the reconstruction. Hence, the chosen truncation order of $m_{\rm max} = 16, n^{\rm e/o}_{\rm max} = 36$ is sufficient for MIRSE convergence.}
    \label{fig:mirse}
\end{figure}

Following F26, we formally evaluate the quality of the reconstruction with the Mean Integrated Relative Square Error (MIRSE) as the statistical measure. The MIRSE is defined as:
\begin{equation} \label{eq:mise}
   {\rm MIRSE} = \frac{1}{N_{\rm grid}}\sum_i\left[\frac{\rho^{\rm BFE}_{\rm disk}(x_i, y_i, z_i) - \rho_{\rm disk} (x_i, y_i, z_i)}{\rho_{\rm disk} (x_i, y_i, z_i)}\right]^2  
\end{equation}
\noindent where $\rho^{\rm BFE}_{\rm disk}$ and $\rho_{\rm disk}$ denote the BFE-reconstructed and simulated disk density field respectively. The index $i$ denotes each grid cell as defined in section \ref{sec:disk_bfe_recon}, and $N_{\rm grid}$ denotes the number of grid cells. Since the MIRSE computation involves a normalization with the density field, the MIRSE measure is sensitive to density perturbations occurring in the disk outskirts or above/below the disk's mid-plane.

In Figure \ref{fig:mirse}, we analyze the MIRSE as a function of the BFE truncation order ($m_{\rm max}, n^{\rm e/o}_{\rm max}$) for the example snapshot shown in Figure \ref{fig:bfe_reconstruction_eg}. Here, $n^{\rm e/o}_{\rm max}$ denotes the maximum number of vertically symmetric (even) and anti-symmetric (odd) pairs of basis functions used. The MIRSE is the worst (largest) close to ($m_{\rm max} = 0$, $n^{\rm e/o}_{\rm max}$). This is expected since the lowest order basis functions are designed to match the unperturbed disk density profile, and higher order basis is required to capture the perturbations. Note that even the worst MIRSE value is order unity, indicating that the lowest order basis function reasonably captures the bulk of the simulated disk. The MIRSE becomes better (smaller) as more orders are added in $m$ and $n$. After $m_{\rm max} = 5$, $n^{\rm e/o}_{\rm max} = 25$, the MIRSE does not change significantly with the addition of more orders, implying convergence. Hence, our chosen basis order of $m_{\rm max} = 16$, $n^{\rm e/o}_{\rm max} = 36$ (or equivalently $n_{\rm max} = 72$) is sufficient for analyzing the perturbations in the simulated LMC disk.

\bibliography{references}{}

@ARTICLE{Tahmasebzadeh2024b,
       author = {{Tahmasebzadeh}, Behzad and {Zhu}, Ling and {Shen}, Juntai and {Gadotti}, Dimitri A. and {Valluri}, Monica and {Thater}, Sabine and {van de Ven}, Glenn and {Jin}, Yunpeng and {Gerhard}, Ortwin and {Erwin}, Peter and {Jethwa}, Prashin and {Zocchi}, Alice and {Lilley}, Edward J. and {Fragkoudi}, Francesca and {de Lorenzo-C{\'a}ceres}, Adriana and {M{\'e}ndez-Abreu}, Jairo and {Neumann}, Justus and {Guo}, Rui},
        title = "{Schwarzschild modelling of barred s0 galaxy NGC 4371}",
      journal = {\mnras},
         year = 2024,
        month = oct,
       volume = {534},
       number = {1},
        pages = {861-882},
          doi = {10.1093/mnras/stae2109},
archivePrefix = {arXiv},
       eprint = {2310.00497},
 primaryClass = {astro-ph.GA},
       adsurl = {https://ui.adsabs.harvard.edu/abs/2024MNRAS.534..861T}
}

@ARTICLE{SellwoodGerhard2020,
       author = {{Sellwood}, J.~A. and {Gerhard}, Ortwin},
        title = "{Three mechanisms for bar thickening}",
      journal = {\mnras},
         year = 2020,
        month = jul,
       volume = {495},
       number = {3},
        pages = {3175-3191},
          doi = {10.1093/mnras/staa1336},
archivePrefix = {arXiv},
       eprint = {2005.05184},
 primaryClass = {astro-ph.GA},
       adsurl = {https://ui.adsabs.harvard.edu/abs/2020MNRAS.495.3175S}
}

@ARTICLE{BeraldoeSilva2023,
       author = {{Beraldo e Silva}, Leandro and {Debattista}, Victor P. and {Anderson}, Stuart Robert and {Valluri}, Monica and {Erwin}, Peter and {Daniel}, Kathryne J. and {Deg}, Nathan},
        title = "{Orbital Support and Evolution of Flat Profiles of Bars (Shoulders)}",
      journal = {\apj},
         year = 2023,
        month = sep,
       volume = {955},
       number = {1},
          eid = {38},
        pages = {38},
          doi = {10.3847/1538-4357/ace976},
archivePrefix = {arXiv},
       eprint = {2303.04828},
 primaryClass = {astro-ph.GA},
       adsurl = {https://ui.adsabs.harvard.edu/abs/2023ApJ...955...38B}
}

@ARTICLE{Quillen2014,
       author = {{Quillen}, Alice C. and {Minchev}, Ivan and {Sharma}, Sanjib and {Qin}, Yu-Jing and {Di Matteo}, Paola},
        title = "{A vertical resonance heating model for X- or peanut-shaped galactic bulges}",
      journal = {\mnras},
         year = 2014,
        month = jan,
       volume = {437},
       number = {2},
        pages = {1284-1307},
          doi = {10.1093/mnras/stt1972},
archivePrefix = {arXiv},
       eprint = {1307.8441},
 primaryClass = {astro-ph.GA},
       adsurl = {https://ui.adsabs.harvard.edu/abs/2014MNRAS.437.1284Q}
}

@ARTICLE{Choi2018a,
       author = {{Choi}, Yumi and {Nidever}, David L. and {Olsen}, Knut and {Blum}, Robert D. and {Besla}, Gurtina and {Zaritsky}, Dennis and {van der Marel}, Roeland P. and {Bell}, Eric F. and {Gallart}, Carme and {Cioni}, Maria-Rosa L. and {Johnson}, L. Clifton and {Vivas}, A. Katherina and {Saha}, Abhijit and {de Boer}, Thomas J.~L. and {No{\"e}l}, Noelia E.~D. and {Monachesi}, Antonela and {Massana}, Pol and {Conn}, Blair C. and {Martinez-Delgado}, David and {Mu{\~n}oz}, Ricardo R. and {Stringfellow}, Guy S.},
        title = "{SMASHing the LMC: A Tidally Induced Warp in the Outer LMC and a Large-scale Reddening Map}",
      journal = {\apj},
         year = 2018,
        month = oct,
       volume = {866},
       number = {2},
          eid = {90},
        pages = {90},
          doi = {10.3847/1538-4357/aae083},
archivePrefix = {arXiv},
       eprint = {1804.07765},
 primaryClass = {astro-ph.GA},
       adsurl = {https://ui.adsabs.harvard.edu/abs/2018ApJ...866...90C}
}

@ARTICLE{Choi2018b,
       author = {{Choi}, Yumi and {Nidever}, David L. and {Olsen}, Knut and {Besla}, Gurtina and {Blum}, Robert D. and {Zaritsky}, Dennis and {Cioni}, Maria-Rosa L. and {van der Marel}, Roeland P. and {Bell}, Eric F. and {Johnson}, L. Clifton and {Vivas}, A. Katherina and {Walker}, Alistair R. and {de Boer}, Thomas J.~L. and {No{\"e}l}, Noelia E.~D. and {Monachesi}, Antonela and {Gallart}, Carme and {Monelli}, Matteo and {Stringfellow}, Guy S. and {Massana}, Pol and {Martinez-Delgado}, David and {Mu{\~n}oz}, Ricardo R.},
        title = "{SMASHing the LMC: Mapping a Ring-like Stellar Overdensity in the LMC Disk}",
      journal = {\apj},
         year = 2018,
        month = dec,
       volume = {869},
       number = {2},
          eid = {125},
        pages = {125},
          doi = {10.3847/1538-4357/aaed1f},
archivePrefix = {arXiv},
       eprint = {1805.00481},
 primaryClass = {astro-ph.GA},
       adsurl = {https://ui.adsabs.harvard.edu/abs/2018ApJ...869..125C}
}

@ARTICLE{Oden2025,
       author = {{Oden}, Slater J. and {Nidever}, David L. and {Povick}, Joshua and {Massana}, Pol and {Choi}, Yumi and {van der Marel}, Roeland P. and {Cioni}, Maria-Rosa L. and {Sakowska}, Joanna and {Olsen}, Knut A.~G. and {Cullinane}, Lara and {Carballo-Bello}, J.~A. and {Crnojevi{\'c}}, D. and {Ferguson}, P.~S. and {Mart{\'\i}nez-V{\'a}zquez}, C.~E. and {Medina}, G.~E. and {Mutlu-Pakdil}, B. and {Navabi}, M. and {Pace}, A.~B. and {Riley}, A.~H. and {Stringfellow}, Guy S. and {Vivas}, A.~K.},
        title = "{Warped \& Hooked: Mapping the Magellanic Clouds in 3D using Red Clump stars}",
      journal = {arXiv e-prints},
         year = 2025,
        month = dec,
          eid = {arXiv:2512.04200},
        pages = {arXiv:2512.04200},
          doi = {10.48550/arXiv.2512.04200},
archivePrefix = {arXiv},
       eprint = {2512.04200},
 primaryClass = {astro-ph.GA},
       adsurl = {https://ui.adsabs.harvard.edu/abs/2025arXiv251204200O}
}

@ARTICLE{Arranz2025,
       author = {{Jim{\'e}nez-Arranz}, {\'O}. and {Horta}, D. and {van der Marel}, R.~P. and {Nidever}, D. and {Laporte}, C.~F.~P. and {Patel}, E. and {Rix}, H.-W.},
        title = "{Vertical structure and kinematics of the LMC disc from SDSS/Gaia}",
      journal = {\aap},
         year = 2025,
        month = jun,
       volume = {698},
          eid = {A88},
        pages = {A88},
          doi = {10.1051/0004-6361/202553705},
archivePrefix = {arXiv},
       eprint = {2501.04616},
 primaryClass = {astro-ph.GA},
       adsurl = {https://ui.adsabs.harvard.edu/abs/2025A&A...698A..88J}
}

@ARTICLE{Saroon2022,
       author = {{Saroon}, S. and {Subramanian}, S.},
        title = "{Shape of the outer stellar warp in the Large Magellanic Cloud disk}",
      journal = {\aap},
         year = 2022,
        month = oct,
       volume = {666},
          eid = {A103},
        pages = {A103},
          doi = {10.1051/0004-6361/202141435},
archivePrefix = {arXiv},
       eprint = {2207.13269},
 primaryClass = {astro-ph.GA},
       adsurl = {https://ui.adsabs.harvard.edu/abs/2022A&A...666A.103S}
}

@ARTICLE{Subramaniam2009,
       author = {{Subramaniam}, Annapurni and {Subramanian}, Smitha},
        title = "{The Mysterious Bar of the Large Magellanic Cloud: What Is It?}",
      journal = {\apjl},
         year = 2009,
        month = sep,
       volume = {703},
       number = {1},
        pages = {L37-L40},
          doi = {10.1088/0004-637X/703/1/L37},
archivePrefix = {arXiv},
       eprint = {0908.0177},
 primaryClass = {astro-ph.CO},
       adsurl = {https://ui.adsabs.harvard.edu/abs/2009ApJ...703L..37S}
}

@ARTICLE{Besla2016,
       author = {{Besla}, Gurtina and {Mart{\'\i}nez-Delgado}, David and {van der Marel}, Roeland P. and {Beletsky}, Yuri and {Seibert}, Mark and {Schlafly}, Edward F. and {Grebel}, Eva K. and {Neyer}, Fabian},
        title = "{Low Surface Brightness Imaging of the Magellanic System: Imprints of Tidal Interactions between the Clouds in the Stellar Periphery}",
      journal = {\apj},
         year = 2016,
        month = jul,
       volume = {825},
       number = {1},
          eid = {20},
        pages = {20},
          doi = {10.3847/0004-637X/825/1/20},
archivePrefix = {arXiv},
       eprint = {1602.04222},
 primaryClass = {astro-ph.GA},
       adsurl = {https://ui.adsabs.harvard.edu/abs/2016ApJ...825...20B}
}

@ARTICLE{Haschke2012,
       author = {{Haschke}, Raoul and {Grebel}, Eva K. and {Duffau}, Sonia},
        title = "{Three-dimensional Maps of the Magellanic Clouds using RR Lyrae Stars and Cepheids. I. The Large Magellanic Cloud}",
      journal = {\aj},
         year = 2012,
        month = oct,
       volume = {144},
       number = {4},
          eid = {106},
        pages = {106},
          doi = {10.1088/0004-6256/144/4/106},
archivePrefix = {arXiv},
       eprint = {1207.5791},
 primaryClass = {astro-ph.GA},
       adsurl = {https://ui.adsabs.harvard.edu/abs/2012AJ....144..106H}
}

@ARTICLE{Cioni2011,
       author = {{Cioni}, M.-R.~L. and {Clementini}, G. and {Girardi}, L. and {Guandalini}, R. and {Gullieuszik}, M. and {Miszalski}, B. and {Moretti}, M.-I. and {Ripepi}, V. and {Rubele}, S. and {Bagheri}, G. and {Bekki}, K. and {Cross}, N. and {de Blok}, W.~J.~G. and {de Grijs}, R. and {Emerson}, J.~P. and {Evans}, C.~J. and {Gibson}, B. and {Gonzales-Solares}, E. and {Groenewegen}, M.~A.~T. and {Irwin}, M. and {Ivanov}, V.~D. and {Lewis}, J. and {Marconi}, M. and {Marquette}, J.-B. and {Mastropietro}, C. and {Moore}, B. and {Napiwotzki}, R. and {Naylor}, T. and {Oliveira}, J.~M. and {Read}, M. and {Sutorius}, E. and {van Loon}, J. Th. and {Wilkinson}, M.~I. and {Wood}, P.~R.},
        title = "{The VMC survey. I. Strategy and first data}",
      journal = {\aap},
         year = 2011,
        month = mar,
       volume = {527},
          eid = {A116},
        pages = {A116},
          doi = {10.1051/0004-6361/201016137},
archivePrefix = {arXiv},
       eprint = {1012.5193},
 primaryClass = {astro-ph.CO},
       adsurl = {https://ui.adsabs.harvard.edu/abs/2011A&A...527A.116C}
}

@ARTICLE{Cusano2021,
       author = {{Cusano}, F. and {Moretti}, M.~I. and {Clementini}, G. and {Ripepi}, V. and {Marconi}, M. and {Cioni}, M.-R.~L. and {Rubele}, S. and {Garofalo}, A. and {de Grijs}, R. and {Groenewegen}, M.~A.~T. and {Oliveira}, J.~M. and {Subramanian}, S. and {Sun}, N.-C. and {van Loon}, J. Th},
        title = "{The VMC Survey - XLII. Near-infrared period-luminosity relations for RR Lyrae stars and the structure of the Large Magellanic Cloud}",
      journal = {\mnras},
         year = 2021,
        month = jun,
       volume = {504},
       number = {1},
        pages = {1-15},
          doi = {10.1093/mnras/stab901},
archivePrefix = {arXiv},
       eprint = {2103.15492},
 primaryClass = {astro-ph.GA},
       adsurl = {https://ui.adsabs.harvard.edu/abs/2021MNRAS.504....1C}
}

@ARTICLE{Rathore2025a,
       author = {{Rathore}, Himansh and {Choi}, Yumi and {Olsen}, Knut A.~G. and {Besla}, Gurtina},
        title = "{Precise Measurements of the LMC Bar's Geometry with Gaia DR3 and a Novel Solution to Crowding-induced Incompleteness in Star Counting}",
      journal = {\apj},
         year = 2025,
        month = jan,
       volume = {978},
       number = {1},
          eid = {55},
        pages = {55},
          doi = {10.3847/1538-4357/ad93ae},
archivePrefix = {arXiv},
       eprint = {2410.18182},
 primaryClass = {astro-ph.GA},
       adsurl = {https://ui.adsabs.harvard.edu/abs/2025ApJ...978...55R}
}

@ARTICLE{Marel2001,
       author = {{van der Marel}, Roeland P. and {Cioni}, Maria-Rosa L.},
        title = "{Magellanic Cloud Structure from Near-Infrared Surveys. I. The Viewing Angles of the Large Magellanic Cloud}",
      journal = {\aj},
         year = 2001,
        month = oct,
       volume = {122},
       number = {4},
        pages = {1807-1826},
          doi = {10.1086/323099},
archivePrefix = {arXiv},
       eprint = {astro-ph/0105339},
 primaryClass = {astro-ph},
       adsurl = {https://ui.adsabs.harvard.edu/abs/2001AJ....122.1807V}
}

@ARTICLE{Olsen2002,
       author = {{Olsen}, K.~A.~G. and {Salyk}, C.},
        title = "{A Warp in the LMC Disk?}",
      journal = {arXiv e-prints},
         year = 2002,
        month = jul,
          eid = {astro-ph/0207077},
        pages = {astro-ph/0207077},
          doi = {10.48550/arXiv.astro-ph/0207077},
archivePrefix = {arXiv},
       eprint = {astro-ph/0207077},
 primaryClass = {astro-ph},
       adsurl = {https://ui.adsabs.harvard.edu/abs/2002astro.ph..7077O}
}

@ARTICLE{Rathore2025b,
       author = {{Rathore}, Himansh and {Besla}, Gurtina and {Daniel}, Kathryne J. and {Beraldo e Silva}, Leandro},
        title = "{Response of the LMC's Bar to a Recent SMC Collision and Implications for the SMC's Dark Matter Profile}",
      journal = {\apj},
         year = 2025,
        month = jul,
       volume = {988},
       number = {1},
          eid = {79},
        pages = {79},
          doi = {10.3847/1538-4357/ade0ae},
archivePrefix = {arXiv},
       eprint = {2504.16163},
 primaryClass = {astro-ph.GA},
       adsurl = {https://ui.adsabs.harvard.edu/abs/2025ApJ...988...79R}
}

@ARTICLE{Arranz2024,
       author = {{Jim{\'e}nez-Arranz}, {\'O}. and {Roca-F{\`a}brega}, S. and {Romero-G{\'o}mez}, M. and {Luri}, X. and {Bernet}, M. and {McMillan}, P.~J. and {Chemin}, L.},
        title = "{KRATOS: A large suite of N-body simulations to interpret the stellar kinematics of LMC-like discs}",
      journal = {\aap},
         year = 2024,
        month = aug,
       volume = {688},
          eid = {A51},
        pages = {A51},
          doi = {10.1051/0004-6361/202349058},
archivePrefix = {arXiv},
       eprint = {2404.04061},
 primaryClass = {astro-ph.GA},
       adsurl = {https://ui.adsabs.harvard.edu/abs/2024A&A...688A..51J}
}

@misc{Garver2026,
      title={Modeling the recent interactions between the Magellanic Clouds and Milky Way}, 
      author={Bethany Garver and David Nidever and Victor Debattista and Nathan Deg},
      year={2026},
      eprint={2602.05021},
      archivePrefix={arXiv},
      primaryClass={astro-ph.GA},
      url={https://arxiv.org/abs/2602.05021}, 
}

@ARTICLE{Pardy2016,
       author = {{Pardy}, Stephen A. and {D'Onghia}, Elena and {Athanassoula}, E. and {Wilcots}, Eric M. and {Sheth}, Kartik},
        title = "{Tidally Induced Offset Disks in Magellanic Spiral Galaxies}",
      journal = {\apj},
         year = 2016,
        month = aug,
       volume = {827},
       number = {2},
          eid = {149},
        pages = {149},
          doi = {10.3847/0004-637X/827/2/149},
archivePrefix = {arXiv},
       eprint = {1602.07689},
 primaryClass = {astro-ph.GA},
       adsurl = {https://ui.adsabs.harvard.edu/abs/2016ApJ...827..149P}
}

@ARTICLE{Foote2026,
       author = {{Foote}, Hayden R. and {Rathore}, Himansh and {Besla}, Gurtina and {Garavito-Camargo}, Nicol{\'a}s and {Patel}, Ekta and {Petersen}, Michael S. and {Weinberg}, Martin D. and {G{\'o}mez}, Facundo A. and {Laporte}, Chervin F.~P.},
        title = "{Mapping the Distorted Dark Matter Distribution of the LMC-SMC System Prior to Milky Way Infall with Basis Function Expansions}",
      journal = {arXiv e-prints},
         year = 2026,
        month = jan,
          eid = {arXiv:2601.00946},
        pages = {arXiv:2601.00946},
          doi = {10.48550/arXiv.2601.00946},
archivePrefix = {arXiv},
       eprint = {2601.00946},
 primaryClass = {astro-ph.GA},
       adsurl = {https://ui.adsabs.harvard.edu/abs/2026arXiv260100946F}
}

@ARTICLE{Chandrasekhar1943,
       author = {{Chandrasekhar}, S.},
        title = "{Dynamical Friction. I. General Considerations: the Coefficient of Dynamical Friction.}",
      journal = {\apj},
         year = 1943,
        month = mar,
       volume = {97},
        pages = {255},
          doi = {10.1086/144517},
       adsurl = {https://ui.adsabs.harvard.edu/abs/1943ApJ....97..255C}
}

@BOOK{BT2008,
       author = {{Binney}, James and {Tremaine}, Scott},
        title = "{Galactic Dynamics: Second Edition}",
         year = 2008,
       adsurl = {https://ui.adsabs.harvard.edu/abs/2008gady.book.....B}
}

@ARTICLE{TW1984,
       author = {{Tremaine}, S. and {Weinberg}, M.~D.},
        title = "{Dynamical friction in spherical systems.}",
      journal = {\mnras},
         year = 1984,
        month = aug,
       volume = {209},
        pages = {729-757},
          doi = {10.1093/mnras/209.4.729},
       adsurl = {https://ui.adsabs.harvard.edu/abs/1984MNRAS.209..729T}
}

@ARTICLE{White1983,
       author = {{White}, S.~D.~M.},
        title = "{Simulations of sinking satellites}",
      journal = {\apj},
         year = 1983,
        month = nov,
       volume = {274},
        pages = {53-61},
          doi = {10.1086/161425},
       adsurl = {https://ui.adsabs.harvard.edu/abs/1983ApJ...274...53W}
}

@ARTICLE{Choi2009,
       author = {{Choi}, Jun-Hwan and {Weinberg}, Martin D. and {Katz}, Neal},
        title = "{The dynamics of satellite disruption in cold dark matter haloes}",
      journal = {\mnras},
         year = 2009,
        month = dec,
       volume = {400},
       number = {3},
        pages = {1247-1263},
          doi = {10.1111/j.1365-2966.2009.15556.x},
archivePrefix = {arXiv},
       eprint = {0812.0009},
 primaryClass = {astro-ph},
       adsurl = {https://ui.adsabs.harvard.edu/abs/2009MNRAS.400.1247C}
}

@ARTICLE{GC2019,
       author = {{Garavito-Camargo}, Nicolas and {Besla}, Gurtina and {Laporte}, Chervin F.~P. and {Johnston}, Kathryn V. and {G{\'o}mez}, Facundo A. and {Watkins}, Laura L.},
        title = "{Hunting for the Dark Matter Wake Induced by the Large Magellanic Cloud}",
      journal = {\apj},
         year = 2019,
        month = oct,
       volume = {884},
       number = {1},
          eid = {51},
        pages = {51},
          doi = {10.3847/1538-4357/ab32eb},
archivePrefix = {arXiv},
       eprint = {1902.05089},
 primaryClass = {astro-ph.GA},
       adsurl = {https://ui.adsabs.harvard.edu/abs/2019ApJ...884...51G}
}

@ARTICLE{Weinberg1998a,
       author = {{Weinberg}, Martin D.},
        title = "{Fluctuations in finite-N equilibrium stellar systems}",
      journal = {\mnras},
         year = 1998,
        month = jun,
       volume = {297},
       number = {1},
        pages = {101-107},
          doi = {10.1046/j.1365-8711.1998.01456.x},
archivePrefix = {arXiv},
       eprint = {astro-ph/9707206},
 primaryClass = {astro-ph},
       adsurl = {https://ui.adsabs.harvard.edu/abs/1998MNRAS.297..101W}
}

@ARTICLE{Weinberg1998,
       author = {{Weinberg}, Martin D.},
        title = "{Dynamics of an interacting luminous disc, dark halo and satellite companion}",
      journal = {\mnras},
         year = 1998,
        month = sep,
       volume = {299},
       number = {2},
        pages = {499-514},
          doi = {10.1046/j.1365-8711.1998.01790.x},
       adsurl = {https://ui.adsabs.harvard.edu/abs/1998MNRAS.299..499W}
}

@ARTICLE{Vesperini2000,
       author = {{Vesperini}, E. and {Weinberg}, Martin D.},
        title = "{Perturbations of Spherical Stellar Systems during Flyby Encounters}",
      journal = {\apj},
         year = 2000,
        month = may,
       volume = {534},
       number = {2},
        pages = {598-623},
          doi = {10.1086/308788},
archivePrefix = {arXiv},
       eprint = {astro-ph/9908099},
 primaryClass = {astro-ph},
       adsurl = {https://ui.adsabs.harvard.edu/abs/2000ApJ...534..598V}
}

@ARTICLE{Gomez2016,
       author = {{G{\'o}mez}, Facundo A. and {White}, Simon D.~M. and {Marinacci}, Federico and {Slater}, Colin T. and {Grand}, Robert J.~J. and {Springel}, Volker and {Pakmor}, R{\"u}diger},
        title = "{A fully cosmological model of a Monoceros-like ring}",
      journal = {\mnras},
         year = 2016,
        month = mar,
       volume = {456},
       number = {3},
        pages = {2779-2793},
          doi = {10.1093/mnras/stv2786},
archivePrefix = {arXiv},
       eprint = {1509.08459},
 primaryClass = {astro-ph.GA},
       adsurl = {https://ui.adsabs.harvard.edu/abs/2016MNRAS.456.2779G}
}

@ARTICLE{Laporte2018,
       author = {{Laporte}, Chervin F.~P. and {Johnston}, Kathryn V. and {G{\'o}mez}, Facundo A. and {Garavito-Camargo}, Nicolas and {Besla}, Gurtina},
        title = "{The influence of Sagittarius and the Large Magellanic Cloud on the stellar disc of the Milky Way Galaxy}",
      journal = {\mnras},
         year = 2018,
        month = nov,
       volume = {481},
       number = {1},
        pages = {286-306},
          doi = {10.1093/mnras/sty1574},
archivePrefix = {arXiv},
       eprint = {1710.02538},
 primaryClass = {astro-ph.GA},
       adsurl = {https://ui.adsabs.harvard.edu/abs/2018MNRAS.481..286L}
}

@ARTICLE{Poggio2021,
       author = {{Poggio}, Eloisa and {Laporte}, Chervin F.~P. and {Johnston}, Kathryn V. and {D'Onghia}, Elena and {Drimmel}, Ronald and {Grion Filho}, Douglas},
        title = "{Measuring the vertical response of the Galactic disc to an infalling satellite}",
      journal = {\mnras},
         year = 2021,
        month = nov,
       volume = {508},
       number = {1},
        pages = {541-559},
          doi = {10.1093/mnras/stab2245},
archivePrefix = {arXiv},
       eprint = {2011.11642},
 primaryClass = {astro-ph.GA},
       adsurl = {https://ui.adsabs.harvard.edu/abs/2021MNRAS.508..541P}
}

@ARTICLE{Wille2024,
       author = {{Wille}, A. and {Machado}, R.~E.~G.},
        title = "{Warps induced by satellites on barred and non-barred galaxies}",
      journal = {\mnras},
         year = 2024,
        month = sep,
       volume = {533},
       number = {3},
        pages = {2997-3007},
          doi = {10.1093/mnras/stae2004},
archivePrefix = {arXiv},
       eprint = {2408.09932},
 primaryClass = {astro-ph.GA},
       adsurl = {https://ui.adsabs.harvard.edu/abs/2024MNRAS.533.2997W}
}

@ARTICLE{DB2012,
       author = {{Diaz}, Jonathan D. and {Bekki}, Kenji},
        title = "{The Tidal Origin of the Magellanic Stream and the Possibility of a Stellar Counterpart}",
      journal = {\apj},
         year = 2012,
        month = may,
       volume = {750},
       number = {1},
          eid = {36},
        pages = {36},
          doi = {10.1088/0004-637X/750/1/36},
archivePrefix = {arXiv},
       eprint = {1112.6191},
 primaryClass = {astro-ph.GA},
       adsurl = {https://ui.adsabs.harvard.edu/abs/2012ApJ...750...36D}
}

@ARTICLE{Besla2012,
       author = {{Besla}, Gurtina and {Kallivayalil}, Nitya and {Hernquist}, Lars and {van der Marel}, Roeland P. and {Cox}, T.~J. and {Kere{\v{s}}}, Du{\v{s}}an},
        title = "{The role of dwarf galaxy interactions in shaping the Magellanic System and implications for Magellanic Irregulars}",
      journal = {\mnras},
         year = 2012,
        month = apr,
       volume = {421},
       number = {3},
        pages = {2109-2138},
          doi = {10.1111/j.1365-2966.2012.20466.x},
archivePrefix = {arXiv},
       eprint = {1201.1299},
 primaryClass = {astro-ph.GA},
       adsurl = {https://ui.adsabs.harvard.edu/abs/2012MNRAS.421.2109B}
}

@ARTICLE{Lowing2011,
       author = {{Lowing}, Ben and {Jenkins}, Adrian and {Eke}, Vincent and {Frenk}, Carlos},
        title = "{A halo expansion technique for approximating simulated dark matter haloes}",
      journal = {\mnras},
         year = 2011,
        month = oct,
       volume = {416},
       number = {4},
        pages = {2697-2711},
          doi = {10.1111/j.1365-2966.2011.19222.x},
archivePrefix = {arXiv},
       eprint = {1010.6197},
 primaryClass = {astro-ph.CO},
       adsurl = {https://ui.adsabs.harvard.edu/abs/2011MNRAS.416.2697L}
}

@ARTICLE{CluttonBrock73,
       author = {{Clutton-Brock}, M.},
        title = "{The Gravitational Field of Three Dimensional Galaxies}",
      journal = {\apss},
         year = 1973,
        month = jul,
       volume = {23},
       number = {1},
        pages = {55-69},
          doi = {10.1007/BF00647652},
       adsurl = {https://ui.adsabs.harvard.edu/abs/1973Ap&SS..23...55C}
}

@ARTICLE{Ostriker1992,
       author = {{Hernquist}, Lars and {Ostriker}, Jeremiah P.},
        title = "{A Self-consistent Field Method for Galactic Dynamics}",
      journal = {\apj},
         year = 1992,
        month = feb,
       volume = {386},
        pages = {375},
          doi = {10.1086/171025},
       adsurl = {https://ui.adsabs.harvard.edu/abs/1992ApJ...386..375H}
}

@ARTICLE{Weinberg1999,
       author = {{Weinberg}, Martin D.},
        title = "{An Adaptive Algorithm for N-Body Field Expansions}",
      journal = {\aj},
         year = 1999,
        month = jan,
       volume = {117},
       number = {1},
        pages = {629-637},
          doi = {10.1086/300669},
archivePrefix = {arXiv},
       eprint = {astro-ph/9805357},
 primaryClass = {astro-ph},
       adsurl = {https://ui.adsabs.harvard.edu/abs/1999AJ....117..629W}
}

@ARTICLE{GC2021,
       author = {{Garavito-Camargo}, Nicol{\'a}s and {Besla}, Gurtina and {Laporte}, Chervin F.~P. and {Price-Whelan}, Adrian M. and {Cunningham}, Emily C. and {Johnston}, Kathryn V. and {Weinberg}, Martin and {G{\'o}mez}, Facundo A.},
        title = "{Quantifying the Impact of the Large Magellanic Cloud on the Structure of the Milky Way's Dark Matter Halo Using Basis Function Expansions}",
      journal = {\apj},
         year = 2021,
        month = oct,
       volume = {919},
       number = {2},
          eid = {109},
        pages = {109},
          doi = {10.3847/1538-4357/ac0b44},
archivePrefix = {arXiv},
       eprint = {2010.00816},
 primaryClass = {astro-ph.GA},
       adsurl = {https://ui.adsabs.harvard.edu/abs/2021ApJ...919..109G}
}

@ARTICLE{Gomez2021,
       author = {{G{\'o}mez}, Facundo A. and {Torres-Flores}, Sergio and {Mora-Urrejola}, Catalina and {Monachesi}, Antonela and {White}, Simon D.~M. and {Maffione}, Nicolas P. and {Grand}, Robert J.~J. and {Marinacci}, Federico and {Pakmor}, R{\"u}diger and {Springel}, Volker and {Frenk}, Carlos S. and {Amram}, Philippe and {Epinat}, Beno{\^\i}t and {Mendes de Oliveira}, Claudia},
        title = "{A Tidally Induced Global Corrugation Pattern in an External Disk Galaxy Similar to the Milky Way}",
      journal = {\apj},
         year = 2021,
        month = feb,
       volume = {908},
       number = {1},
          eid = {27},
        pages = {27},
          doi = {10.3847/1538-4357/abcd97},
archivePrefix = {arXiv},
       eprint = {2011.12323},
 primaryClass = {astro-ph.GA},
       adsurl = {https://ui.adsabs.harvard.edu/abs/2021ApJ...908...27G}
}

@ARTICLE{Gomez2017,
       author = {{G{\'o}mez}, Facundo A. and {White}, Simon D.~M. and {Grand}, Robert J.~J. and {Marinacci}, Federico and {Springel}, Volker and {Pakmor}, R{\"u}diger},
        title = "{Warps and waves in the stellar discs of the Auriga cosmological simulations}",
      journal = {\mnras},
         year = 2017,
        month = mar,
       volume = {465},
       number = {3},
        pages = {3446-3460},
          doi = {10.1093/mnras/stw2957},
archivePrefix = {arXiv},
       eprint = {1606.06295},
 primaryClass = {astro-ph.GA},
       adsurl = {https://ui.adsabs.harvard.edu/abs/2017MNRAS.465.3446G}
}

@ARTICLE{GarciaConde2024,
       author = {{Garc{\'\i}a-Conde}, B. and {Antoja}, T. and {Roca-F{\`a}brega}, S. and {G{\'o}mez}, F. and {Ramos}, P. and {Garavito-Camargo}, N. and {G{\'o}mez-Flechoso}, M.~A.},
        title = "{Galactoseismology in cosmological simulations. Vertical perturbations by dark matter, satellite galaxies, and gas}",
      journal = {\aap},
         year = 2024,
        month = mar,
       volume = {683},
          eid = {A47},
        pages = {A47},
          doi = {10.1051/0004-6361/202347446},
archivePrefix = {arXiv},
       eprint = {2311.07137},
 primaryClass = {astro-ph.GA},
       adsurl = {https://ui.adsabs.harvard.edu/abs/2024A&A...683A..47G}
}

@ARTICLE{Lilleengen2023,
       author = {{Lilleengen}, Sophia and {Petersen}, Michael S. and {Erkal}, Denis and {Pe{\~n}arrubia}, Jorge and {Koposov}, Sergey E. and {Li}, Ting S. and {Cullinane}, Lara R. and {Ji}, Alexander P. and {Kuehn}, Kyler and {Lewis}, Geraint F. and {Mackey}, Dougal and {Pace}, Andrew B. and {Shipp}, Nora and {Zucker}, Daniel B. and {Bland-Hawthorn}, Joss and {Hilmi}, Tariq and {S5 Collaboration}},
        title = "{The effect of the deforming dark matter haloes of the Milky Way and the Large Magellanic Cloud on the Orphan-Chenab stream}",
      journal = {\mnras},
         year = 2023,
        month = jan,
       volume = {518},
       number = {1},
        pages = {774-790},
          doi = {10.1093/mnras/stac3108},
archivePrefix = {arXiv},
       eprint = {2205.01688},
 primaryClass = {astro-ph.GA},
       adsurl = {https://ui.adsabs.harvard.edu/abs/2023MNRAS.518..774L}
}

@ARTICLE{Brooks2025,
       author = {{Brooks}, Richard A.~N. and {Garavito-Camargo}, Nicol{\'a}s and {Johnston}, Kathryn V. and {Price-Whelan}, Adrian M. and {Sanders}, Jason L. and {Lilleengen}, Sophia},
        title = "{LMC Calls, Milky Way Halo Answers: Disentangling the Effects of the MW─LMC Interaction on Stellar Stream Populations}",
      journal = {\apj},
         year = 2025,
        month = jan,
       volume = {978},
       number = {1},
          eid = {79},
        pages = {79},
          doi = {10.3847/1538-4357/ad93a7},
archivePrefix = {arXiv},
       eprint = {2410.02574},
 primaryClass = {astro-ph.GA},
       adsurl = {https://ui.adsabs.harvard.edu/abs/2025ApJ...978...79B}
}

@ARTICLE{Darragh-Ford2025,
       author = {{Darragh-Ford}, Elise and {Garavito-Camargo}, Nicolas and {Arora}, Arpit and {Wechsler}, Risa H. and {Mansfield}, Phil and {Besla}, Gurtina and {Petersen}, Michael S. and {Weinberg}, Martin D. and {Varela-Lavin}, Silvio and {Buch}, Deveshi and {Cunningham}, Emily C. and {Daniel}, Kathryne J. and {Gomez}, Facundo A. and {Johnston}, Kathryn V. and {Laporte}, Chervin F.~P. and {Mao}, Yao-Yuan and {Nadler}, Ethan O. and {Sanderson}, Robyn},
        title = "{Shaping the Milky Way. II. The dark matter halo's response to the LMC's passage in a cosmological context}",
      journal = {arXiv e-prints},
         year = 2025,
        month = nov,
          eid = {arXiv:2511.02031},
        pages = {arXiv:2511.02031},
          doi = {10.48550/arXiv.2511.02031},
archivePrefix = {arXiv},
       eprint = {2511.02031},
 primaryClass = {astro-ph.GA},
       adsurl = {https://ui.adsabs.harvard.edu/abs/2025arXiv251102031D}
}

@ARTICLE{Khachaturyants2022,
       author = {{Khachaturyants}, Tigran and {Debattista}, Victor P. and {Ghosh}, Soumavo and {Beraldo e Silva}, Leandro and {Daniel}, Kathryne J.},
        title = "{The pattern speeds of vertical breathing waves}",
      journal = {\mnras},
         year = 2022,
        month = nov,
       volume = {517},
       number = {1},
        pages = {L55-L59},
          doi = {10.1093/mnrasl/slac112},
archivePrefix = {arXiv},
       eprint = {2210.00341},
 primaryClass = {astro-ph.GA},
       adsurl = {https://ui.adsabs.harvard.edu/abs/2022MNRAS.517L..55K}
}

@ARTICLE{Grand2023,
       author = {{Grand}, Robert J.~J. and {Pakmor}, R{\"u}diger and {Fragkoudi}, Francesca and {G{\'o}mez}, Facundo A. and {Trick}, Wilma and {Simpson}, Christine M. and {van de Voort}, Freeke and {Bieri}, Rebekka},
        title = "{An ever-present Gaia snail shell triggered by a dark matter wake}",
      journal = {\mnras},
         year = 2023,
        month = sep,
       volume = {524},
       number = {1},
        pages = {801-816},
          doi = {10.1093/mnras/stad1969},
archivePrefix = {arXiv},
       eprint = {2211.08437},
 primaryClass = {astro-ph.GA},
       adsurl = {https://ui.adsabs.harvard.edu/abs/2023MNRAS.524..801G}
}

@ARTICLE{AM2002,
       author = {{Athanassoula}, E. and {Misiriotis}, A.},
        title = "{Morphology, photometry and kinematics of N -body bars - I. Three models with different halo central concentrations}",
      journal = {\mnras},
         year = 2002,
        month = feb,
       volume = {330},
       number = {1},
        pages = {35-52},
          doi = {10.1046/j.1365-8711.2002.05028.x},
archivePrefix = {arXiv},
       eprint = {astro-ph/0111449},
 primaryClass = {astro-ph},
       adsurl = {https://ui.adsabs.harvard.edu/abs/2002MNRAS.330...35A}
}

@ARTICLE{Silva2023,
       author = {{Beraldo e Silva}, Leandro and {Debattista}, Victor P. and {Anderson}, Stuart Robert and {Valluri}, Monica and {Erwin}, Peter and {Daniel}, Kathryne J. and {Deg}, Nathan},
        title = "{Orbital Support and Evolution of Flat Profiles of Bars (Shoulders)}",
      journal = {\apj},
         year = 2023,
        month = sep,
       volume = {955},
       number = {1},
          eid = {38},
        pages = {38},
          doi = {10.3847/1538-4357/ace976},
archivePrefix = {arXiv},
       eprint = {2303.04828},
 primaryClass = {astro-ph.GA},
       adsurl = {https://ui.adsabs.harvard.edu/abs/2023ApJ...955...38B}
}

@ARTICLE{Ghil1991,
       author = {{Ghil}, M. and {Vautard}, R.},
        title = "{Interdecadal oscillations and the warming trend in global temperature time series}",
      journal = {\nat},
         year = 1991,
        month = mar,
       volume = {350},
       number = {6316},
        pages = {324-327},
          doi = {10.1038/350324a0},
       adsurl = {https://ui.adsabs.harvard.edu/abs/1991Natur.350..324G}
}

@book{Golyandina2013,
    author = {{Golyandina}, N. and {Zhigljavsky}, A.},
    title = "{Singular Spectrum Analysis for Time Series}",
    publisher = {Springer Berlin, Heidelberg},
    year = 2013,
    doi = {https://doi.org/10.1007/978-3-642-34913-3}
}

@ARTICLE{Weinberg2021,
       author = {{Weinberg}, Martin D. and {Petersen}, Michael S.},
        title = "{Using multichannel singular spectrum analysis to study galaxy dynamics}",
      journal = {\mnras},
         year = 2021,
        month = mar,
       volume = {501},
       number = {4},
        pages = {5408-5423},
          doi = {10.1093/mnras/staa3997},
archivePrefix = {arXiv},
       eprint = {2009.07870},
 primaryClass = {astro-ph.GA},
       adsurl = {https://ui.adsabs.harvard.edu/abs/2021MNRAS.501.5408W}
}

@ARTICLE{Hunt2026,
       author = {{Hunt}, Jason A.~S. and {Petersen}, Michael S. and {Weinberg}, Martin D. and {Johnston}, Kathryn V. and {Bernet}, Marcel and {Daniel}, Kathryne J. and {Hyman}, S{\'o}ley {\'O}. and {Price-Whelan}, Adrian M. and {Arora}, Arpit},
        title = "{The dark matter wake of a galactic bar revealed by multichannel singular spectral analysis}",
      journal = {\mnras},
         year = 2026,
        month = jan,
       volume = {545},
       number = {3},
          eid = {staf2118},
        pages = {staf2118},
          doi = {10.1093/mnras/staf2118},
archivePrefix = {arXiv},
       eprint = {2510.09751},
 primaryClass = {astro-ph.GA},
       adsurl = {https://ui.adsabs.harvard.edu/abs/2026MNRAS.545f2118H}
}

@ARTICLE{Johnson2023,
       author = {{Johnson}, Alexander C. and {Petersen}, Michael S. and {Johnston}, Kathryn V. and {Weinberg}, Martin D.},
        title = "{Dynamical data mining captures disc-halo couplings that structure galaxies}",
      journal = {\mnras},
         year = 2023,
        month = may,
       volume = {521},
       number = {2},
        pages = {1757-1774},
          doi = {10.1093/mnras/stad485},
archivePrefix = {arXiv},
       eprint = {2301.02256},
 primaryClass = {astro-ph.GA},
       adsurl = {https://ui.adsabs.harvard.edu/abs/2023MNRAS.521.1757J}
}

@ARTICLE{Arora2025,
       author = {{Arora}, Arpit and {Garavito-Camargo}, Nicol{\'a}s and {Sanderson}, Robyn E. and {Weinberg}, Martin D. and {Petersen}, Michael S. and {Varela-Lavin}, Silvio and {G{\'o}mez}, Facundo A. and {Johnston}, Kathryn V. and {Laporte}, Chervin F.~P. and {Shipp}, Nora and {Hunt}, Jason A.~S. and {Besla}, Gurtina and {Darragh-Ford}, Elise and {Panithanpaisal}, Nondh and {Daniel}, Kathryne J. and {EXP Collaboration}},
        title = "{Shaping the Milky Way: The Interplay of Mergers and Cosmic Filaments}",
      journal = {\apj},
         year = 2025,
        month = aug,
       volume = {988},
       number = {2},
          eid = {190},
        pages = {190},
          doi = {10.3847/1538-4357/ade30d},
archivePrefix = {arXiv},
       eprint = {2504.20133},
 primaryClass = {astro-ph.GA},
       adsurl = {https://ui.adsabs.harvard.edu/abs/2025ApJ...988..190A}
}

@ARTICLE{Besla2018,
       author = {{Besla}, Gurtina and {Patton}, David R. and {Stierwalt}, Sabrina and {Rodriguez-Gomez}, Vicente and {Patel}, Ekta and {Kallivayalil}, Nitya J. and {Johnson}, Kelsey E. and {Pearson}, Sarah and {Privon}, George C. and {Putman}, Mary E.},
        title = "{The frequency of dwarf galaxy multiples at low redshift in SDSS versus cosmological expectations}",
      journal = {\mnras},
         year = 2018,
        month = nov,
       volume = {480},
       number = {3},
        pages = {3376-3396},
          doi = {10.1093/mnras/sty2041},
archivePrefix = {arXiv},
       eprint = {1807.06673},
 primaryClass = {astro-ph.GA},
       adsurl = {https://ui.adsabs.harvard.edu/abs/2018MNRAS.480.3376B}
}

@ARTICLE{Chamberlain2024,
       author = {{Chamberlain}, Katie and {Besla}, Gurtina and {Patel}, Ekta and {Rodriguez-Gomez}, Vicente and {Torrey}, Paul and {Martin}, Garreth and {Johnson}, Kelsey and {Kallivayalil}, Nitya and {Patton}, David and {Pearson}, Sarah and {Privon}, George and {Stierwalt}, Sabrina},
        title = "{A Physically Motivated Framework to Compare Pair Fractions of Isolated Low- and High-mass Galaxies across Cosmic Time}",
      journal = {\apj},
         year = 2024,
        month = feb,
       volume = {962},
       number = {2},
          eid = {162},
        pages = {162},
          doi = {10.3847/1538-4357/ad19d0},
archivePrefix = {arXiv},
       eprint = {2309.13228},
 primaryClass = {astro-ph.GA},
       adsurl = {https://ui.adsabs.harvard.edu/abs/2024ApJ...962..162C}
}

@ARTICLE{Zaritsky1997,
       author = {{Zaritsky}, Dennis and {Smith}, Rodney and {Frenk}, Carlos and {White}, Simon D.~M.},
        title = "{More Satellites of Spiral Galaxies}",
      journal = {\apj},
         year = 1997,
        month = mar,
       volume = {478},
       number = {1},
        pages = {39-48},
          doi = {10.1086/303784},
archivePrefix = {arXiv},
       eprint = {astro-ph/9611199},
 primaryClass = {astro-ph},
       adsurl = {https://ui.adsabs.harvard.edu/abs/1997ApJ...478...39Z}
}

@ARTICLE{Zaritsky1993,
       author = {{Zaritsky}, Dennis and {Smith}, Rodney and {Frenk}, Carlos and {White}, Simon D.~M.},
        title = "{Satellites of Spiral Galaxies}",
      journal = {\apj},
         year = 1993,
        month = mar,
       volume = {405},
        pages = {464},
          doi = {10.1086/172379},
       adsurl = {https://ui.adsabs.harvard.edu/abs/1993ApJ...405..464Z}
}

@ARTICLE{Hernquist1990,
       author = {{Hernquist}, Lars},
        title = "{An Analytical Model for Spherical Galaxies and Bulges}",
      journal = {\apj},
         year = 1990,
        month = jun,
       volume = {356},
        pages = {359},
          doi = {10.1086/168845},
       adsurl = {https://ui.adsabs.harvard.edu/abs/1990ApJ...356..359H}
}

@ARTICLE{Erkal2019,
       author = {{Erkal}, D. and {Belokurov}, V. and {Laporte}, C.~F.~P. and {Koposov}, S.~E. and {Li}, T.~S. and {Grillmair}, C.~J. and {Kallivayalil}, N. and {Price-Whelan}, A.~M. and {Evans}, N.~W. and {Hawkins}, K. and {Hendel}, D. and {Mateu}, C. and {Navarro}, J.~F. and {del Pino}, A. and {Slater}, C.~T. and {Sohn}, S.~T. and {Orphan Aspen Treasury Collaboration}},
        title = "{The total mass of the Large Magellanic Cloud from its perturbation on the Orphan stream}",
      journal = {\mnras},
         year = 2019,
        month = aug,
       volume = {487},
       number = {2},
        pages = {2685-2700},
          doi = {10.1093/mnras/stz1371},
archivePrefix = {arXiv},
       eprint = {1812.08192},
 primaryClass = {astro-ph.GA},
       adsurl = {https://ui.adsabs.harvard.edu/abs/2019MNRAS.487.2685E}
}

@ARTICLE{Watkins2024,
       author = {{Watkins}, Laura L. and {van der Marel}, Roeland P. and {Bennet}, Paul},
        title = "{The Mass of the Large Magellanic Cloud from the Three-dimensional Kinematics of Its Globular Clusters}",
      journal = {\apj},
         year = 2024,
        month = mar,
       volume = {963},
       number = {2},
          eid = {84},
        pages = {84},
          doi = {10.3847/1538-4357/ad1f58},
archivePrefix = {arXiv},
       eprint = {2401.14458},
 primaryClass = {astro-ph.GA},
       adsurl = {https://ui.adsabs.harvard.edu/abs/2024ApJ...963...84W}
}

@ARTICLE{Penarrubia2016,
       author = {{Pe{\~n}arrubia}, Jorge and {G{\'o}mez}, Facundo A. and {Besla}, Gurtina and {Erkal}, Denis and {Ma}, Yin-Zhe},
        title = "{A timing constraint on the (total) mass of the Large Magellanic Cloud}",
      journal = {\mnras},
         year = 2016,
        month = feb,
       volume = {456},
       number = {1},
        pages = {L54-L58},
          doi = {10.1093/mnrasl/slv160},
archivePrefix = {arXiv},
       eprint = {1507.03594},
 primaryClass = {astro-ph.GA},
       adsurl = {https://ui.adsabs.harvard.edu/abs/2016MNRAS.456L..54P}
}

@ARTICLE{DeLeo2024,
       author = {{De Leo}, Michele and {Read}, Justin I. and {No{\"e}l}, Noelia E.~D. and {Erkal}, Denis and {Massana}, Pol and {Carrera}, Ricardo},
        title = "{Surviving the waves: evidence for a dark matter cusp in the tidally disrupting Small Magellanic Cloud}",
      journal = {\mnras},
         year = 2024,
        month = nov,
       volume = {535},
       number = {1},
        pages = {1015-1034},
          doi = {10.1093/mnras/stae2428},
archivePrefix = {arXiv},
       eprint = {2303.08838},
 primaryClass = {astro-ph.GA},
       adsurl = {https://ui.adsabs.harvard.edu/abs/2024MNRAS.535.1015D}
}

@ARTICLE{Marel2002,
       author = {{van der Marel}, Roeland P. and {Alves}, David R. and {Hardy}, Eduardo and {Suntzeff}, Nicholas B.},
        title = "{New Understanding of Large Magellanic Cloud Structure, Dynamics, and Orbit from Carbon Star Kinematics}",
      journal = {\aj},
         year = 2002,
        month = nov,
       volume = {124},
       number = {5},
        pages = {2639-2663},
          doi = {10.1086/343775},
archivePrefix = {arXiv},
       eprint = {astro-ph/0205161},
 primaryClass = {astro-ph},
       adsurl = {https://ui.adsabs.harvard.edu/abs/2002AJ....124.2639V}
}

@ARTICLE{Pardy2018,
       author = {{Pardy}, Stephen A. and {D'Onghia}, Elena and {Fox}, Andrew J.},
        title = "{Models of Tidally Induced Gas Filaments in the Magellanic Stream}",
      journal = {\apj},
         year = 2018,
        month = apr,
       volume = {857},
       number = {2},
          eid = {101},
        pages = {101},
          doi = {10.3847/1538-4357/aab95b},
archivePrefix = {arXiv},
       eprint = {1802.01600},
 primaryClass = {astro-ph.GA},
       adsurl = {https://ui.adsabs.harvard.edu/abs/2018ApJ...857..101P}
}

@ARTICLE{Lucchini2024,
       author = {{Lucchini}, Scott and {D'Onghia}, Elena and {Fox}, Andrew J.},
        title = "{Properties of the Magellanic Corona}",
      journal = {\apj},
         year = 2024,
        month = may,
       volume = {967},
       number = {1},
          eid = {16},
        pages = {16},
          doi = {10.3847/1538-4357/ad3c3b},
archivePrefix = {arXiv},
       eprint = {2311.16221},
 primaryClass = {astro-ph.GA},
       adsurl = {https://ui.adsabs.harvard.edu/abs/2024ApJ...967...16L}
}

@ARTICLE{Rathore2026,
       author = {{Rathore}, Himansh and {Besla}, Gurtina and {van der Marel}, Roeland P. and {Kallivayalil}, Nitya},
        title = "{A Galactic Transformation{\textemdash}Understanding the SMC's Structural and Kinematic Disequilibrium}",
      journal = {\apj},
         year = 2026,
        month = mar,
       volume = {1000},
       number = {1},
          eid = {50},
        pages = {50},
          doi = {10.3847/1538-4357/ae4507},
archivePrefix = {arXiv},
       eprint = {2512.06075},
 primaryClass = {astro-ph.GA},
       adsurl = {https://ui.adsabs.harvard.edu/abs/2026ApJ..1000...50R}
}

@ARTICLE{Kreckel2011,
       author = {{Kreckel}, K. and {Platen}, E. and {Arag{\'o}n-Calvo}, M.~A. and {van Gorkom}, J.~H. and {van de Weygaert}, R. and {van der Hulst}, J.~M. and {Kova{\v{c}}}, K. and {Yip}, C.-W. and {Peebles}, P.~J.~E.},
        title = "{Only the Lonely: H I Imaging of Void Galaxies}",
      journal = {\aj},
         year = 2011,
        month = jan,
       volume = {141},
       number = {1},
          eid = {4},
        pages = {4},
          doi = {10.1088/0004-6256/141/1/4},
archivePrefix = {arXiv},
       eprint = {1008.4616},
 primaryClass = {astro-ph.CO},
       adsurl = {https://ui.adsabs.harvard.edu/abs/2011AJ....141....4K}
}

@ARTICLE{Swaters2002,
       author = {{Swaters}, R.~A. and {van Albada}, T.~S. and {van der Hulst}, J.~M. and {Sancisi}, R.},
        title = "{The Westerbork HI survey of spiral and irregular galaxies. I. HI imaging of late-type dwarf galaxies}",
      journal = {\aap},
         year = 2002,
        month = aug,
       volume = {390},
        pages = {829-861},
          doi = {10.1051/0004-6361:20011755},
archivePrefix = {arXiv},
       eprint = {astro-ph/0204525},
 primaryClass = {astro-ph},
       adsurl = {https://ui.adsabs.harvard.edu/abs/2002A&A...390..829S}
}

@ARTICLE{Hunter2012,
       author = {{Hunter}, Deidre A. and {Ficut-Vicas}, Dana and {Ashley}, Trisha and {Brinks}, Elias and {Cigan}, Phil and {Elmegreen}, Bruce G. and {Heesen}, Volker and {Herrmann}, Kimberly A. and {Johnson}, Megan and {Oh}, Se-Heon and {Rupen}, Michael P. and {Schruba}, Andreas and {Simpson}, Caroline E. and {Walter}, Fabian and {Westpfahl}, David J. and {Young}, Lisa M. and {Zhang}, Hong-Xin},
        title = "{Little Things}",
      journal = {\aj},
         year = 2012,
        month = nov,
       volume = {144},
       number = {5},
          eid = {134},
        pages = {134},
          doi = {10.1088/0004-6256/144/5/134},
archivePrefix = {arXiv},
       eprint = {1208.5834},
 primaryClass = {astro-ph.GA},
       adsurl = {https://ui.adsabs.harvard.edu/abs/2012AJ....144..134H}
}

@ARTICLE{Springel2021,
       author = {{Springel}, Volker and {Pakmor}, R{\"u}diger and {Zier}, Oliver and {Reinecke}, Martin},
        title = "{Simulating cosmic structure formation with the GADGET-4 code}",
      journal = {\mnras},
         year = 2021,
        month = sep,
       volume = {506},
       number = {2},
        pages = {2871-2949},
          doi = {10.1093/mnras/stab1855},
archivePrefix = {arXiv},
       eprint = {2010.03567},
 primaryClass = {astro-ph.IM},
       adsurl = {https://ui.adsabs.harvard.edu/abs/2021MNRAS.506.2871S}
}

@ARTICLE{Noel2007,
       author = {{No{\"e}l}, Noelia E.~D. and {Gallart}, Carme and {Costa}, Edgardo and {M{\'e}ndez}, Ren{\'e} A.},
        title = "{Old Main-Sequence Turnoff Photometry in the Small Magellanic Cloud. I. Constraints on the Star Formation History in Different Fields}",
      journal = {\aj},
         year = 2007,
        month = may,
       volume = {133},
       number = {5},
        pages = {2037-2052},
          doi = {10.1086/512668},
archivePrefix = {arXiv},
       eprint = {0704.1151},
 primaryClass = {astro-ph},
       adsurl = {https://ui.adsabs.harvard.edu/abs/2007AJ....133.2037N}
}

@ARTICLE{Noel2009,
       author = {{No{\"e}l}, Noelia E.~D. and {Aparicio}, Antonio and {Gallart}, Carme and {Hidalgo}, Sebasti{\'a}n L. and {Costa}, Edgardo and {M{\'e}ndez}, Ren{\'e} A.},
        title = "{Old Main-Sequence Turnoff Photometry in the Small Magellanic Cloud. II. Star Formation History and its Spatial Gradients}",
      journal = {\apj},
         year = 2009,
        month = nov,
       volume = {705},
       number = {2},
        pages = {1260-1274},
          doi = {10.1088/0004-637X/705/2/1260},
archivePrefix = {arXiv},
       eprint = {0909.2873},
 primaryClass = {astro-ph.CO},
       adsurl = {https://ui.adsabs.harvard.edu/abs/2009ApJ...705.1260N}
}

@ARTICLE{Weisz2013,
       author = {{Weisz}, Daniel R. and {Dolphin}, Andrew E. and {Skillman}, Evan D. and {Holtzman}, Jon and {Dalcanton}, Julianne J. and {Cole}, Andrew A. and {Neary}, Kyle},
        title = "{Comparing the ancient star formation histories of the Magellanic Clouds}",
      journal = {\mnras},
         year = 2013,
        month = may,
       volume = {431},
       number = {1},
        pages = {364-371},
          doi = {10.1093/mnras/stt165},
archivePrefix = {arXiv},
       eprint = {1301.7422},
 primaryClass = {astro-ph.CO},
       adsurl = {https://ui.adsabs.harvard.edu/abs/2013MNRAS.431..364W}
}

@ARTICLE{Massana2022,
       author = {{Massana}, P. and {Ruiz-Lara}, T. and {No{\"e}l}, N.~E.~D. and {Gallart}, C. and {Nidever}, D.~L. and {Choi}, Y. and {Sakowska}, J.~D. and {Besla}, G. and {Olsen}, K.~A.~G. and {Monelli}, M. and {Dorta}, A. and {Stringfellow}, G.~S. and {Cassisi}, S. and {Bernard}, E.~J. and {Zaritsky}, D. and {Cioni}, M.-R.~L. and {Monachesi}, A. and {van der Marel}, R.~P. and {de Boer}, T.~J.~L. and {Walker}, A.~R.},
        title = "{The synchronized dance of the magellanic clouds' star formation history}",
      journal = {\mnras},
         year = 2022,
        month = jun,
       volume = {513},
       number = {1},
        pages = {L40-L45},
          doi = {10.1093/mnrasl/slac030},
archivePrefix = {arXiv},
       eprint = {2203.09523},
 primaryClass = {astro-ph.GA},
       adsurl = {https://ui.adsabs.harvard.edu/abs/2022MNRAS.513L..40M}
}

@ARTICLE{Burhene2026,
       author = {{Burhenne}, Clare and {McQuinn}, Kristen B.~W. and {Cohen}, Roger E. and {Murray}, Claire E. and {Patel}, Ekta and {Williams}, Benjamin F. and {Lindberg}, Christina W. and {Merica-Jones}, Petia Yanchulova and {Gordon}, Karl D. and {Choi}, Yumi and {Dolphin}, Andrew E. and {Roman-Duval}, Julia C.},
        title = "{Scylla. V. Constraints on the Spatial and Temporal Distribution of Bursts and the Interaction History of the Magellanic Clouds from Their Resolved Stellar Populations}",
      journal = {\apj},
         year = 2026,
        month = jan,
       volume = {997},
       number = {1},
          eid = {23},
        pages = {23},
          doi = {10.3847/1538-4357/ae13af},
archivePrefix = {arXiv},
       eprint = {2511.02947},
 primaryClass = {astro-ph.GA},
       adsurl = {https://ui.adsabs.harvard.edu/abs/2026ApJ...997...23B}
}

@ARTICLE{Kallivayalil2006,
       author = {{Kallivayalil}, Nitya and {van der Marel}, Roeland P. and {Alcock}, Charles},
        title = "{Is the SMC Bound to the LMC? The Hubble Space Telescope Proper Motion of the SMC}",
      journal = {\apj},
         year = 2006,
        month = dec,
       volume = {652},
       number = {2},
        pages = {1213-1229},
          doi = {10.1086/508014},
archivePrefix = {arXiv},
       eprint = {astro-ph/0606240},
 primaryClass = {astro-ph},
       adsurl = {https://ui.adsabs.harvard.edu/abs/2006ApJ...652.1213K}
}

@ARTICLE{Kallivayalil2006b,
       author = {{Kallivayalil}, Nitya and {van der Marel}, Roeland P. and {Alcock}, Charles and {Axelrod}, Tim and {Cook}, Kem H. and {Drake}, A.~J. and {Geha}, M.},
        title = "{The Proper Motion of the Large Magellanic Cloud Using HST}",
      journal = {\apj},
         year = 2006,
        month = feb,
       volume = {638},
       number = {2},
        pages = {772-785},
          doi = {10.1086/498972},
archivePrefix = {arXiv},
       eprint = {astro-ph/0508457},
 primaryClass = {astro-ph},
       adsurl = {https://ui.adsabs.harvard.edu/abs/2006ApJ...638..772K}
}

@ARTICLE{Mathewson1974,
       author = {{Mathewson}, D.~S. and {Cleary}, M.~N. and {Murray}, J.~D.},
        title = "{The Magellanic Stream.}",
      journal = {\apj},
         year = 1974,
        month = jun,
       volume = {190},
        pages = {291-296},
          doi = {10.1086/152875},
       adsurl = {https://ui.adsabs.harvard.edu/abs/1974ApJ...190..291M}
}

@ARTICLE{Braun2004,
       author = {{Braun}, R. and {Thilker}, D.~A.},
        title = "{The WSRT wide-field H I survey. II. Local Group features}",
      journal = {\aap},
         year = 2004,
        month = apr,
       volume = {417},
        pages = {421-435},
          doi = {10.1051/0004-6361:20034423},
archivePrefix = {arXiv},
       eprint = {astro-ph/0312323},
 primaryClass = {astro-ph},
       adsurl = {https://ui.adsabs.harvard.edu/abs/2004A&A...417..421B}
}

@ARTICLE{Nidever2010,
       author = {{Nidever}, David L. and {Majewski}, Steven R. and {Butler Burton}, W. and {Nigra}, Lou},
        title = "{The 200{\textdegree} Long Magellanic Stream System}",
      journal = {\apj},
         year = 2010,
        month = nov,
       volume = {723},
       number = {2},
        pages = {1618-1631},
          doi = {10.1088/0004-637X/723/2/1618},
archivePrefix = {arXiv},
       eprint = {1009.0001},
 primaryClass = {astro-ph.GA},
       adsurl = {https://ui.adsabs.harvard.edu/abs/2010ApJ...723.1618N}
}

@ARTICLE{Chandra2023,
       author = {{Chandra}, Vedant and {Naidu}, Rohan P. and {Conroy}, Charlie and {Bonaca}, Ana and {Zaritsky}, Dennis and {Cargile}, Phillip A. and {Caldwell}, Nelson and {Johnson}, Benjamin D. and {Han}, Jiwon Jesse and {Ting}, Yuan-Sen},
        title = "{Discovery of the Magellanic Stellar Stream Out to 100 kpc}",
      journal = {\apj},
         year = 2023,
        month = oct,
       volume = {956},
       number = {2},
          eid = {110},
        pages = {110},
          doi = {10.3847/1538-4357/acf7bf},
archivePrefix = {arXiv},
       eprint = {2306.15719},
 primaryClass = {astro-ph.GA},
       adsurl = {https://ui.adsabs.harvard.edu/abs/2023ApJ...956..110C}
}

@ARTICLE{Zaritsky2025,
       author = {{Zaritsky}, Dennis and {Chandra}, Vedant and {Conroy}, Charlie and {Bonaca}, Ana and {Cargile}, Phillip A. and {Naidu}, Rohan P.},
        title = "{Untangling Magellanic Streams}",
      journal = {The Open Journal of Astrophysics},
         year = 2025,
        month = feb,
       volume = {8},
          eid = {16},
        pages = {16},
          doi = {10.33232/001c.129885},
archivePrefix = {arXiv},
       eprint = {2411.15044},
 primaryClass = {astro-ph.GA},
       adsurl = {https://ui.adsabs.harvard.edu/abs/2025OJAp....8E..16Z}
}

@ARTICLE{Zaritsky2020,
       author = {{Zaritsky}, Dennis and {Conroy}, Charlie and {Naidu}, Rohan P. and {Cargile}, Phillip A. and {Putman}, Mary and {Besla}, Gurtina and {Bonaca}, Ana and {Caldwell}, Nelson and {Han}, Jiwon Jesse and {Johnson}, Benjamin D. and {Speagle}, Joshua S. and {Ting}, Yuan-Sen},
        title = "{Discovery of Magellanic Stellar Debris in the H3 Survey}",
      journal = {\apjl},
         year = 2020,
        month = dec,
       volume = {905},
       number = {1},
          eid = {L3},
        pages = {L3},
          doi = {10.3847/2041-8213/abcb83},
archivePrefix = {arXiv},
       eprint = {2011.09395},
 primaryClass = {astro-ph.GA},
       adsurl = {https://ui.adsabs.harvard.edu/abs/2020ApJ...905L...3Z}
}

@ARTICLE{Kerr1957,
       author = {{Kerr}, F.~J. and {Hindman}, J.~V.},
        title = "{Mass Distribution of Galactic Neutral Hydrogen}",
      journal = {\pasp},
         year = 1957,
        month = dec,
       volume = {69},
       number = {411},
        pages = {558},
          doi = {10.1086/127147},
       adsurl = {https://ui.adsabs.harvard.edu/abs/1957PASP...69..558K}
}

@ARTICLE{Putman2003,
       author = {{Putman}, Mary E. and {Staveley-Smith}, Lister and {Freeman}, Kenneth C. and {Gibson}, Brad K. and {Barnes}, David G.},
        title = "{The Magellanic Stream, High-Velocity Clouds, and the Sculptor Group}",
      journal = {\apj},
         year = 2003,
        month = mar,
       volume = {586},
       number = {1},
        pages = {170-194},
          doi = {10.1086/344477},
archivePrefix = {arXiv},
       eprint = {astro-ph/0209127},
 primaryClass = {astro-ph},
       adsurl = {https://ui.adsabs.harvard.edu/abs/2003ApJ...586..170P}
}

@ARTICLE{Bruns2005,
       author = {{Br{\"u}ns}, C. and {Kerp}, J. and {Staveley-Smith}, L. and {Mebold}, U. and {Putman}, M.~E. and {Haynes}, R.~F. and {Kalberla}, P.~M.~W. and {Muller}, E. and {Filipovic}, M.~D.},
        title = "{The Parkes H I Survey of the Magellanic System}",
      journal = {\aap},
         year = 2005,
        month = mar,
       volume = {432},
       number = {1},
        pages = {45-67},
          doi = {10.1051/0004-6361:20040321},
archivePrefix = {arXiv},
       eprint = {astro-ph/0411453},
 primaryClass = {astro-ph},
       adsurl = {https://ui.adsabs.harvard.edu/abs/2005A&A...432...45B}
}

@ARTICLE{Zivick2019,
       author = {{Zivick}, Paul and {Kallivayalil}, Nitya and {Besla}, Gurtina and {Sohn}, Sangmo Tony and {van der Marel}, Roeland P. and {del Pino}, Andr{\'e}s and {Linden}, Sean T. and {Fritz}, Tobias K. and {Anderson}, J.},
        title = "{The Proper-motion Field along the Magellanic Bridge: A New Probe of the LMC-SMC Interaction}",
      journal = {\apj},
         year = 2019,
        month = mar,
       volume = {874},
       number = {1},
          eid = {78},
        pages = {78},
          doi = {10.3847/1538-4357/ab0554},
archivePrefix = {arXiv},
       eprint = {1811.09318},
 primaryClass = {astro-ph.GA},
       adsurl = {https://ui.adsabs.harvard.edu/abs/2019ApJ...874...78Z}
}

@ARTICLE{Choi2022,
       author = {{Choi}, Yumi and {Olsen}, Knut A.~G. and {Besla}, Gurtina and {van der Marel}, Roeland P. and {Zivick}, Paul and {Kallivayalil}, Nitya and {Nidever}, David L.},
        title = "{The Recent LMC-SMC Collision: Timing and Impact Parameter Constraints from Comparison of Gaia LMC Disk Kinematics and N-body Simulations}",
      journal = {\apj},
         year = 2022,
        month = mar,
       volume = {927},
       number = {2},
          eid = {153},
        pages = {153},
          doi = {10.3847/1538-4357/ac4e90},
archivePrefix = {arXiv},
       eprint = {2201.04648},
 primaryClass = {astro-ph.GA},
       adsurl = {https://ui.adsabs.harvard.edu/abs/2022ApJ...927..153C}
}

@ARTICLE{Besla2010,
       author = {{Besla}, G. and {Kallivayalil}, N. and {Hernquist}, L. and {van der Marel}, R.~P. and {Cox}, T.~J. and {Kere{\v{s}}}, D.},
        title = "{Simulations of the Magellanic Stream in a First Infall Scenario}",
      journal = {\apjl},
         year = 2010,
        month = oct,
       volume = {721},
       number = {2},
        pages = {L97-L101},
          doi = {10.1088/2041-8205/721/2/L97},
archivePrefix = {arXiv},
       eprint = {1008.2210},
 primaryClass = {astro-ph.GA},
       adsurl = {https://ui.adsabs.harvard.edu/abs/2010ApJ...721L..97B}
}

@ARTICLE{Power2003,
       author = {{Power}, C. and {Navarro}, J.~F. and {Jenkins}, A. and {Frenk}, C.~S. and {White}, S.~D.~M. and {Springel}, V. and {Stadel}, J. and {Quinn}, T.},
        title = "{The inner structure of {\ensuremath{\Lambda}}CDM haloes - I. A numerical convergence study}",
      journal = {\mnras},
         year = 2003,
        month = jan,
       volume = {338},
       number = {1},
        pages = {14-34},
          doi = {10.1046/j.1365-8711.2003.05925.x},
archivePrefix = {arXiv},
       eprint = {astro-ph/0201544},
 primaryClass = {astro-ph},
       adsurl = {https://ui.adsabs.harvard.edu/abs/2003MNRAS.338...14P}
}

@ARTICLE{Hernquist1993,
       author = {{Hernquist}, Lars},
        title = "{N-Body Realizations of Compound Galaxies}",
      journal = {\apjs},
         year = 1993,
        month = jun,
       volume = {86},
        pages = {389},
          doi = {10.1086/191784},
       adsurl = {https://ui.adsabs.harvard.edu/abs/1993ApJS...86..389H}
}

@ARTICLE{Lucey2023,
       author = {{Lucey}, Madeline and {Pearson}, Sarah and {Hunt}, Jason A.~S. and {Hawkins}, Keith and {Ness}, Melissa and {Petersen}, Michael S. and {Price-Whelan}, Adrian M. and {Weinberg}, Martin D.},
        title = "{Dynamically constraining the length of the Milky way bar}",
      journal = {\mnras},
         year = 2023,
        month = apr,
       volume = {520},
       number = {3},
        pages = {4779-4792},
          doi = {10.1093/mnras/stad406},
archivePrefix = {arXiv},
       eprint = {2206.01798},
 primaryClass = {astro-ph.GA},
       adsurl = {https://ui.adsabs.harvard.edu/abs/2023MNRAS.520.4779L}
}

@ARTICLE{Varela-Lavin2023,
       author = {{Varela-Lavin}, Silvio and {G{\'o}mez}, Facundo A. and {Tissera}, Patricia B. and {Besla}, Gurtina and {Garavito-Camargo}, Nicol{\'a}s and {Marinacci}, Federico and {Laporte}, Chervin F.~P.},
        title = "{Lopsided galaxies in a cosmological context: a new galaxy-halo connection}",
      journal = {\mnras},
         year = 2023,
        month = aug,
       volume = {523},
       number = {4},
        pages = {5853-5868},
          doi = {10.1093/mnras/stad1724},
archivePrefix = {arXiv},
       eprint = {2211.16577},
 primaryClass = {astro-ph.GA},
       adsurl = {https://ui.adsabs.harvard.edu/abs/2023MNRAS.523.5853V}
}

@ARTICLE{Kalnajs1977,
       author = {{Kalnajs}, A.~J.},
        title = "{Dynamics of flat galaxies. IV. The integral equation for normal modes in matrix form.}",
      journal = {\apj},
         year = 1977,
        month = mar,
       volume = {212},
        pages = {637-644},
          doi = {10.1086/155086},
       adsurl = {https://ui.adsabs.harvard.edu/abs/1977ApJ...212..637K}
}

@ARTICLE{Weinberg1991,
       author = {{Weinberg}, Martin D.},
        title = "{A Search for Instability in Two Families of Spherical Stellar Models}",
      journal = {\apj},
         year = 1991,
        month = feb,
       volume = {368},
        pages = {66},
          doi = {10.1086/169671},
       adsurl = {https://ui.adsabs.harvard.edu/abs/1991ApJ...368...66W}
}

@ARTICLE{Sellwood2024,
       author = {{Sellwood}, J.~A.},
        title = "{Particle selection from an equilibrium DF}",
      journal = {\mnras},
         year = 2024,
        month = apr,
       volume = {529},
       number = {3},
        pages = {3035-3043},
          doi = {10.1093/mnras/stae595},
archivePrefix = {arXiv},
       eprint = {2402.15435},
 primaryClass = {astro-ph.GA},
       adsurl = {https://ui.adsabs.harvard.edu/abs/2024MNRAS.529.3035S}
}

@ARTICLE{Petersen2025,
       author = {{Petersen}, Michael and {Weinberg}, Martin},
        title = "{EXP: a Python/C++ package for basis function expansion methods in galactic dynamics}",
      journal = {The Journal of Open Source Software},
         year = 2025,
        month = may,
       volume = {10},
       number = {109},
          eid = {7302},
        pages = {7302},
          doi = {10.21105/joss.07302},
archivePrefix = {arXiv},
       eprint = {2505.06403},
 primaryClass = {astro-ph.GA},
       adsurl = {https://ui.adsabs.harvard.edu/abs/2025JOSS...10.7302P}
}

@ARTICLE{Earn1996,
       author = {{Earn}, David J.~D.},
        title = "{Potential-Density Basis Sets for Galactic Disks}",
      journal = {\apj},
         year = 1996,
        month = jul,
       volume = {465},
        pages = {91},
          doi = {10.1086/177404},
archivePrefix = {arXiv},
       eprint = {astro-ph/9601101},
 primaryClass = {astro-ph},
       adsurl = {https://ui.adsabs.harvard.edu/abs/1996ApJ...465...91E}
}

@BOOK{Fridman1984,
       author = {{Fridman}, A.~M. and {Polyachenko}, V.~L. and {Aries}, A.~B. and {Poliakoff}, I.~N.},
        title = "{Physics of gravitating systems. I. Equilibrium and stability.}",
         year = 1984,
       adsurl = {https://ui.adsabs.harvard.edu/abs/1984pgs1.book.....F}
}

@ARTICLE{Valluri2016,
       author = {{Valluri}, Monica and {Shen}, Juntai and {Abbott}, Caleb and {Debattista}, Victor P.},
        title = "{A Unified Framework for the Orbital Structure of Bars and Triaxial Ellipsoids}",
      journal = {\apj},
         year = 2016,
        month = feb,
       volume = {818},
       number = {2},
          eid = {141},
        pages = {141},
          doi = {10.3847/0004-637X/818/2/141},
archivePrefix = {arXiv},
       eprint = {1512.03467},
 primaryClass = {astro-ph.GA},
       adsurl = {https://ui.adsabs.harvard.edu/abs/2016ApJ...818..141V}
}

@ARTICLE{Dattathri2024,
       author = {{Dattathri}, Shashank and {Valluri}, Monica and {Vasiliev}, Eugene and {Wheeler}, Vance and {Erwin}, Peter},
        title = "{Deprojection and stellar dynamical modelling of boxy/peanut bars in edge-on discs}",
      journal = {\mnras},
         year = 2024,
        month = may,
       volume = {530},
       number = {1},
        pages = {1195-1217},
          doi = {10.1093/mnras/stae802},
archivePrefix = {arXiv},
       eprint = {2309.11557},
 primaryClass = {astro-ph.GA},
       adsurl = {https://ui.adsabs.harvard.edu/abs/2024MNRAS.530.1195D}
}

@ARTICLE{Tahmasebzadeh2024,
       author = {{Tahmasebzadeh}, Behzad and {Dattathri}, Shashank and {Valluri}, Monica and {Shen}, Juntai and {Zhu}, Ling and {Wheeler}, Vance and {Gerhard}, Ortwin and {Kataria}, Sandeep Kumar and {Beraldo e Silva}, Leandro and {Daniel}, Kathryne J.},
        title = "{Orbital Support and Evolution of CX/OX Structures in Boxy/Peanut Bars}",
      journal = {\apj},
         year = 2024,
        month = nov,
       volume = {975},
       number = {1},
          eid = {120},
        pages = {120},
          doi = {10.3847/1538-4357/ad77c8},
archivePrefix = {arXiv},
       eprint = {2409.03746},
 primaryClass = {astro-ph.GA},
       adsurl = {https://ui.adsabs.harvard.edu/abs/2024ApJ...975..120T}
}

@article{Pryce1991,
title = {A new multi-purpose software package for Schrödinger and Sturm-Liouville computations},
journal = {Computer Physics Communications},
volume = {62},
number = {1},
pages = {42-52},
year = {1991},
issn = {0010-4655},
doi = {https://doi.org/10.1016/0010-4655(91)90119-6},
url = {https://www.sciencedirect.com/science/article/pii/0010465591901196},
author = {John D. Pryce and Marco Marletta},
}

@article{Press1993,
author = {Pruess, Steven and Fulton, Charles T.},
title = {Mathematical software for Sturm-Liouville problems},
year = {1993},
issue_date = {Sept. 1993},
publisher = {Association for Computing Machinery},
address = {New York, NY, USA},
volume = {19},
number = {3},
issn = {0098-3500},
url = {https://doi.org/10.1145/155743.155791},
doi = {10.1145/155743.155791},
journal = {ACM Trans. Math. Softw.},
month = sep,
pages = {360–376},
numpages = {17}
}

@ARTICLE{Banik2021,
       author = {{Banik}, Uddipan and {van den Bosch}, Frank C.},
        title = "{A Self-consistent, Time-dependent Treatment of Dynamical Friction: New Insights Regarding Core Stalling and Dynamical Buoyancy}",
      journal = {\apj},
         year = 2021,
        month = may,
       volume = {912},
       number = {1},
          eid = {43},
        pages = {43},
          doi = {10.3847/1538-4357/abeb6d},
archivePrefix = {arXiv},
       eprint = {2103.05004},
 primaryClass = {astro-ph.GA},
       adsurl = {https://ui.adsabs.harvard.edu/abs/2021ApJ...912...43B}
}

@ARTICLE{Sridhar2018,
       author = {{Kaur}, Karamveer and {Sridhar}, S.},
        title = "{Stalling of Globular Cluster Orbits in Dwarf Galaxies}",
      journal = {\apj},
         year = 2018,
        month = dec,
       volume = {868},
       number = {2},
          eid = {134},
        pages = {134},
          doi = {10.3847/1538-4357/aaeacf},
archivePrefix = {arXiv},
       eprint = {1810.00369},
 primaryClass = {astro-ph.GA},
       adsurl = {https://ui.adsabs.harvard.edu/abs/2018ApJ...868..134K}
}

@ARTICLE{Banik2022,
       author = {{Banik}, Uddipan and {van den Bosch}, Frank C.},
        title = "{Dynamical Friction, Buoyancy, and Core-stalling. I. A Nonperturbative Orbit-based Analysis}",
      journal = {\apj},
         year = 2022,
        month = feb,
       volume = {926},
       number = {2},
          eid = {215},
        pages = {215},
          doi = {10.3847/1538-4357/ac4242},
archivePrefix = {arXiv},
       eprint = {2112.06944},
 primaryClass = {astro-ph.GA},
       adsurl = {https://ui.adsabs.harvard.edu/abs/2022ApJ...926..215B}
}

@ARTICLE{Tavangar2026,
       author = {{Tavangar}, Kiyan and {Johnston}, Kathryn V. and {Hunt}, Jason A.~S. and {Widmark}, Axel and {Hamilton}, Chris and {Petersen}, Michael S. and {Weinberg}, Martin D.},
        title = "{Phase spirals across galactic disks I: Exploring dynamical influences on winding}",
      journal = {arXiv e-prints},
         year = 2026,
        month = mar,
          eid = {arXiv:2603.23588},
        pages = {arXiv:2603.23588},
          doi = {10.48550/arXiv.2603.23588},
archivePrefix = {arXiv},
       eprint = {2603.23588},
 primaryClass = {astro-ph.GA},
       adsurl = {https://ui.adsabs.harvard.edu/abs/2026arXiv260323588T}
}

@ARTICLE{Tremaine1999,
       author = {{Tremaine}, Scott},
        title = "{The geometry of phase mixing}",
      journal = {\mnras},
         year = 1999,
        month = aug,
       volume = {307},
       number = {4},
        pages = {877-883},
          doi = {10.1046/j.1365-8711.1999.02690.x},
archivePrefix = {arXiv},
       eprint = {astro-ph/9812146},
 primaryClass = {astro-ph},
       adsurl = {https://ui.adsabs.harvard.edu/abs/1999MNRAS.307..877T}
}

@ARTICLE{Johnston1996,
       author = {{Johnston}, Kathryn V. and {Hernquist}, Lars and {Bolte}, Michael},
        title = "{Fossil Signatures of Ancient Accretion Events in the Halo}",
      journal = {\apj},
         year = 1996,
        month = jul,
       volume = {465},
        pages = {278},
          doi = {10.1086/177418},
archivePrefix = {arXiv},
       eprint = {astro-ph/9602060},
 primaryClass = {astro-ph},
       adsurl = {https://ui.adsabs.harvard.edu/abs/1996ApJ...465..278J}
}

@ARTICLE{Necib2026,
       author = {{Necib}, Lina and {Folsom}, Dylan and {Davies}, Elliot Y. and {Starkman}, Nathaniel and {Thoyas}, Andreas},
        title = "{Galactic Amnesia: The Information Washout of the Milky Way Merger History}",
      journal = {arXiv e-prints},
         year = 2026,
        month = may,
          eid = {arXiv:2605.04138},
        pages = {arXiv:2605.04138},
archivePrefix = {arXiv},
       eprint = {2605.04138},
 primaryClass = {astro-ph.GA},
       adsurl = {https://ui.adsabs.harvard.edu/abs/2026arXiv260504138N}
}

@ARTICLE{Lokas2025,
       author = {{{\L}okas}, Ewa L.},
        title = "{On the nature of buckling instability in galactic bars}",
      journal = {\aap},
         year = 2025,
        month = jul,
       volume = {699},
          eid = {A234},
        pages = {A234},
          doi = {10.1051/0004-6361/202553817},
archivePrefix = {arXiv},
       eprint = {2501.12308},
 primaryClass = {astro-ph.GA},
       adsurl = {https://ui.adsabs.harvard.edu/abs/2025A&A...699A.234L}
}

@ARTICLE{Athanassoula2005,
       author = {{Athanassoula}, E.},
        title = "{On the nature of bulges in general and of box/peanut bulges in particular: input from N-body simulations}",
      journal = {\mnras},
         year = 2005,
        month = apr,
       volume = {358},
       number = {4},
        pages = {1477-1488},
          doi = {10.1111/j.1365-2966.2005.08872.x},
archivePrefix = {arXiv},
       eprint = {astro-ph/0502316},
 primaryClass = {astro-ph},
       adsurl = {https://ui.adsabs.harvard.edu/abs/2005MNRAS.358.1477A}
}

@ARTICLE{McClure2025,
       author = {{McClure}, Rachel Lee and {Beane}, Angus and {D'Onghia}, Elena and {Filion}, Carrie and {Daniel}, Kathryne J.},
        title = "{The impact of classical bulges on stellar bars and boxy-peanut-X features in disc galaxies}",
      journal = {\mnras},
         year = 2025,
        month = feb,
       volume = {537},
       number = {2},
        pages = {1475-1488},
          doi = {10.1093/mnras/staf107},
archivePrefix = {arXiv},
       eprint = {2410.08277},
 primaryClass = {astro-ph.GA},
       adsurl = {https://ui.adsabs.harvard.edu/abs/2025MNRAS.537.1475M}
}

@ARTICLE{Raha1991,
       author = {{Raha}, N. and {Sellwood}, J.~A. and {James}, R.~A. and {Kahn}, F.~D.},
        title = "{A dynamical instability of bars in disk galaxies}",
      journal = {\nat},
         year = 1991,
        month = aug,
       volume = {352},
       number = {6334},
        pages = {411-412},
          doi = {10.1038/352411a0},
       adsurl = {https://ui.adsabs.harvard.edu/abs/1991Natur.352..411R}
}

@ARTICLE{Sellwood1994,
       author = {{Merritt}, David and {Sellwood}, J.~A.},
        title = "{Bending Instabilities in Stellar Systems}",
      journal = {\apj},
         year = 1994,
        month = apr,
       volume = {425},
        pages = {551},
          doi = {10.1086/174005},
       adsurl = {https://ui.adsabs.harvard.edu/abs/1994ApJ...425..551M}
}

@ARTICLE{Petersen2019,
       author = {{Petersen}, Michael S. and {Weinberg}, Martin D. and {Katz}, Neal},
        title = "{Using Harmonic Decomposition to Understand Barred Galaxy Evolution}",
      journal = {arXiv e-prints},
         year = 2019,
        month = mar,
          eid = {arXiv:1903.08203},
        pages = {arXiv:1903.08203},
          doi = {10.48550/arXiv.1903.08203},
archivePrefix = {arXiv},
       eprint = {1903.08203},
 primaryClass = {astro-ph.GA},
       adsurl = {https://ui.adsabs.harvard.edu/abs/2019arXiv190308203P}
}

@ARTICLE{Jalali2007,
       author = {{Jalali}, Mir Abbas},
        title = "{Unstable Disk Galaxies. I. Modal Properties}",
      journal = {\apj},
         year = 2007,
        month = nov,
       volume = {669},
       number = {1},
        pages = {218-231},
          doi = {10.1086/521523},
archivePrefix = {arXiv},
       eprint = {0707.1250},
 primaryClass = {astro-ph},
       adsurl = {https://ui.adsabs.harvard.edu/abs/2007ApJ...669..218J}
}

@ARTICLE{Sellwood2014,
       author = {{Sellwood}, J.~A.},
        title = "{Secular evolution in disk galaxies}",
      journal = {Reviews of Modern Physics},
         year = 2014,
        month = jan,
       volume = {86},
       number = {1},
        pages = {1-46},
          doi = {10.1103/RevModPhys.86.1},
archivePrefix = {arXiv},
       eprint = {1310.0403},
 primaryClass = {astro-ph.GA},
       adsurl = {https://ui.adsabs.harvard.edu/abs/2014RvMP...86....1S}
}

@ARTICLE{deVFreeman72,
       author = {{de Vaucouleurs}, G. and {Freeman}, K.~C.},
        title = "{Structure and dynamics of barred spiral galaxies, in particular of the Magellanic type}",
      journal = {Vistas in Astronomy},
         year = 1972,
        month = jan,
       volume = {14},
       number = {1},
        pages = {163-294},
          doi = {10.1016/0083-6656(72)90026-8},
       adsurl = {https://ui.adsabs.harvard.edu/abs/1972VA.....14..163D}
}

@ARTICLE{Stavely-Smith2003,
       author = {{Staveley-Smith}, L. and {Kim}, S. and {Calabretta}, M.~R. and {Haynes}, R.~F. and {Kesteven}, M.~J.},
        title = "{A new look at the large-scale HI structure of the Large Magellanic Cloud}",
      journal = {\mnras},
         year = 2003,
        month = feb,
       volume = {339},
       number = {1},
        pages = {87-104},
          doi = {10.1046/j.1365-8711.2003.06146.x},
archivePrefix = {arXiv},
       eprint = {astro-ph/0210501},
 primaryClass = {astro-ph},
       adsurl = {https://ui.adsabs.harvard.edu/abs/2003MNRAS.339...87S}
}

@ARTICLE{Chiba2021,
       author = {{Chiba}, Rimpei and {Sch{\"o}nrich}, Ralph},
        title = "{Tree-ring structure of Galactic bar resonance}",
      journal = {\mnras},
         year = 2021,
        month = aug,
       volume = {505},
       number = {2},
        pages = {2412-2426},
          doi = {10.1093/mnras/stab1094},
archivePrefix = {arXiv},
       eprint = {2102.08388},
 primaryClass = {astro-ph.GA},
       adsurl = {https://ui.adsabs.harvard.edu/abs/2021MNRAS.505.2412C}
}

@ARTICLE{Hamilton2023,
       author = {{Hamilton}, Chris and {Tolman}, Elizabeth A. and {Arzamasskiy}, Lev and {Duarte}, Vin{\'\i}cius N.},
        title = "{Galactic Bar Resonances with Diffusion: An Analytic Model with Implications for Bar-Dark Matter Halo Dynamical Friction}",
      journal = {\apj},
         year = 2023,
        month = sep,
       volume = {954},
       number = {1},
          eid = {12},
        pages = {12},
          doi = {10.3847/1538-4357/acd69b},
archivePrefix = {arXiv},
       eprint = {2208.03855},
 primaryClass = {astro-ph.GA},
       adsurl = {https://ui.adsabs.harvard.edu/abs/2023ApJ...954...12H}
}

@ARTICLE{vanderWalt2011,
       author = {{van der Walt}, St{\'e}fan and {Colbert}, S. Chris and {Varoquaux}, Ga{\"e}l},
        title = "{The NumPy Array: A Structure for Efficient Numerical Computation}",
      journal = {Computing in Science and Engineering},
         year = 2011,
        month = mar,
       volume = {13},
       number = {2},
        pages = {22-30},
          doi = {10.1109/MCSE.2011.37},
archivePrefix = {arXiv},
       eprint = {1102.1523},
 primaryClass = {cs.MS},
       adsurl = {https://ui.adsabs.harvard.edu/abs/2011CSE....13b..22V}
}

@ARTICLE{Harris2020,
       author = {{Harris}, Charles R. and {Millman}, K. Jarrod and {van der Walt}, St{\'e}fan J. and {Gommers}, Ralf and {Virtanen}, Pauli and {Cournapeau}, David and {Wieser}, Eric and {Taylor}, Julian and {Berg}, Sebastian and {Smith}, Nathaniel J. and {Kern}, Robert and {Picus}, Matti and {Hoyer}, Stephan and {van Kerkwijk}, Marten H. and {Brett}, Matthew and {Haldane}, Allan and {del R{\'\i}o}, Jaime Fern{\'a}ndez and {Wiebe}, Mark and {Peterson}, Pearu and {G{\'e}rard-Marchant}, Pierre and {Sheppard}, Kevin and {Reddy}, Tyler and {Weckesser}, Warren and {Abbasi}, Hameer and {Gohlke}, Christoph and {Oliphant}, Travis E.},
        title = "{Array programming with NumPy}",
      journal = {\nat},
         year = 2020,
        month = sep,
       volume = {585},
       number = {7825},
        pages = {357-362},
          doi = {10.1038/s41586-020-2649-2},
archivePrefix = {arXiv},
       eprint = {2006.10256},
 primaryClass = {cs.MS},
       adsurl = {https://ui.adsabs.harvard.edu/abs/2020Natur.585..357H}
}

@ARTICLE{Virtanen2020,
       author = {{Virtanen}, Pauli and {Gommers}, Ralf and {Oliphant}, Travis E. and {Haberland}, Matt and {Reddy}, Tyler and {Cournapeau}, David and {Burovski}, Evgeni and {Peterson}, Pearu and {Weckesser}, Warren and {Bright}, Jonathan and {van der Walt}, St{\'e}fan J. and {Brett}, Matthew and {Wilson}, Joshua and {Millman}, K. Jarrod and {Mayorov}, Nikolay and {Nelson}, Andrew R.~J. and {Jones}, Eric and {Kern}, Robert and {Larson}, Eric and {Carey}, C.~J. and {Polat}, {\.I}lhan and {Feng}, Yu and {Moore}, Eric W. and {VanderPlas}, Jake and {Laxalde}, Denis and {Perktold}, Josef and {Cimrman}, Robert and {Henriksen}, Ian and {Quintero}, E.~A. and {Harris}, Charles R. and {Archibald}, Anne M. and {Ribeiro}, Ant{\^o}nio H. and {Pedregosa}, Fabian and {van Mulbregt}, Paul and {SciPy 1. 0 Contributors}},
        title = "{SciPy 1.0: fundamental algorithms for scientific computing in Python}",
      journal = {Nature Methods},
         year = 2020,
        month = feb,
       volume = {17},
        pages = {261-272},
          doi = {10.1038/s41592-019-0686-2},
archivePrefix = {arXiv},
       eprint = {1907.10121},
 primaryClass = {cs.MS},
       adsurl = {https://ui.adsabs.harvard.edu/abs/2020NatMe..17..261V}
}

@ARTICLE{Hunter2007,
  author={Hunter, John D.},
  journal={Computing in Science \& Engineering}, 
  title={Matplotlib: A 2D Graphics Environment}, 
  year={2007},
  volume={9},
  number={3},
  pages={90-95},
  doi={10.1109/MCSE.2007.55}}

@inproceedings{Kluyver2016,
       booktitle = {Positioning and Power in Academic Publishing: Players, Agents and Agendas},
          editor = {Fernando Loizides and Birgit Scmidt},
           title = {Jupyter Notebooks - a publishing format for reproducible computational workflows},
          author = {Thomas Kluyver and Benjamin Ragan-Kelley and Fernando P{\'e}rez and Brian Granger and Matthias Bussonnier and Jonathan Frederic and Kyle Kelley and Jessica Hamrick and Jason Grout and Sylvain Corlay and Paul Ivanov and Dami{\'a}n Avila and Safia Abdalla and Carol Willing and  Jupyter development team},
       publisher = {IOS Press},
         address = {Netherlands},
            year = {2016},
           pages = {87--90},
             url = {https://eprints.soton.ac.uk/403913/}
}

@article{Granger2021,
	title = {Jupyter: {Thinking} and {Storytelling} {With} {Code} and {Data}},
	volume = {23},
	issn = {1558-366X},
	shorttitle = {Jupyter},
	url = {https://ieeexplore.ieee.org/document/9387490},
	doi = {10.1109/MCSE.2021.3059263},
	number = {2},
	urldate = {2025-07-10},
	journal = {Computing in Science \& Engineering},
	author = {Granger, Brian E. and Pérez, Fernando},
	month = mar,
	year = {2021},
	pages = {7--14},
}

@ARTICLE{Dhanush2024,
       author = {{Dhanush}, S.~R. and {Subramaniam}, A. and {Subramanian}, S.},
        title = "{A Comprehensive Kinematic Model of the Large Magellanic Cloud Disk from Star Clusters and Field Stars using Gaia DR3: Tracing the Disk Characteristics, Rotation, Bar, and Outliers}",
      journal = {\apj},
         year = 2024,
        month = jun,
       volume = {968},
       number = {2},
          eid = {103},
        pages = {103},
          doi = {10.3847/1538-4357/ad4453},
archivePrefix = {arXiv},
       eprint = {2404.18658},
 primaryClass = {astro-ph.GA},
       adsurl = {https://ui.adsabs.harvard.edu/abs/2024ApJ...968..103D}
}

@ARTICLE{Arranz2025b,
       author = {{Jim{\'e}nez-Arranz}, {\'O}. and {Roca-F{\`a}brega}, S.},
        title = "{Tidal interaction can stop galactic bars: On the LMC non-rotating bar}",
      journal = {\aap},
         year = 2025,
        month = jun,
       volume = {698},
          eid = {L7},
        pages = {L7},
          doi = {10.1051/0004-6361/202555019},
archivePrefix = {arXiv},
       eprint = {2504.01870},
 primaryClass = {astro-ph.GA},
       adsurl = {https://ui.adsabs.harvard.edu/abs/2025A&A...698L...7J}
}

@ARTICLE{Patel2020,
       author = {{Patel}, Ekta and {Kallivayalil}, Nitya and {Garavito-Camargo}, Nicolas and {Besla}, Gurtina and {Weisz}, Daniel R. and {van der Marel}, Roeland P. and {Boylan-Kolchin}, Michael and {Pawlowski}, Marcel S. and {G{\'o}mez}, Facundo A.},
        title = "{The Orbital Histories of Magellanic Satellites Using Gaia DR2 Proper Motions}",
      journal = {\apj},
         year = 2020,
        month = apr,
       volume = {893},
       number = {2},
          eid = {121},
        pages = {121},
          doi = {10.3847/1538-4357/ab7b75},
archivePrefix = {arXiv},
       eprint = {2001.01746},
 primaryClass = {astro-ph.GA},
       adsurl = {https://ui.adsabs.harvard.edu/abs/2020ApJ...893..121P}
}

@ARTICLE{Zivick2018,
       author = {{Zivick}, Paul and {Kallivayalil}, Nitya and {van der Marel}, Roeland P. and {Besla}, Gurtina and {Linden}, Sean T. and {Koz{\l}owski}, Szymon and {Fritz}, Tobias K. and {Kochanek}, C.~S. and {Anderson}, J. and {Sohn}, Sangmo Tony and {Geha}, Marla C. and {Alcock}, Charles R.},
        title = "{The Proper Motion Field of the Small Magellanic Cloud: Kinematic Evidence for Its Tidal Disruption}",
      journal = {\apj},
         year = 2018,
        month = sep,
       volume = {864},
       number = {1},
          eid = {55},
        pages = {55},
          doi = {10.3847/1538-4357/aad4b0},
archivePrefix = {arXiv},
       eprint = {1804.04110},
 primaryClass = {astro-ph.GA},
       adsurl = {https://ui.adsabs.harvard.edu/abs/2018ApJ...864...55Z}
}

@ARTICLE{Vasiliev2024,
       author = {{Vasiliev}, Eugene},
        title = "{Dear Magellanic Clouds, welcome back!}",
      journal = {\mnras},
         year = 2024,
        month = jan,
       volume = {527},
       number = {1},
        pages = {437-456},
          doi = {10.1093/mnras/stad2612},
archivePrefix = {arXiv},
       eprint = {2306.04837},
 primaryClass = {astro-ph.GA},
       adsurl = {https://ui.adsabs.harvard.edu/abs/2024MNRAS.527..437V}
}

@ARTICLE{Dillamore2026,
       author = {{Dillamore}, Adam M. and {Sanders}, Jason L. and {Brooks}, Richard A.~N.},
        title = "{GSE vs. LMC: reshaping of radially biased stellar haloes by satellites}",
      journal = {arXiv e-prints},
         year = 2026,
        month = mar,
          eid = {arXiv:2603.11159},
        pages = {arXiv:2603.11159},
          doi = {10.48550/arXiv.2603.11159},
archivePrefix = {arXiv},
       eprint = {2603.11159},
 primaryClass = {astro-ph.GA},
       adsurl = {https://ui.adsabs.harvard.edu/abs/2026arXiv260311159D}
}

@ARTICLE{Du2024,
       author = {{Du}, Xiaolong and {Benson}, Andrew and {Zeng}, Zhichao Carton and {Treu}, Tommaso and {Peter}, Annika H.~G. and {Mace}, Charlie and {Jiang}, Fangzhou and {Yang}, Shengqi and {Gannon}, Charles and {Gilman}, Daniel and {Nierenberg}, Anna. M. and {Nadler}, Ethan O.},
        title = "{Tidal evolution of cored and cuspy dark matter halos}",
      journal = {\prd},
         year = 2024,
        month = jul,
       volume = {110},
       number = {2},
          eid = {023019},
        pages = {023019},
          doi = {10.1103/PhysRevD.110.023019},
archivePrefix = {arXiv},
       eprint = {2403.09597},
 primaryClass = {astro-ph.GA},
       adsurl = {https://ui.adsabs.harvard.edu/abs/2024PhRvD.110b3019D}
}

@ARTICLE{Glennon2024,
       author = {{Glennon}, Noah and {Musoke}, Nathan and {Nadler}, Ethan O. and {Prescod-Weinstein}, Chanda and {Wechsler}, Risa H.},
        title = "{Dynamical friction in self-interacting ultralight dark matter}",
      journal = {\prd},
         year = 2024,
        month = mar,
       volume = {109},
       number = {6},
          eid = {063501},
        pages = {063501},
          doi = {10.1103/PhysRevD.109.063501},
archivePrefix = {arXiv},
       eprint = {2312.07684},
 primaryClass = {astro-ph.CO},
       adsurl = {https://ui.adsabs.harvard.edu/abs/2024PhRvD.109f3501G}
}

@ARTICLE{Nadler2020,
       author = {{Nadler}, Ethan O. and {Banerjee}, Arka and {Adhikari}, Susmita and {Mao}, Yao-Yuan and {Wechsler}, Risa H.},
        title = "{Signatures of Velocity-dependent Dark Matter Self-interactions in Milky Way-mass Halos}",
      journal = {\apj},
         year = 2020,
        month = jun,
       volume = {896},
       number = {2},
          eid = {112},
        pages = {112},
          doi = {10.3847/1538-4357/ab94b0},
archivePrefix = {arXiv},
       eprint = {2001.08754},
 primaryClass = {astro-ph.CO},
       adsurl = {https://ui.adsabs.harvard.edu/abs/2020ApJ...896..112N}
}

@ARTICLE{Zeng2022,
       author = {{Zeng}, Zhichao Carton and {Peter}, Annika H.~G. and {Du}, Xiaolong and {Benson}, Andrew and {Kim}, Stacy and {Jiang}, Fangzhou and {Cyr-Racine}, Francis-Yan and {Vogelsberger}, Mark},
        title = "{Core-collapse, evaporation, and tidal effects: the life story of a self-interacting dark matter subhalo}",
      journal = {\mnras},
         year = 2022,
        month = jul,
       volume = {513},
       number = {4},
        pages = {4845-4868},
          doi = {10.1093/mnras/stac1094},
archivePrefix = {arXiv},
       eprint = {2110.00259},
 primaryClass = {astro-ph.CO},
       adsurl = {https://ui.adsabs.harvard.edu/abs/2022MNRAS.513.4845Z}
}

@ARTICLE{Zhang2024,
       author = {{Zhang}, Xingyu and {Yu}, Hai-Bo and {Yang}, Daneng and {An}, Haipeng},
        title = "{Self-interacting Dark Matter Interpretation of Crater II}",
      journal = {\apjl},
         year = 2024,
        month = jun,
       volume = {968},
       number = {1},
          eid = {L13},
        pages = {L13},
          doi = {10.3847/2041-8213/ad50cd},
archivePrefix = {arXiv},
       eprint = {2401.04985},
 primaryClass = {astro-ph.GA},
       adsurl = {https://ui.adsabs.harvard.edu/abs/2024ApJ...968L..13Z}
}

@ARTICLE{NFW1997,
       author = {{Navarro}, Julio F. and {Frenk}, Carlos S. and {White}, Simon D.~M.},
        title = "{A Universal Density Profile from Hierarchical Clustering}",
      journal = {\apj},
         year = 1997,
        month = dec,
       volume = {490},
       number = {2},
        pages = {493-508},
          doi = {10.1086/304888},
archivePrefix = {arXiv},
       eprint = {astro-ph/9611107},
 primaryClass = {astro-ph},
       adsurl = {https://ui.adsabs.harvard.edu/abs/1997ApJ...490..493N}
}

@ARTICLE{Foote2023,
       author = {{Foote}, Hayden R. and {Besla}, Gurtina and {Mocz}, Philip and {Garavito-Camargo}, Nicol{\'a}s and {Lancaster}, Lachlan and {Sparre}, Martin and {Cunningham}, Emily C. and {Vogelsberger}, Mark and {G{\'o}mez}, Facundo A. and {Laporte}, Chervin F.~P.},
        title = "{Structure, Kinematics, and Observability of the Large Magellanic Cloud's Dynamical Friction Wake in Cold versus Fuzzy Dark Matter}",
      journal = {\apj},
         year = 2023,
        month = sep,
       volume = {954},
       number = {2},
          eid = {163},
        pages = {163},
          doi = {10.3847/1538-4357/ace533},
archivePrefix = {arXiv},
       eprint = {2307.00053},
 primaryClass = {astro-ph.GA},
       adsurl = {https://ui.adsabs.harvard.edu/abs/2023ApJ...954..163F}
}

@ARTICLE{Lancaster2020,
       author = {{Lancaster}, Lachlan and {Giovanetti}, Cara and {Mocz}, Philip and {Kahn}, Yonatan and {Lisanti}, Mariangela and {Spergel}, David N.},
        title = "{Dynamical friction in a Fuzzy Dark Matter universe}",
      journal = {\jcap},
         year = 2020,
        month = jan,
       volume = {2020},
       number = {1},
          eid = {001},
        pages = {001},
          doi = {10.1088/1475-7516/2020/01/001},
archivePrefix = {arXiv},
       eprint = {1909.06381},
 primaryClass = {astro-ph.CO},
       adsurl = {https://ui.adsabs.harvard.edu/abs/2020JCAP...01..001L}
}

@ARTICLE{Luri2021,
       author = {{Gaia Collaboration} and {Luri}, X. and {Chemin}, L. and {Clementini}, G. and {Delgado}, H.~E. and {McMillan}, P.~J. and {Romero-G{\'o}mez}, M. and {Balbinot}, E. and {Castro-Ginard}, A. and {Mor}, R. and {Ripepi}, V. and {Sarro}, L.~M. and {Cioni}, M.-R.~L. and {Fabricius}, C. and {Garofalo}, A. and {Helmi}, A. and {Muraveva}, T. and {Brown}, A.~G.~A. and {Vallenari}, A. and {Prusti}, T. and {de Bruijne}, J.~H.~J. and {Babusiaux}, C. and {Biermann}, M. and {Creevey}, O.~L. and {Evans}, D.~W. and {Eyer}, L. and {Hutton}, A. and {Jansen}, F. and {Jordi}, C. and {Klioner}, S.~A. and {Lammers}, U. and {Lindegren}, L. and {Mignard}, F. and {Panem}, C. and {Pourbaix}, D. and {Randich}, S. and {Sartoretti}, P. and {Soubiran}, C. and {Walton}, N.~A. and {Arenou}, F. and {Bailer-Jones}, C.~A.~L. and {Bastian}, U. and {Cropper}, M. and {Drimmel}, R. and {Katz}, D. and {Lattanzi}, M.~G. and {van Leeuwen}, F. and {Bakker}, J. and {Casta{\~n}eda}, J. and {De Angeli}, F. and {Ducourant}, C. and {Fouesneau}, M. and {Fr{\'e}mat}, Y. and {Guerra}, R. and {Guerrier}, A. and {Guiraud}, J. and {Jean-Antoine Piccolo}, A. and {Masana}, E. and {Messineo}, R. and {Mowlavi}, N. and {Nicolas}, C. and {Nienartowicz}, K. and {Pailler}, F. and {Panuzzo}, P. and {Riclet}, F. and {Roux}, W. and {Seabroke}, G.~M. and {Sordo}, R. and {Tanga}, P. and {Th{\'e}venin}, F. and {Gracia-Abril}, G. and {Portell}, J. and {Teyssier}, D. and {Altmann}, M. and {Andrae}, R. and {Bellas-Velidis}, I. and {Benson}, K. and {Berthier}, J. and {Blomme}, R. and {Brugaletta}, E. and {Burgess}, P.~W. and {Busso}, G. and {Carry}, B. and {Cellino}, A. and {Cheek}, N. and {Damerdji}, Y. and {Davidson}, M. and {Delchambre}, L. and {Dell'Oro}, A. and {Fern{\'a}ndez-Hern{\'a}ndez}, J. and {Galluccio}, L. and {Garc{\'\i}a-Lario}, P. and {Garcia-Reinaldos}, M. and {Gonz{\'a}lez-N{\'u}{\~n}ez}, J. and {Gosset}, E. and {Haigron}, R. and {Halbwachs}, J.-L. and {Hambly}, N.~C. and {Harrison}, D.~L. and {Hatzidimitriou}, D. and {Heiter}, U. and {Hern{\'a}ndez}, J. and {Hestroffer}, D. and {Hodgkin}, S.~T. and {Holl}, B. and {Jan{\ss}en}, K. and {Jevardat de Fombelle}, G. and {Jordan}, S. and {Krone-Martins}, A. and {Lanzafame}, A.~C. and {L{\"o}ffler}, W. and {Lorca}, A. and {Manteiga}, M. and {Marchal}, O. and {Marrese}, P.~M. and {Moitinho}, A. and {Mora}, A. and {Muinonen}, K. and {Osborne}, P. and {Pancino}, E. and {Pauwels}, T. and {Recio-Blanco}, A. and {Richards}, P.~J. and {Riello}, M. and {Rimoldini}, L. and {Robin}, A.~C. and {Roegiers}, T. and {Rybizki}, J. and {Siopis}, C. and {Smith}, M. and {Sozzetti}, A. and {Ulla}, A. and {Utrilla}, E. and {van Leeuwen}, M. and {van Reeven}, W. and {Abbas}, U. and {Abreu Aramburu}, A. and {Accart}, S. and {Aerts}, C. and {Aguado}, J.~J. and {Ajaj}, M. and {Altavilla}, G. and {{\'A}lvarez}, M.~A. and {{\'A}lvarez Cid-Fuentes}, J. and {Alves}, J. and {Anderson}, R.~I. and {Anglada Varela}, E. and {Antoja}, T. and {Audard}, M. and {Baines}, D. and {Baker}, S.~G. and {Balaguer-N{\'u}{\~n}ez}, L. and {Balog}, Z. and {Barache}, C. and {Barbato}, D. and {Barros}, M. and {Barstow}, M.~A. and {Bartolom{\'e}}, S. and {Bassilana}, J.-L. and {Bauchet}, N. and {Baudesson-Stella}, A. and {Becciani}, U. and {Bellazzini}, M. and {Bernet}, M. and {Bertone}, S. and {Bianchi}, L. and {Blanco-Cuaresma}, S. and {Boch}, T. and {Bombrun}, A. and {Bossini}, D. and {Bouquillon}, S. and {Bragaglia}, A. and {Bramante}, L. and {Breedt}, E. and {Bressan}, A. and {Brouillet}, N. and {Bucciarelli}, B. and {Burlacu}, A. and {Busonero}, D. and {Butkevich}, A.~G. and {Buzzi}, R. and {Caffau}, E. and {Cancelliere}, R. and {C{\'a}novas}, H. and {Cantat-Gaudin}, T. and {Carballo}, R. and {Carlucci}, T. and {Carnerero}, M.~I. and {Carrasco}, J.~M. and {Casamiquela}, L. and {Castellani}, M. and {Castro Sampol}, P. and {Chaoul}, L. and {Charlot}, P. and {Chiavassa}, A. and {Comoretto}, G. and {Cooper}, W.~J. and {Cornez}, T. and {Cowell}, S. and {Crifo}, F.},
        title = "{Gaia Early Data Release 3. Structure and properties of the Magellanic Clouds}",
      journal = {\aap},
         year = 2021,
        month = may,
       volume = {649},
          eid = {A7},
        pages = {A7},
          doi = {10.1051/0004-6361/202039588},
archivePrefix = {arXiv},
       eprint = {2012.01771},
 primaryClass = {astro-ph.GA},
       adsurl = {https://ui.adsabs.harvard.edu/abs/2021A&A...649A...7G}
}

@ARTICLE{Nidever2026,
       author = {{Nidever}, David L. and {Horta}, Danny and {Majewski}, Steven R. and {Almeida}, Andres and {Povick}, Joshua T. and {Oden}, Slater J. and {Jimenez-Arranz}, Oscar and {Stringfellow}, Guy and {Chojnowski}, S. Drew and {van der Marel}, Roeland and {Cullinane}, Lara and {Dias}, Bruno and {Johnson}, Jennifer and {Donor}, John and {Cioni}, Maria-Rosa and {Kollmeier}, Juna and {Tkachenko}, Andrew},
        title = "{Overview of The SDSS-V Magellanic Genesis Survey}",
      journal = {arXiv e-prints},
         year = 2026,
        month = feb,
          eid = {arXiv:2602.02693},
        pages = {arXiv:2602.02693},
          doi = {10.48550/arXiv.2602.02693},
archivePrefix = {arXiv},
       eprint = {2602.02693},
 primaryClass = {astro-ph.GA},
       adsurl = {https://ui.adsabs.harvard.edu/abs/2026arXiv260202693N}
}

@ARTICLE{Nidever2017,
       author = {{Nidever}, David L. and {Olsen}, Knut and {Walker}, Alistair R. and {Vivas}, A. Katherina and {Blum}, Robert D. and {Kaleida}, Catherine and {Choi}, Yumi and {Conn}, Blair C. and {Gruendl}, Robert A. and {Bell}, Eric F. and {Besla}, Gurtina and {Mu{\~n}oz}, Ricardo R. and {Gallart}, Carme and {Martin}, Nicolas F. and {Olszewski}, Edward W. and {Saha}, Abhijit and {Monachesi}, Antonela and {Monelli}, Matteo and {de Boer}, Thomas J.~L. and {Johnson}, L. Clifton and {Zaritsky}, Dennis and {Stringfellow}, Guy S. and {van der Marel}, Roeland P. and {Cioni}, Maria-Rosa L. and {Jin}, Shoko and {Majewski}, Steven R. and {Martinez-Delgado}, David and {Monteagudo}, Lara and {No{\"e}l}, Noelia E.~D. and {Bernard}, Edouard J. and {Kunder}, Andrea and {Chu}, You-Hua and {Bell}, Cameron P.~M. and {Santana}, Felipe and {Frechem}, Joshua and {Medina}, Gustavo E. and {Parkash}, Vaishali and {Navarrete}, J.~C. Ser{\'o}n and {Hayes}, Christian},
        title = "{SMASH: Survey of the MAgellanic Stellar History}",
      journal = {\aj},
         year = 2017,
        month = nov,
       volume = {154},
       number = {5},
          eid = {199},
        pages = {199},
          doi = {10.3847/1538-3881/aa8d1c},
archivePrefix = {arXiv},
       eprint = {1701.00502},
 primaryClass = {astro-ph.GA},
       adsurl = {https://ui.adsabs.harvard.edu/abs/2017AJ....154..199N}
}

@ARTICLE{Boylan-Kolchin2011,
       author = {{Boylan-Kolchin}, Michael and {Besla}, Gurtina and {Hernquist}, Lars},
        title = "{Dynamics of the Magellanic Clouds in a Lambda cold dark matter universe}",
      journal = {\mnras},
         year = 2011,
        month = jun,
       volume = {414},
       number = {2},
        pages = {1560-1572},
          doi = {10.1111/j.1365-2966.2011.18495.x},
archivePrefix = {arXiv},
       eprint = {1010.4797},
 primaryClass = {astro-ph.GA},
       adsurl = {https://ui.adsabs.harvard.edu/abs/2011MNRAS.414.1560B}
}

@ARTICLE{Mayer2001,
       author = {{Mayer}, Lucio and {Governato}, Fabio and {Colpi}, Monica and {Moore}, Ben and {Quinn}, Thomas and {Wadsley}, James and {Stadel}, Joachim and {Lake}, George},
        title = "{Tidal Stirring and the Origin of Dwarf Spheroidals in the Local Group}",
      journal = {\apjl},
         year = 2001,
        month = feb,
       volume = {547},
       number = {2},
        pages = {L123-L127},
          doi = {10.1086/318898},
archivePrefix = {arXiv},
       eprint = {astro-ph/0011041},
 primaryClass = {astro-ph},
       adsurl = {https://ui.adsabs.harvard.edu/abs/2001ApJ...547L.123M}
}

@ARTICLE{Kazantzidis2017,
       author = {{Kazantzidis}, Stelios and {Mayer}, Lucio and {Callegari}, Simone and {Dotti}, Massimo and {Moustakas}, Leonidas A.},
        title = "{The Effects of Ram-pressure Stripping and Supernova Winds on the Tidal Stirring of Disky Dwarfs: Enhanced Transformation into Dwarf Spheroidals}",
      journal = {\apjl},
         year = 2017,
        month = feb,
       volume = {836},
       number = {1},
          eid = {L13},
        pages = {L13},
          doi = {10.3847/2041-8213/aa5b8f},
archivePrefix = {arXiv},
       eprint = {1703.08381},
 primaryClass = {astro-ph.GA},
       adsurl = {https://ui.adsabs.harvard.edu/abs/2017ApJ...836L..13K}
}

@ARTICLE{Lokas2015,
       author = {{{\L}okas}, Ewa L. and {Semczuk}, Marcin and {Gajda}, Grzegorz and {D'Onghia}, Elena},
        title = "{The Resonant Nature of Tidal Stirring of Disky Dwarf Galaxies Orbiting the Milky Way}",
      journal = {\apj},
         year = 2015,
        month = sep,
       volume = {810},
       number = {2},
          eid = {100},
        pages = {100},
          doi = {10.1088/0004-637X/810/2/100},
archivePrefix = {arXiv},
       eprint = {1505.00951},
 primaryClass = {astro-ph.GA},
       adsurl = {https://ui.adsabs.harvard.edu/abs/2015ApJ...810..100L}
}

@ARTICLE{Semczuk2018,
       author = {{Semczuk}, Marcin and {{\L}okas}, Ewa L. and {Salomon}, Jean-Baptiste and {Athanassoula}, E. and {D'Onghia}, Elena},
        title = "{Tidally Induced Morphology of M33 in Hydrodynamical Simulations of Its Recent Interaction with M31}",
      journal = {\apj},
         year = 2018,
        month = sep,
       volume = {864},
       number = {1},
          eid = {34},
        pages = {34},
          doi = {10.3847/1538-4357/aad4ae},
archivePrefix = {arXiv},
       eprint = {1804.04536},
 primaryClass = {astro-ph.GA},
       adsurl = {https://ui.adsabs.harvard.edu/abs/2018ApJ...864...34S}
}

@ARTICLE{Zhou2026,
       author = {{Zhou}, Yufan and {Li}, Zhiyuan and {Jim{\'e}nez-Arranz}, {\'O}scar and {Roca-F{\`a}brega}, Santi},
        title = "{Survival or Destruction: Effects of Spheroidal Satellite Collisions on Bars in Milky Way-Like Galaxies}",
      journal = {arXiv e-prints},
         year = 2026,
        month = mar,
          eid = {arXiv:2603.26077},
        pages = {arXiv:2603.26077},
          doi = {10.48550/arXiv.2603.26077},
archivePrefix = {arXiv},
       eprint = {2603.26077},
 primaryClass = {astro-ph.GA},
       adsurl = {https://ui.adsabs.harvard.edu/abs/2026arXiv260326077Z}
}

@ARTICLE{Niederhofer2022,
       author = {{Niederhofer}, Florian and {Cioni}, Maria-Rosa L. and {Schmidt}, Thomas and {Bekki}, Kenji and {de Grijs}, Richard and {Ivanov}, Valentin D. and {Oliveira}, Joana M. and {Ripepi}, Vincenzo and {Subramanian}, Smitha and {van Loon}, Jacco Th},
        title = "{The VMC survey - XLVI. Stellar proper motions in the centre of the Large Magellanic Cloud}",
      journal = {\mnras},
         year = 2022,
        month = jun,
       volume = {512},
       number = {4},
        pages = {5423-5439},
          doi = {10.1093/mnras/stac712},
archivePrefix = {arXiv},
       eprint = {2203.14369},
 primaryClass = {astro-ph.GA},
       adsurl = {https://ui.adsabs.harvard.edu/abs/2022MNRAS.512.5423N}
}

@ARTICLE{Arranz2024b,
       author = {{Jim{\'e}nez-Arranz}, {\'O}. and {Chemin}, L. and {Romero-G{\'o}mez}, M. and {Luri}, X. and {Adamczyk}, P. and {Castro-Ginard}, A. and {Roca-F{\`a}brega}, S. and {McMillan}, P.~J. and {Cioni}, M.-R.~L.},
        title = "{The bar pattern speed of the Large Magellanic Cloud}",
      journal = {\aap},
         year = 2024,
        month = mar,
       volume = {683},
          eid = {A102},
        pages = {A102},
          doi = {10.1051/0004-6361/202347266},
archivePrefix = {arXiv},
       eprint = {2312.11192},
 primaryClass = {astro-ph.GA},
       adsurl = {https://ui.adsabs.harvard.edu/abs/2024A&A...683A.102J}
}

@ARTICLE{Scholch2025,
       author = {{Sch{\"o}lch}, M. and {Jim{\'e}nez-Arranz}, {\'O}. and {Romero-G{\'o}mez}, M. and {Luri}, X. and {Hobbs}, D. and {Salmer{\'o}n-Larraz}, D. and {L{\'o}pez Vilamaj{\'o}}, M.},
        title = "{Asymmetries in the LMC velocity maps}",
      journal = {\aap},
         year = 2025,
        month = sep,
       volume = {701},
          eid = {A227},
        pages = {A227},
          doi = {10.1051/0004-6361/202556309},
archivePrefix = {arXiv},
       eprint = {2508.01434},
 primaryClass = {astro-ph.GA},
       adsurl = {https://ui.adsabs.harvard.edu/abs/2025A&A...701A.227S}
}

@ARTICLE{Pearson2018,
       author = {{Pearson}, Sarah and {Privon}, George C. and {Besla}, Gurtina and {Putman}, Mary E. and {Mart{\'\i}nez-Delgado}, David and {Johnston}, Kathryn V. and {Gabany}, R. Jay and {Patton}, David R. and {Kallivayalil}, Nitya},
        title = "{Modelling the baryon cycle in low-mass galaxy encounters: the case of NGC 4490 and NGC 4485}",
      journal = {\mnras},
         year = 2018,
        month = nov,
       volume = {480},
       number = {3},
        pages = {3069-3090},
          doi = {10.1093/mnras/sty2052},
archivePrefix = {arXiv},
       eprint = {1807.03791},
 primaryClass = {astro-ph.GA},
       adsurl = {https://ui.adsabs.harvard.edu/abs/2018MNRAS.480.3069P}
}

@ARTICLE{Pearson2016,
       author = {{Pearson}, Sarah and {Besla}, Gurtina and {Putman}, Mary E. and {Lutz}, Katharina A. and {Fern{\'a}ndez}, Ximena and {Stierwalt}, Sabrina and {Patton}, David R. and {Kim}, Jinhyub and {Kallivayalil}, Nitya and {Johnson}, Kelsey and {Sung}, Eon-Chang},
        title = "{Local Volume TiNy Titans: gaseous dwarf-dwarf interactions in the Local Universe}",
      journal = {\mnras},
         year = 2016,
        month = jun,
       volume = {459},
       number = {2},
        pages = {1827-1846},
          doi = {10.1093/mnras/stw757},
archivePrefix = {arXiv},
       eprint = {1603.09342},
 primaryClass = {astro-ph.GA},
       adsurl = {https://ui.adsabs.harvard.edu/abs/2016MNRAS.459.1827P}
}

@ARTICLE{Phookun1992,
       author = {{Phookun}, Bikram and {Mundy}, Lee G. and {Teuben}, Peter J. and {Wainscoat}, Richard J.},
        title = "{NGC 4027: an Interacting One-armed Spiral Galaxy with a Warped Ring}",
      journal = {\apj},
         year = 1992,
        month = dec,
       volume = {400},
        pages = {516},
          doi = {10.1086/172014},
       adsurl = {https://ui.adsabs.harvard.edu/abs/1992ApJ...400..516P}
}

@ARTICLE{Wilcots2004,
       author = {{Wilcots}, Eric M. and {Prescott}, Moire K.~M.},
        title = "{H I Observations of Barred Magellanic Spirals. II. The Frequency and Impact of Companions}",
      journal = {\aj},
         year = 2004,
        month = apr,
       volume = {127},
       number = {4},
        pages = {1900-1916},
          doi = {10.1086/381293},
       adsurl = {https://ui.adsabs.harvard.edu/abs/2004AJ....127.1900W}
}

@ARTICLE{Danieli2025,
       author = {{Danieli}, Shany and {Kado-Fong}, Erin and {Huang}, Song and {Luo}, Yifei and {Li}, Ting S. and {Kelvin}, Lee S. and {Leauthaud}, Alexie and {Greene}, Jenny E. and {Mintz}, Abby and {Lin}, Xiaojing and {Li}, Jiaxuan and {Baldassare}, Vivienne and {Banerjee}, Arka and {Bhattacharyya}, Joy and {Blanco}, Diana and {Brooks}, Alyson and {Cai}, Zheng and {Chen}, Xinjun and {Cruz}, Akaxia and {Geda}, Robel and {Guan}, Runquan and {Johnson}, Sean and {Kannawadi}, Arun and {Kim}, Stacy Y. and {Li}, Mingyu and {Lupton}, Robert and {Mace}, Charlie and {Medina}, Gustavo E. and {Pan}, Yue and {Peter}, Annika H.~G. and {Read}, Justin I. and {C{\'o}rdova Rosado}, Rodrigo and {Seifert}, Allen and {Wasleske}, Erik J. and {Wick}, Joseph},
        title = "{First Data Release of the Merian Survey: A Wide-field Imaging Survey of Dwarf Galaxies at z {\ensuremath{\sim}} 0.06─0.10}",
      journal = {\apj},
         year = 2025,
        month = nov,
       volume = {993},
       number = {1},
          eid = {110},
        pages = {110},
          doi = {10.3847/1538-4357/ae003e},
archivePrefix = {arXiv},
       eprint = {2410.01884},
 primaryClass = {astro-ph.GA},
       adsurl = {https://ui.adsabs.harvard.edu/abs/2025ApJ...993..110D}
}

@ARTICLE{Drlica-Wagner2021,
       author = {{Drlica-Wagner}, A. and {Carlin}, J.~L. and {Nidever}, D.~L. and {Ferguson}, P.~S. and {Kuropatkin}, N. and {Adam{\'o}w}, M. and {Cerny}, W. and {Choi}, Y. and {Esteves}, J.~H. and {Mart{\'\i}nez-V{\'a}zquez}, C.~E. and {Mau}, S. and {Miller}, A.~E. and {Mutlu-Pakdil}, B. and {Neilsen}, E.~H. and {Olsen}, K.~A.~G. and {Pace}, A.~B. and {Riley}, A.~H. and {Sakowska}, J.~D. and {Sand}, D.~J. and {Santana-Silva}, L. and {Tollerud}, E.~J. and {Tucker}, D.~L. and {Vivas}, A.~K. and {Zaborowski}, E. and {Zenteno}, A. and {Abbott}, T.~M.~C. and {Allam}, S. and {Bechtol}, K. and {Bell}, C.~P.~M. and {Bell}, E.~F. and {Bilaji}, P. and {Bom}, C.~R. and {Carballo-Bello}, J.~A. and {Crnojevi{\'c}}, D. and {Cioni}, M.-R.~L. and {Diaz-Ocampo}, A. and {de Boer}, T.~J.~L. and {Erkal}, D. and {Gruendl}, R.~A. and {Hernandez-Lang}, D. and {Hughes}, A.~K. and {James}, D.~J. and {Johnson}, L.~C. and {Li}, T.~S. and {Mao}, Y.-Y. and {Mart{\'\i}nez-Delgado}, D. and {Massana}, P. and {McNanna}, M. and {Morgan}, R. and {Nadler}, E.~O. and {No{\"e}l}, N.~E.~D. and {Palmese}, A. and {Peter}, A.~H.~G. and {Rykoff}, E.~S. and {S{\'a}nchez}, J. and {Shipp}, N. and {Simon}, J.~D. and {Smercina}, A. and {Soares-Santos}, M. and {Stringfellow}, G.~S. and {Tavangar}, K. and {van der Marel}, R.~P. and {Walker}, A.~R. and {Wechsler}, R.~H. and {Wu}, J.~F. and {Yanny}, B. and {Fitzpatrick}, M. and {Huang}, L. and {Jacques}, A. and {Nikutta}, R. and {Scott}, A. and {Astro Data Lab}},
        title = "{The DECam Local Volume Exploration Survey: Overview and First Data Release}",
      journal = {\apjs},
         year = 2021,
        month = sep,
       volume = {256},
       number = {1},
          eid = {2},
        pages = {2},
          doi = {10.3847/1538-4365/ac079d},
archivePrefix = {arXiv},
       eprint = {2103.07476},
 primaryClass = {astro-ph.GA},
       adsurl = {https://ui.adsabs.harvard.edu/abs/2021ApJS..256....2D}
}

@ARTICLE{Patel2026,
       author = {{Patel}, Ekta and {Bennet}, Paul and {Sohn}, Sangmo Tony and {Fardal}, Mark A. and {van der Marel}, Roeland P.},
        title = "{Encounters between M33 and Present-day M31 Satellites Hint at a Previous Group Accretion}",
      journal = {\apj},
         year = 2026,
        month = mar,
       volume = {999},
       number = {2},
          eid = {174},
        pages = {174},
          doi = {10.3847/1538-4357/ae3d98},
archivePrefix = {arXiv},
       eprint = {2601.20837},
 primaryClass = {astro-ph.GA},
       adsurl = {https://ui.adsabs.harvard.edu/abs/2026ApJ...999..174P}
}

@ARTICLE{Savino2025,
       author = {{Savino}, Alessandro and {Weisz}, Daniel R. and {Dolphin}, Andrew E. and {Durbin}, Meredith J. and {Kallivayalil}, Nitya and {Wetzel}, Andrew and {Anderson}, Jay and {Besla}, Gurtina and {Boylan-Kolchin}, Michael and {Brown}, Thomas M. and {Bullock}, James S. and {Cole}, Andrew A. and {Collins}, Michelle L.~M. and {Cooper}, M.~C. and {Deason}, Alis J. and {Dotter}, Aaron L. and {Fardal}, Mark and {Ferguson}, Annette M.~N. and {Fritz}, Tobias K. and {Geha}, Marla C. and {Gilbert}, Karoline M. and {Guhathakurta}, Puragra and {Ibata}, Rodrigo and {Irwin}, Michael J. and {Jeon}, Myoungwon and {Kirby}, Evan N. and {Lewis}, Geraint F. and {Mackey}, Dougal and {Majewski}, Steven R. and {Martin}, Nicolas and {McConnachie}, Alan and {Patel}, Ekta and {Rich}, R. Michael and {Skillman}, Evan D. and {Simon}, Joshua D. and {Sohn}, Sangmo Tony and {Tollerud}, Erik J. and {van der Marel}, Roeland P.},
        title = "{The Hubble Space Telescope Survey of M31 Satellite Galaxies. IV. Survey Overview and Lifetime Star Formation Histories}",
      journal = {\apj},
         year = 2025,
        month = feb,
       volume = {979},
       number = {2},
          eid = {205},
        pages = {205},
          doi = {10.3847/1538-4357/ada24f},
archivePrefix = {arXiv},
       eprint = {2501.13152},
 primaryClass = {astro-ph.GA},
       adsurl = {https://ui.adsabs.harvard.edu/abs/2025ApJ...979..205S}
}

@ARTICLE{Mao2024,
       author = {{Mao}, Yao-Yuan and {Geha}, Marla and {Wechsler}, Risa H. and {Asali}, Yasmeen and {Wang}, Yunchong and {Kado-Fong}, Erin and {Kallivayalil}, Nitya and {Nadler}, Ethan O. and {Tollerud}, Erik J. and {Weiner}, Benjamin and {de los Reyes}, Mithi A.~C. and {Wu}, John F.},
        title = "{The SAGA Survey. III. A Census of 101 Satellite Systems around Milky Way─mass Galaxies}",
      journal = {\apj},
         year = 2024,
        month = nov,
       volume = {976},
       number = {1},
          eid = {117},
        pages = {117},
          doi = {10.3847/1538-4357/ad64c4},
archivePrefix = {arXiv},
       eprint = {2404.14498},
 primaryClass = {astro-ph.GA},
       adsurl = {https://ui.adsabs.harvard.edu/abs/2024ApJ...976..117M}
}

@ARTICLE{Corbelli2024,
       author = {{Corbelli}, Edvige and {Burkert}, Andreas},
        title = "{The outskirts of M33: Tidally induced distortions versus signatures of gas accretion}",
      journal = {\aap},
         year = 2024,
        month = may,
       volume = {685},
          eid = {A38},
        pages = {A38},
          doi = {10.1051/0004-6361/202348910},
archivePrefix = {arXiv},
       eprint = {2402.16957},
 primaryClass = {astro-ph.GA},
       adsurl = {https://ui.adsabs.harvard.edu/abs/2024A&A...685A..38C}
}

@ARTICLE{Weinberg2015,
       author = {{Weinberg}, David H. and {Bullock}, James S. and {Governato}, Fabio and {Kuzio de Naray}, Rachel and {Peter}, Annika H.~G.},
        title = "{Cold dark matter: Controversies on small scales}",
      journal = {Proceedings of the National Academy of Science},
         year = 2015,
        month = oct,
       volume = {112},
       number = {40},
        pages = {12249-12255},
          doi = {10.1073/pnas.1308716112},
archivePrefix = {arXiv},
       eprint = {1306.0913},
 primaryClass = {astro-ph.CO},
       adsurl = {https://ui.adsabs.harvard.edu/abs/2015PNAS..11212249W}
}

@article{Price-Whelan2017,
  doi = {10.21105/joss.00388},
  url = {https://doi.org/10.21105%2Fjoss.00388},
  year = 2017,
  month = {oct},
  publisher = {The Open Journal},
  volume = {2},
  number = {18},
  author = {Adrian M. Price-Whelan},
  title = {Gala: A Python package for galactic dynamics},
  journal = {The Journal of Open Source Software}}

@ARTICLE{Bovy2015,
       author = {{Bovy}, Jo},
        title = "{galpy: A python Library for Galactic Dynamics}",
      journal = {\apjs},
         year = 2015,
        month = feb,
       volume = {216},
       number = {2},
          eid = {29},
        pages = {29},
          doi = {10.1088/0067-0049/216/2/29},
archivePrefix = {arXiv},
       eprint = {1412.3451},
 primaryClass = {astro-ph.GA},
       adsurl = {https://ui.adsabs.harvard.edu/abs/2015ApJS..216...29B}
}

@ARTICLE{Vasiliev2019,
       author = {{Vasiliev}, Eugene},
        title = "{AGAMA: action-based galaxy modelling architecture}",
      journal = {\mnras},
         year = 2019,
        month = jan,
       volume = {482},
       number = {2},
        pages = {1525-1544},
          doi = {10.1093/mnras/sty2672},
archivePrefix = {arXiv},
       eprint = {1802.08239},
 primaryClass = {astro-ph.GA},
       adsurl = {https://ui.adsabs.harvard.edu/abs/2019MNRAS.482.1525V}
}

@ARTICLE{Jog2000,
       author = {{Jog}, Chanda J.},
        title = "{Self-consistent Response of a Galactic Disk to an Elliptical Perturbation Halo Potential}",
      journal = {\apj},
         year = 2000,
        month = oct,
       volume = {542},
       number = {1},
        pages = {216-223},
          doi = {10.1086/309521},
archivePrefix = {arXiv},
       eprint = {astro-ph/0007364},
 primaryClass = {astro-ph},
       adsurl = {https://ui.adsabs.harvard.edu/abs/2000ApJ...542..216J}
}

@ARTICLE{Ghosh2022,
       author = {{Ghosh}, Soumavo and {Saha}, Kanak and {Jog}, Chanda J. and {Combes}, Francoise and {Di Matteo}, Paola},
        title = "{Genesis of morpho-kinematic lopsidedness in minor merger of galaxies}",
      journal = {\mnras},
         year = 2022,
        month = apr,
       volume = {511},
       number = {4},
        pages = {5878-5896},
          doi = {10.1093/mnras/stac461},
archivePrefix = {arXiv},
       eprint = {2105.05270},
 primaryClass = {astro-ph.GA},
       adsurl = {https://ui.adsabs.harvard.edu/abs/2022MNRAS.511.5878G}
}

@ARTICLE{Arora2026,
       author = {{Arora}, Arpit and {Ferguson}, Peter S. and {Nibauer}, Jacob and {Shipp}, Nora and {Reddy}, Videep and {Vasiliev}, Eugene and {Kohm}, Jack and {Marin}, Laurella C. and {Price-Whelan}, Adrian M. and {Erkal}, Denis and {Pearson}, Sarah and {Wetzel}, Andrew and {Bailin}, Jeremy and {Feldmann}, Robert},
        title = "{No Stream Left Unscathed: The imprint of a host galaxy}",
      journal = {arXiv e-prints},
         year = 2026,
        month = may,
          eid = {arXiv:2605.16200},
        pages = {arXiv:2605.16200},
archivePrefix = {arXiv},
       eprint = {2605.16200},
 primaryClass = {astro-ph.GA},
       adsurl = {https://ui.adsabs.harvard.edu/abs/2026arXiv260516200A}
}

@ARTICLE{vdM2014,
       author = {{van der Marel}, Roeland P. and {Kallivayalil}, Nitya},
        title = "{Third-epoch Magellanic Cloud Proper Motions. II. The Large Magellanic Cloud Rotation Field in Three Dimensions}",
      journal = {\apj},
         year = 2014,
        month = feb,
       volume = {781},
       number = {2},
          eid = {121},
        pages = {121},
          doi = {10.1088/0004-637X/781/2/121},
archivePrefix = {arXiv},
       eprint = {1305.4641},
 primaryClass = {astro-ph.CO},
       adsurl = {https://ui.adsabs.harvard.edu/abs/2014ApJ...781..121V}
}

@ARTICLE{Harris2006,
       author = {{Harris}, Jason and {Zaritsky}, Dennis},
        title = "{Spectroscopic Survey of Red Giants in the Small Magellanic Cloud. I. Kinematics}",
      journal = {\aj},
         year = 2006,
        month = may,
       volume = {131},
       number = {5},
        pages = {2514-2524},
          doi = {10.1086/500974},
archivePrefix = {arXiv},
       eprint = {astro-ph/0601025},
 primaryClass = {astro-ph},
       adsurl = {https://ui.adsabs.harvard.edu/abs/2006AJ....131.2514H}
}

@ARTICLE{Zivick2021,
       author = {{Zivick}, Paul and {Kallivayalil}, Nitya and {van der Marel}, Roeland P.},
        title = "{Deciphering the Kinematic Structure of the Small Magellanic Cloud through Its Red Giant Population}",
      journal = {\apj},
         year = 2021,
        month = mar,
       volume = {910},
       number = {1},
          eid = {36},
        pages = {36},
          doi = {10.3847/1538-4357/abe1bb},
archivePrefix = {arXiv},
       eprint = {2011.02525},
 primaryClass = {astro-ph.GA},
       adsurl = {https://ui.adsabs.harvard.edu/abs/2021ApJ...910...36Z}
}

@ARTICLE{Salem2015,
       author = {{Salem}, Munier and {Besla}, Gurtina and {Bryan}, Greg and {Putman}, Mary and {van der Marel}, Roeland P. and {Tonnesen}, Stephanie},
        title = "{Ram Pressure Stripping of the Large Magellanic Cloud's Disk as a Probe of the Milky Way's Circumgalactic Medium}",
      journal = {\apj},
         year = 2015,
        month = dec,
       volume = {815},
       number = {1},
          eid = {77},
        pages = {77},
          doi = {10.1088/0004-637X/815/1/77},
archivePrefix = {arXiv},
       eprint = {1507.07935},
 primaryClass = {astro-ph.GA},
       adsurl = {https://ui.adsabs.harvard.edu/abs/2015ApJ...815...77S}
}

@ARTICLE{Besla2007,
       author = {{Besla}, Gurtina and {Kallivayalil}, Nitya and {Hernquist}, Lars and {Robertson}, Brant and {Cox}, T.~J. and {van der Marel}, Roeland P. and {Alcock}, Charles},
        title = "{Are the Magellanic Clouds on Their First Passage about the Milky Way?}",
      journal = {\apj},
         year = 2007,
        month = oct,
       volume = {668},
       number = {2},
        pages = {949-967},
          doi = {10.1086/521385},
archivePrefix = {arXiv},
       eprint = {astro-ph/0703196},
 primaryClass = {astro-ph},
       adsurl = {https://ui.adsabs.harvard.edu/abs/2007ApJ...668..949B}
}

@ARTICLE{TepperGarcia2019,
       author = {{Tepper-Garc{\'\i}a}, Thor and {Bland-Hawthorn}, Joss and {Pawlowski}, Marcel S. and {Fritz}, Tobias K.},
        title = "{The Magellanic System: the puzzle of the leading gas stream}",
      journal = {\mnras},
         year = 2019,
        month = sep,
       volume = {488},
       number = {1},
        pages = {918-938},
          doi = {10.1093/mnras/stz1659},
archivePrefix = {arXiv},
       eprint = {1901.05636},
 primaryClass = {astro-ph.GA},
       adsurl = {https://ui.adsabs.harvard.edu/abs/2019MNRAS.488..918T}
}

@ARTICLE{Craig2021,
       author = {{Craig}, Peter and {Chakrabarti}, Sukanya and {Baum}, Stefi and {Lewis}, Benjamin T.},
        title = "{A Dynamical Mass Estimate from the Magellanic Stream}",
      journal = {arXiv e-prints},
         year = 2021,
        month = jul,
          eid = {arXiv:2107.09791},
        pages = {arXiv:2107.09791},
          doi = {10.48550/arXiv.2107.09791},
archivePrefix = {arXiv},
       eprint = {2107.09791},
 primaryClass = {astro-ph.GA},
       adsurl = {https://ui.adsabs.harvard.edu/abs/2021arXiv210709791C}
}

@ARTICLE{Ferreira2025,
       author = {{Ferreira}, Bernardo P.~L. and {Santos}, Jr., Jo{\~a}o F.~C. and {Dias}, Bruno and {Maia}, Francisco F.~S. and {Sasi}, Saroon and {Westera}, Pieter and {Oliveira}, Raphael A.~P. and {Souza}, Stefano O. and {Fern{\'a}ndez-Trincado}, Jos{\'e} G. and {Kerber}, Leandro and {Quint}, Bruno and {Sanmartim}, David and {Fraga}, Luciano and {Armond}, Tina and {Viscacha Team}},
        title = "{The VISCACHA Survey XV: Tracing the 3D Structure and Chemical Evolution of the Outskirts of the Large Magellanic Cloud from Star Clusters}",
      journal = {\apj},
         year = 2025,
        month = dec,
       volume = {995},
       number = {2},
          eid = {169},
        pages = {169},
          doi = {10.3847/1538-4357/ae1959},
       adsurl = {https://ui.adsabs.harvard.edu/abs/2025ApJ...995..169F}
}

@ARTICLE{Zhao2000,
       author = {{Zhao}, HongSheng and {Evans}, N. Wyn},
        title = "{The So-called ``Bar'' in the Large Magellanic Cloud}",
      journal = {\apjl},
         year = 2000,
        month = dec,
       volume = {545},
       number = {1},
        pages = {L35-L38},
          doi = {10.1086/317324},
archivePrefix = {arXiv},
       eprint = {astro-ph/0009155},
 primaryClass = {astro-ph},
       adsurl = {https://ui.adsabs.harvard.edu/abs/2000ApJ...545L..35Z}
}

@ARTICLE{Zaritsky2004,
       author = {{Zaritsky}, Dennis},
        title = "{The Case of the Off-Center, Levitating Bar in the Large Magellanic Cloud}",
      journal = {\apjl},
         year = 2004,
        month = oct,
       volume = {614},
       number = {1},
        pages = {L37-L40},
          doi = {10.1086/425312},
archivePrefix = {arXiv},
       eprint = {astro-ph/0408533},
 primaryClass = {astro-ph},
       adsurl = {https://ui.adsabs.harvard.edu/abs/2004ApJ...614L..37Z}
}

@ARTICLE{Dhanush2025,
       author = {{Dhanush}, S.~R. and {Subramaniam}, A. and {Subramanian}, S.},
        title = "{Unraveling the Kinematic and Morphological Evolution of the Small Magellanic Cloud}",
      journal = {\apj},
         year = 2025,
        month = feb,
       volume = {980},
       number = {1},
          eid = {73},
        pages = {73},
          doi = {10.3847/1538-4357/ada55f},
archivePrefix = {arXiv},
       eprint = {2501.00788},
 primaryClass = {astro-ph.GA},
       adsurl = {https://ui.adsabs.harvard.edu/abs/2025ApJ...980...73D}
}

@ARTICLE{Niederhofer2021,
       author = {{Niederhofer}, Florian and {Cioni}, Maria-Rosa L. and {Rubele}, Stefano and {Schmidt}, Thomas and {Diaz}, Jonathan D. and {Matijev{\u{i}}c}, Gal and {Bekki}, Kenji and {Bell}, Cameron P.~M. and {de Grijs}, Richard and {El Youssoufi}, Dalal and {Ivanov}, Valentin D. and {Oliveira}, Joana M. and {Ripepi}, Vincenzo and {Subramanian}, Smitha and {Sun}, Ning-Chen and {van Loon}, Jacco Th},
        title = "{The VMC survey - XLI. Stellar proper motions within the Small Magellanic Cloud}",
      journal = {\mnras},
         year = 2021,
        month = apr,
       volume = {502},
       number = {2},
        pages = {2859-2878},
          doi = {10.1093/mnras/stab206},
archivePrefix = {arXiv},
       eprint = {2101.09099},
 primaryClass = {astro-ph.GA},
       adsurl = {https://ui.adsabs.harvard.edu/abs/2021MNRAS.502.2859N}
}

@ARTICLE{Vijaysree2026,
       author = {{Vijayasree}, S. and {Niederhofer}, F. and {Cioni}, M.-R.~L. and {van Loon}, J. Th. and {Bekki}, K. and {de Grijs}, R. and {Subramanian}, S. and {Kacharov}, N. and {Omkumar}, A.~O. and {Cullinane}, L.~R. and {Ivanov}, V.~D.},
        title = "{The VMC survey -- LV. The coherent expansion of the SMC}",
      journal = {arXiv e-prints},
         year = 2026,
        month = jun,
          eid = {arXiv:2606.03901},
        pages = {arXiv:2606.03901},
archivePrefix = {arXiv},
       eprint = {2606.03901},
 primaryClass = {astro-ph.GA},
       adsurl = {https://ui.adsabs.harvard.edu/abs/2026arXiv260603901V}
}

@ARTICLE{DiTeodoro2019,
       author = {{Di Teodoro}, E.~M. and {McClure-Griffiths}, N.~M. and {Jameson}, K.~E. and {D{\'e}nes}, H. and {Dickey}, John M. and {Stanimirovi{\'c}}, S. and {Staveley-Smith}, L. and {Anderson}, C. and {Bunton}, J.~D. and {Chippendale}, A. and {Lee-Waddell}, K. and {MacLeod}, A. and {Voronkov}, M.~A.},
        title = "{On the dynamics of the Small Magellanic Cloud through high-resolution ASKAP H I observations}",
      journal = {\mnras},
         year = 2019,
        month = feb,
       volume = {483},
       number = {1},
        pages = {392-406},
          doi = {10.1093/mnras/sty3095},
archivePrefix = {arXiv},
       eprint = {1811.09627},
 primaryClass = {astro-ph.GA},
       adsurl = {https://ui.adsabs.harvard.edu/abs/2019MNRAS.483..392D}
}

@ARTICLE{Stanimirovic1999,
       author = {{Stanimirovic}, S. and {Staveley-Smith}, L. and {Dickey}, J.~M. and {Sault}, R.~J. and {Snowden}, S.~L.},
        title = "{The large-scale HI structure of the Small Magellanic Cloud}",
      journal = {\mnras},
         year = 1999,
        month = jan,
       volume = {302},
       number = {3},
        pages = {417-436},
          doi = {10.1046/j.1365-8711.1999.02013.x},
       adsurl = {https://ui.adsabs.harvard.edu/abs/1999MNRAS.302..417S}
}

@ARTICLE{Murray2019,
       author = {{Murray}, Claire E. and {Peek}, J.~E.~G. and {Di Teodoro}, Enrico M. and {McClure-Griffiths}, N.~M. and {Dickey}, John M. and {D{\'e}nes}, Helga},
        title = "{The 3D Kinematics of Gas in the Small Magellanic Cloud}",
      journal = {\apj},
         year = 2019,
        month = dec,
       volume = {887},
       number = {2},
          eid = {267},
        pages = {267},
          doi = {10.3847/1538-4357/ab510f},
archivePrefix = {arXiv},
       eprint = {1910.11283},
 primaryClass = {astro-ph.GA},
       adsurl = {https://ui.adsabs.harvard.edu/abs/2019ApJ...887..267M}
}

@ARTICLE{Nakano2025b,
       author = {{Nakano}, Satoya and {Tachihara}, Kengo},
        title = "{Dual Directional Expansion of Classical Cepheids in the Small Magellanic Cloud Revealed by Gaia Data Release 3}",
      journal = {\apjl},
         year = 2025,
        month = may,
       volume = {985},
       number = {1},
          eid = {L5},
        pages = {L5},
          doi = {10.3847/2041-8213/adce0b},
archivePrefix = {arXiv},
       eprint = {2503.22866},
 primaryClass = {astro-ph.GA},
       adsurl = {https://ui.adsabs.harvard.edu/abs/2025ApJ...985L...5N}
}

@ARTICLE{Dolfi2026,
       author = {{Dolfi}, Arianna and {G{\'o}mez}, Facundo A. and {Bieri}, Rebekka and {Fragkoudi}, Francesca and {Grand}, Robert J.~J. and {Monachesi}, Antonela and {Pakmor}, Ruediger and {van de Voort}, Freeke},
        title = "{Coupling between stellar and HI lopsidedness in Milky Way-type galaxies from the Auriga Superstars cosmological simulations}",
      journal = {arXiv e-prints},
         year = 2026,
        month = apr,
          eid = {arXiv:2604.27057},
        pages = {arXiv:2604.27057},
          doi = {10.48550/arXiv.2604.27057},
archivePrefix = {arXiv},
       eprint = {2604.27057},
 primaryClass = {astro-ph.GA},
       adsurl = {https://ui.adsabs.harvard.edu/abs/2026arXiv260427057D}
}

@ARTICLE{Purcell2011,
       author = {{Purcell}, Chris W. and {Bullock}, James S. and {Tollerud}, Erik J. and {Rocha}, Miguel and {Chakrabarti}, Sukanya},
        title = "{The Sagittarius impact as an architect of spirality and outer rings in the Milky Way}",
      journal = {\nat},
         year = 2011,
        month = sep,
       volume = {477},
       number = {7364},
        pages = {301-303},
          doi = {10.1038/nature10417},
archivePrefix = {arXiv},
       eprint = {1109.2918},
 primaryClass = {astro-ph.GA},
       adsurl = {https://ui.adsabs.harvard.edu/abs/2011Natur.477..301P}
}

@ARTICLE{Berentzen2003,
       author = {{Berentzen}, I. and {Athanassoula}, E. and {Heller}, C.~H. and {Fricke}, K.~J.},
        title = "{Numerical simulations of interacting gas-rich barred galaxies: vertical impact of small companions}",
      journal = {\mnras},
         year = 2003,
        month = may,
       volume = {341},
       number = {1},
        pages = {343-360},
          doi = {10.1046/j.1365-8711.2003.06417.x},
archivePrefix = {arXiv},
       eprint = {astro-ph/0301300},
 primaryClass = {astro-ph},
       adsurl = {https://ui.adsabs.harvard.edu/abs/2003MNRAS.341..343B}
}

@ARTICLE{Weinberg2023,
       author = {{Weinberg}, Martin D.},
        title = "{New dipole instabilities in spherical stellar systems}",
      journal = {\mnras},
         year = 2023,
        month = nov,
       volume = {525},
       number = {4},
        pages = {4962-4975},
          doi = {10.1093/mnras/stad2591},
archivePrefix = {arXiv},
       eprint = {2209.06846},
 primaryClass = {astro-ph.GA},
       adsurl = {https://ui.adsabs.harvard.edu/abs/2023MNRAS.525.4962W}
}

@ARTICLE{Petersen2022,
       author = {{Petersen}, Michael S. and {Weinberg}, Martin D. and {Katz}, Neal},
        title = "{EXP: N-body integration using basis function expansions}",
      journal = {\mnras},
         year = 2022,
        month = mar,
       volume = {510},
       number = {4},
        pages = {6201-6217},
          doi = {10.1093/mnras/stab3639},
archivePrefix = {arXiv},
       eprint = {2104.14577},
 primaryClass = {astro-ph.GA},
       adsurl = {https://ui.adsabs.harvard.edu/abs/2022MNRAS.510.6201P}
}

@software{snapAnalysis,
  author       = {Hayden Foote},
  title        = {hfoote/snapAnalysis: Version 0.1.1},
  month        = may,
  year         = 2026,
  publisher    = {Zenodo},
  version      = {v0.1.1},
  doi          = {10.5281/zenodo.20386578},
  url          = {https://doi.org/10.5281/zenodo.20386578},
}

@ARTICLE{Belokurov2019,
       author = {{Belokurov}, Vasily A. and {Erkal}, Denis},
        title = "{Clouds in arms}",
      journal = {\mnras},
         year = 2019,
        month = jan,
       volume = {482},
       number = {1},
        pages = {L9-L13},
          doi = {10.1093/mnrasl/sly178},
archivePrefix = {arXiv},
       eprint = {1808.00462},
 primaryClass = {astro-ph.GA},
       adsurl = {https://ui.adsabs.harvard.edu/abs/2019MNRAS.482L...9B}
}

@software{combine_ics,
  author       = {Himansh Rathore},
  title        = {combine\_ics code for combining N-body simulation
                   snapshots of Gadget-like HDF5 format
                  },
  month        = aug,
  year         = 2026,
  publisher    = {Zenodo},
  version      = {v1.0.0-beta},
  doi          = {10.5281/zenodo.22156368},
  url          = {https://doi.org/10.5281/zenodo.22156368},
}
\bibliographystyle{aasjournalv7}

\end{document}